\documentclass[prc, floatfix, twoside, nofootinbib,showpacs]{revtex4-1}
\usepackage[bookmarks,pdfhighlight=/O,colorlinks=false,pdfstartview=FitH]{hyperref}
\usepackage{amsmath,amssymb,mathrsfs,slashed,bm}
\usepackage[dvipsnames]{xcolor}
\usepackage{graphicx}
\usepackage{epstopdf}
\usepackage{natbib}

\newcommand{\gsim}{\gtrsim}
\newcommand{\ii}{\mathrm{i}}
\newcommand{\dd}{\mathrm{d}}
\newcommand{\D}{\mathrm{D}}
\newcommand{\ee}{\mathrm{e}}
\newcommand{\MeV}{\mathrm{MeV}}
 \newcommand{\GeV}{\mathrm{GeV}}

\newcommand{\AG}{\mathrm{AGeV}}
\newcommand{\fm}{\mathrm{fm}}
\newcommand{\bvec}[1]{\ensuremath{\boldsymbol{#1}}}
\newcommand{\erw}[1]{\ensuremath { %
    \left \langle {#1} \right \rangle}}

\newcommand{\Lag}{\ensuremath{\mathscr{L}}}
\newcommand{\fslash}{\slashed}

\newcommand{\tildeint}[1]{\ensuremath{\int_{\R^3} \frac{\mathrm{d}^{3} #1}{2
      E(#1) \, (2\pi)^{3}}}}
\newcommand{\R}{\ensuremath{\mathbb{R}}}

\def\x{{\boldsymbol x}}

\begin{document}

\preprint{}

\title{Elliptic flow and $R_{AA}$ of $\D$ mesons at FAIR comparing the\\UrQMD hybrid model and the coarse-graining approach\\}

\author{Gabriele Inghirami$^{1,2,3,4,5,6}$}
\author{Hendrik van Hees$^{1,2}$} 
\author{\\Stephan Endres$^{1,2}$}
\author{Juan M. Torres-Rincon$^{7}$}
\author{Marcus Bleicher$^{1,2,3,4}$}

\affiliation{
  $^{1}\,$Frankfurt Institute for Advanced Studies
  (FIAS),Ruth-Moufang-Str. 1, 60438 Frankfurt am Main, Germany }
\affiliation{
  $^{2}\,$Institut f\"ur Theoretische Physik, Johann Wolfgang
  Goethe-Universit\"at, Max-von-Laue-Str. 1, 60438 Frankfurt am Main,
  Germany }
\affiliation{
	$^{3}\,$GSI Helmholtzzentrum f\"ur Schwerionenforschung GmbH,
	Planckstra{\ss}e 1,
	64291 Darmstadt, Germany}
\affiliation{
	$^{4}\,$John von Neumann Institute for Computing,
	Forschungszentrum J\"ulich,
	52425 J\"ulich, Germany}
\affiliation{
	$^{5}\,$Current address and affiliation: University of Jyvaskyla, Department of Physics, P.O. Box 35, FI-40014 University of Jyvaskyla, Finland}
\affiliation{
	$^{6}\,$Current second affiliation: Helsinki Institute of Physics, P.O. Box 64, FI-00014 University of Helsinki, Finland}
\affiliation{
	$^{7}\,$Department of Physics and Astronomy,
	Stony Brook University, Stony Brook,
	New York 11794, USA
	}

\date{\today}

\begin{abstract}
	
We present a study of the elliptic flow and $R_{AA}$ of $\D$ and
$\bar{\D}$ mesons in Au+Au collisions at FAIR energies. We propagate the
charm quarks and the $\D$ mesons following a previously applied Langevin
dynamics. The evolution of the background medium is modeled in two
different ways: (I) we use the UrQMD hydrodynamics + Boltzmann transport
hybrid approach including a phase transition to QGP and (II) with the
coarse-graining approach employing also an equation of state with
QGP. The latter approach has previously been used to describe di-lepton
data at various energies very successfully. This comparison allows us to
explore the effects of partial thermalization and viscous effects on the
charm propagation. We explore the centrality dependencies of the
collisions, the variation of the decoupling temperature and various
hadronization parameters. We find that the initial partonic phase is
responsible for the creation of most of the $\D/\bar{\D}$ mesons elliptic
flow and that the subsequent hadronic interactions seem to play only a
minor role. This indicates that $\D/\bar{\D}$ mesons elliptic flow is a
smoking gun for a partonic phase at FAIR energies. However, the results
suggest that the magnitude and the details of the elliptic flow strongly
depend on the dynamics of the medium and on the hadronization procedure,
which is related to the medium properties as well. Therefore, even at
FAIR energies the charm quark might constitute a very useful tool to
probe the Quark-Gluon Plasma and investigate its physics.

\end{abstract}


\maketitle

\section{Introduction} 

Heavy quarks represent an excellent method to probe the hot and dense
medium which is supposed to form in heavy ion
collisions\cite{Rapp:2009my}. Their mass $M_{\mathrm{HF}}$ is much larger than
$\Lambda_{\mathrm{QCD}}$ and $T_{\mathrm{QGP}}$, therefore we can use
perturbative QCD (pQCD)\cite{Brock:1994er} to model their production as a hard process
\cite{Beraudo:2014iva} which happens mostly during the initial collision
processes and almost negligibly by thermal production, except at early
times at LHC energies\cite{Levai:1997bi}. Once formed, since the strong
interaction conserves the flavour quantum number, the heavy quarks
maintain their identity until the hadrons they form decay by weak (or in
the case of the $\mathrm{J}/\Psi$ by electro-magnetic)
interaction. Moreover, since the energy loss in the medium due to
multiple scattering and induced gluon bremsstrahlung depends on the mass
of the propagating
particle\cite{Gyulassy:1993hr,Wang:1994fx,Zhang:2003wk}, heavy quarks
are less affected than light quarks by the interactions with the medium
and they convey information about the whole system evolution. At high
transverse momenta the interest is oriented toward studying the opacity
of the medium through the particle suppression in the high
$p_{\mathrm{T}}$ range, as observed in the experimental nuclear
modification factor\cite{ALICE:2012ab,Chatrchyan:2012np}. In the
low-$p_{\mathrm{T}}$ range the focus is on in-medium hadronization and
thermalization\cite{vanHees:2005wb,Zhu:2006er}, reached by charm quarks
at LHC energies, as theoretical considerations
suggest\cite{Moore:2004tg,Scardina:2017ipo,Graf:2018lok} and as the
observed experimental elliptic flow
proves\cite{Abelev:2013lca}. Numerical simulations, which are an
essential tool to connect theory with experiments, are continuously
improved to provide a consistent, realistic description of the
heavy-quark propagation\cite{Nahrgang:2016lst}, adopting many different
approaches\cite{Cao:2013ita,Song:2015sfa,Djordjevic:2016dyp,Rapp:2018qla} and
investigating also small systems\cite{Beraudo:2016pqv}. 

In this paper we study the elliptic flow and the $R_{AA}$\footnote{We
  use a non-standard definition of $R_{AA}$ as the ratio between the
  normalized transverse momentum distribution of $\D$ mesons in ion-ion
  collisions and the normalized transverse momentum distribution in
  proton-proton collisions. By this we take out the unknown yields of the
  $\D/\bar{\D}$ mesons in pp and AA collisions at this low
  energy. Moreover, we call collectively $\D$ mesons the $\D^+(c\bar{d})$
  and the $\D^0(c\bar{u})$, we call $\bar{\D}$ the $\D^-(\bar{c}d)$ and
  the $\bar{\D}^0(\bar{c}u)$.} of $\D$ and $\bar{\D}$ mesons in Au+Au
collisions at $\sqrt{s_{NN}}\,\simeq 7 \, \GeV$, a collision energy in
the range of the upcoming FAIR facility\cite{Friman:2011zz}, but also available at RHIC, within the Beam Energy scan program\cite{Odyniec:2013kna}, and at NICA\cite{Batyuk:2016yjp}. We adopt a
Langevin propagation model, implicitly assuming that the heavy quark
momentum transfer is much smaller than for the light partons, an
approximation that at low collision energies should work reasonably
well, while at RHIC and LHC energies it is really consistent only for
bottom quarks\cite{Das:2013aga}. After a brief introduction to the bulk
evolution models that we use, i.e. the UrQMD hybrid
model\cite{Bass:1998ca, Bleicher:1999xi, Petersen:2008dd} and the coarse
graining approach\cite{Endres:2014zua}, we shortly review the formalism
of the relativistic Langevin propagation, then we provide a basic
overview of how we compute the transport coefficients, both for charm
quarks and $\D$ mesons. After showing and commenting the results of the
simulations, we discuss how we might improve them.

\section{Models of the medium bulk evolution}
\subsection{The UrQMD hybrid model} 

The primary bulk evolution of the medium is simulated using the
hydrodynamics + Boltzmann setup the \href{https://urqmd.org}{UrQMD}
hybrid model\cite{Bass:1998ca, Bleicher:1999xi, Petersen:2008dd},
adopting fluctuating initial conditions\cite{Bleicher:1998wd}. In the
initial stage, UrQMD follows the elastic and inelastic collisions
between nucleons, including color-flux-tube excitation and fragmentation
processes. The hydrodynamical phase starts when the two Lorentz
contracted nuclei have completely passed through each other, at
$t=(2R)/(\sqrt{\gamma^2_{\text{CM}}-1})$, where $R$ is the radius of the
two nuclei and $\gamma_{\text{CM}}$ is their Lorentz $\gamma$ factor in the
center of mass frame\cite{Steinheimer:2009nn}. The initial momentum,
energy and baryon density distributions are created by summing the
individual particle distributions, assumed to be three-dimensional
Gaussians, like, e.g. for the energy
density\cite{Steinheimer:2007iy,Petersen:2010zt}:
\begin{equation} \varepsilon_{x,y,z}=
\left(\dfrac{1}{2\pi}\right)^{\frac{3}{2}}\dfrac{\gamma_z
E_p}{\sigma^3}\exp \left
[-\frac{(x-x_p)^2+(y-y_p)^2+\gamma_z^2(z-z_p)^2}{2\sigma^2} \right],
\label{enedens}
\end{equation} 
where $E_p$ and $x_p,y_p,z_p$ are the energy and the coordinates of the
particle in the computational frame, $\sigma$ is the width of the
Gaussian (by default, $1\,\fm$) and $\gamma_z$ is the Lorentz $\gamma$
factor to take into account the Lorentz contraction in the beam
direction.  UrQMD computes the fluid evolution by solving the
differential equations which describe the conservation of total
energy and net-baryon number, i.e.,
\begin{equation}
\partial_{\mu} T^{\mu\nu}=0, \qquad \partial_{\mu}N^{\mu}=0,
\end{equation} 
where $T^{\mu\nu}$ is the energy-momentum tensor and $N^{\mu}$ the
baryon four-current. The hydro evolution is based on the SHASTA (SHarp
And Smooth Transport Algorithm)
algorithm\cite{Rischke:1995ir,Rischke:1995mt}; it exploits the chiral
equation of state (EoS)\cite{Steinheimer:2011ea} and assumes local
thermal equilibrium, i.e. it does not take into account dissipative
effects\cite{Heinz:2005bw,Koide:2006ef,Schenke:2010nt,Karpenko:2013wva,DelZanna:2013eua}. The
hydrodynamical simulation is stopped when the maximum of the energy
density on the grid becomes smaller than a certain value chosen for
\emph{particlization}. In the present work, we stopped the simulations
at this point. Nevertheless in the full UrQMD hybrid model a rather
advanced method to determine the freeze-out hypersurface is
employed\cite{Huovinen:2012is} and the energy density distribution is
converted back to particles\cite{Bass:2000ib} through the
Cooper-Frye\cite{Cooper:1974mv} equation. Afterwards, the hadrons
continue to scatter and strongly decay until no more interactions take
place. To partially take into account this hadronic phase, we adopted a
freeze-out temperature slightly below the standard value.

\subsection{The UrQMD coarse-graining approach} 

The hybrid model uses a microscopic description only for the very
initial collisions and the final state interactions after the
hydrodynamic phase, but it is also possible to extract macroscopic
quantities from an underlying microscopic simulation during the whole
collision evolution, as realized within the coarse-graining approach. It
was first proposed in ref.~\cite{Huovinen:2002im} and has proven to
account for the reaction dynamics and the production of electromagnetic
probes successfully from SIS\,18 to LHC energies \cite{Endres:2014zua,
  Endres:2015fna, Endres:2015egk, Endres:2016tkg,
  Staudenmaier:2017vtq}. In this approach an ensemble of collision
events simulated with a transport model (here: UrQMD in cascade mode) is
put on a grid of small space-time cells. By averaging over a sufficient
number of events the hadronic distribution function
$f(\textbf{\textit{x}},\textbf{\textit{p}},t)$ obtains a smooth form as
\begin{equation}
f(\textbf{\textit{x}},\textbf{\textit{p}},t)=\left\langle \sum_{h}
\delta^{(3)}\left(\textbf{\textit{x}}-\textbf{\textit{x}}_{h}(t)\right)
\delta^{(3)}\left(\textbf{\textit{p}}-\textbf{\textit{p}}_{h}(t)\right)\right\rangle,
\end{equation} 
where the angle brackets denote the ensemble average. It is then
possible to extract the energy momentum tensor and the baryon current
locally in space and time, i.e.~for each cell of the grid. These
quantities are given by the relations
\begin{eqnarray} T^{\mu\nu}(\x,t)&=&\frac{1}{\Delta V}\left\langle
\sum\limits_{i=1}^{N_{h} \in \Delta V} \frac{p^{\mu}_{i}
p^{\nu}_{i}}{p^{0}_{i}}\right\rangle, \\
j^{\mu}_{\mathrm{B}}(\x,t)&=&\frac{1}{\mathrm{\Delta} V}\left\langle
\sum\limits_{i=1}^{N_{\mathrm{B}/\bar{\mathrm{B}}} \in \Delta
V}\pm\frac{p^{\mu}_{i}}{p^{0}_{i}}\right\rangle.
\label{eq:CGtmunubcurr}
\end{eqnarray} 
Here $\Delta V$ denotes the cell volume and the sums are taken over the
numbers of all hadrons $N_{h}$ or (anti-)baryons
$N_{\mathrm{B}/\bar{\mathrm{B}}}$, respectively. In addition to the three components of the fluid velocity (using Eckart's frame definition\cite{Eckart:1940te}) from $j^{\mu}$ in Eq. (\ref{eq:CGtmunubcurr}), the energy and the baryon
densities in the cells can be obtained from the local rest-frame (LRF)
values as
\begin{eqnarray} \varepsilon &=& T^{00}_{\mathrm{LRF}}, \\
\rho_{\mathrm{B}} &=& j_{\mathrm{B,\,LRF}}^{0}.
\end{eqnarray} 
Finally, by applying an EoS the local temperature $T$ and baryon
chemical potential $\mu_{\mathrm{B}}$ are calculated from the energy density and the baryon density.
For the present study a hadron gas EoS \cite{Zschiesche:2002zr} with the same degrees of
freedom as in the UrQMD transport model is applied, providing
consistency with the underlying microscopic description which is purely
hadronic. However, note that this may no longer be a fully valid
picture, if the temperature in the fireball exceeds the critical
temperature $T_{c}$, for which a phase transition to a quark-gluon
plasma is expected. But because the maximum temperatures in collisions
at FAIR energies are not found to be significantly above $T_{c}$, the
differences compared to a full treatment of the phase transition by
using an EoS\cite{He:2011zx,Borsanyi:2010cj} fitted to lattice QCD results in the limit of $\mu_{B}=0$ are rather small (see
comparison in ref.~\cite{Endres:2014zua}).

The determination of thermodynamic quantities for each cell via the
coarse-graining approach requires---as in all macroscopic
descriptions---the assumption of kinetic (and chemical) equilibrium, but
in the underlying microscopic transport model these conditions are not
always completely fulfilled. Therefore, deviations from the equilibrium
state need to be considered. For the present case, the most relevant
non-equilibrium effect shows up in the form of kinetic anisotropies,
especially in the very early stages of the collision, due to the strong
compression of the nuclei in longitudinal direction. Here, this
non-equilibrium effect is eliminated by calculating the ``effective'',
i.e. thermalized, energy density using the framework given in
ref.~\cite{Florkowski:2010cf}.\\
For the sake of clarity, we stress that the UrQMD/coarse-graining approach allows the computation of the same physical quantities as in the UrQMD/hydro model, namely the three components of the fluid velocity, the energy density, the baryon density and, by introducing an EoS, also the temperature and the baryon chemical potential. Therefore the data coming from the UrQMD/coarse-graining approach can be used as a replacement of the UrQMD/hybrid-approach, providing an alternative description of the evolution of the medium based on transport models. However, there is an important difference in the utilization of the two approaches: while in the UrQMD/hybrid model we can perform the propagation of the heavy quarks and the computation of the medium dynamics at the same time, in the UrQMD/coarse-graining model the dynamical evolution of the backround medium is calculated in advance by averaging many events and saved in a file, containing the fluid evolution data at fixed intervals of time. This means that, in the UrQMD/coarse-graining approach, the background medium evolution remains the same for all events in a certain centrality class. Nevertheless, we still have fluctuations in the final results due to the different initial positions and momenta of the heavy quarks, which vary event by event, and to their stochastic equations of motion. In a previous work\cite{Lang:2012cx} we found that the nuclear modification factor and the elliptic flow of $D$ mesons seem to not change appreciably if, instead of averaging the final results of many events, we average the medium evolution, provided that the numerical sample of particles is the same and in the limit of the approximations adopted in our model, described in Sect. (\ref{implementation}). Therefore, we consider our approach reasonable. For the present study, to compute the background medium evolution, we averaged $1.44\cdot 10^5$ events for reactions with impact parameter $b=3\,\fm$ and $2.64\cdot10^5$ events for reactions with $b=7\,\fm$.
 
\section{The Relativistic Langevin propagation of the charm quarks}

Since the mass of the charm quarks is much larger than the mass of up,
down and even strange quarks and since it is also much larger than the
typical temperatures of the system, it is reasonable to assume that each
collision with other particles will change the momenta of the charm
quarks only by a small amount. Under these conditions, the Boltzmann
equation can be approximated by a Fokker-Planck equation, which, in
turn, can be recasted as an equivalent stochastic Langevin
equation\cite{Svet88,GolamMustafa:1997id,vanHees:2004gq,Moore:2004tg,vanHees:2005wb,vanHees:2007me,Gossiaux:2008jv}.

When dealing with relativistic speeds, we can formulate the Langevin
process as:
\begin{equation}
\begin{split}
\label{lang.1} \dd x_j &= \frac{p_j}{E} \dd t, \\ \dd p_j &= -\Gamma p_j
\dd t + \sqrt{\dd t} C_{jk} \rho_k.
\end{split}
\end{equation} 
In Eq.\ (\ref{lang.1}) $E=\sqrt{m^2+\bvec{p}^2}$, $\dd t$ is the
advancement time step, $\dd x_j$ and $\dd p_j$ are the variations of
coordinates and momentum in each time-step, the $\rho_k$ are random
variables distributed according to a normalized Gaussian distribution,
$\Gamma$ and $C_{jk}$ are the drag or friction coefficient and the
covariance matrix of the fluctuating force respectively, both defined in
the local rest frame of the fluid and depending on
$(t,\bvec{x},\bvec{p})$. These parameters of the Langevin process in
Eq.\ (\ref{lang.1}) are related to the drag and diffusion coefficients
$A$, $B_0$ and $B_1$ for an isotropic medium by
\begin{alignat}{2}
\label{lang.7} A p_j &=\Gamma p_j - \xi C_{lk} \frac{\partial
C_{jk}}{\partial p_l}, \\
\label{lang.8} C_{jk} &=\sqrt{2 B_0} P_{jk}^{\perp} + \sqrt{2 B_1}
P_{jk}^{\parallel}, \\
\label{lang.9} \text{with} \quad P_{jk}^{\parallel} &=\frac{p_j
p_k}{\bvec{p}^2}, \quad P_{jk}^{\perp}=\delta_{jk}-\frac{p_j
p_k}{\bvec{p}^2}.
\end{alignat} 
It is known that, modeling the medium in global thermal equilibrium,
i.e. in a homogeneous static background medium, the stationary
equilibrium limit should be a Boltzmann-J\"uttner distribution,
\begin{equation}
\label{lang.10} f_Q^{(\text{eq})}(\bvec{p})=\exp \left(-\frac{E}{T}
\right ).
\end{equation} 
Therefore it is possible to tune the drag coefficient in
Eq.\ (\ref{lang.8}) by choosing the longitudinal diffusion coefficient
$B_1$ such as to satisfy this asymptotic equilibration condition
\cite{He2013PhysRevE.88.032138}, leading to dissipation-fluctuation
relations between this diffusion coefficient and the drag coefficient
\cite{Moore:2004tg,Rapp:2009my}. Essentially, if the
dissipation-fluctuation relation,
\begin{equation}
\label{lang.11} \Gamma(E) E T-D(E)+ T(1-\xi) D'(E)=0,\quad
\textrm{with}\, D(E)=B_1
\end{equation} 
is fulfilled, Eq.\ (\ref{lang.10}) becomes a solution of the
corresponding stationary Fokker-Planck equation. In the post-point Ito
realization \cite{deGroot80,Lang:2012cx} $\xi=1$, and this choice allows
to reduce Eq.\ (\ref{lang.11}) to
\begin{equation}
\label{lang.12} D(E)=\Gamma(E) E T,
\end{equation} 
so that, after $\Gamma$ is computed from underlying microscopic models
for heavy-quark scattering with light quarks and gluons introduced in
the next section, the longitudinal diffusion coefficient $B_1$ is given
by
\begin{equation}
\label{lang.13} B_1=\Gamma E T.
\end{equation}
We remind that in the derivation of the Langevin process we assumed to
be in the rest frame of the background medium, therefore, when this
procedure is applied in a dynamically evolving medium, it is necessary
to first perform a boost to the comoving frame of the medium and then,
after performing the Langevin propagation, to perform another boost back
to the computational frame.

\section{Drag and diffusion coefficients}
\subsection{Drag and diffusion coefficients for charm quarks}

\begin{figure*}[h]
	\begin{minipage}[b]{0.48\textwidth}
		\includegraphics[width=1\textwidth]{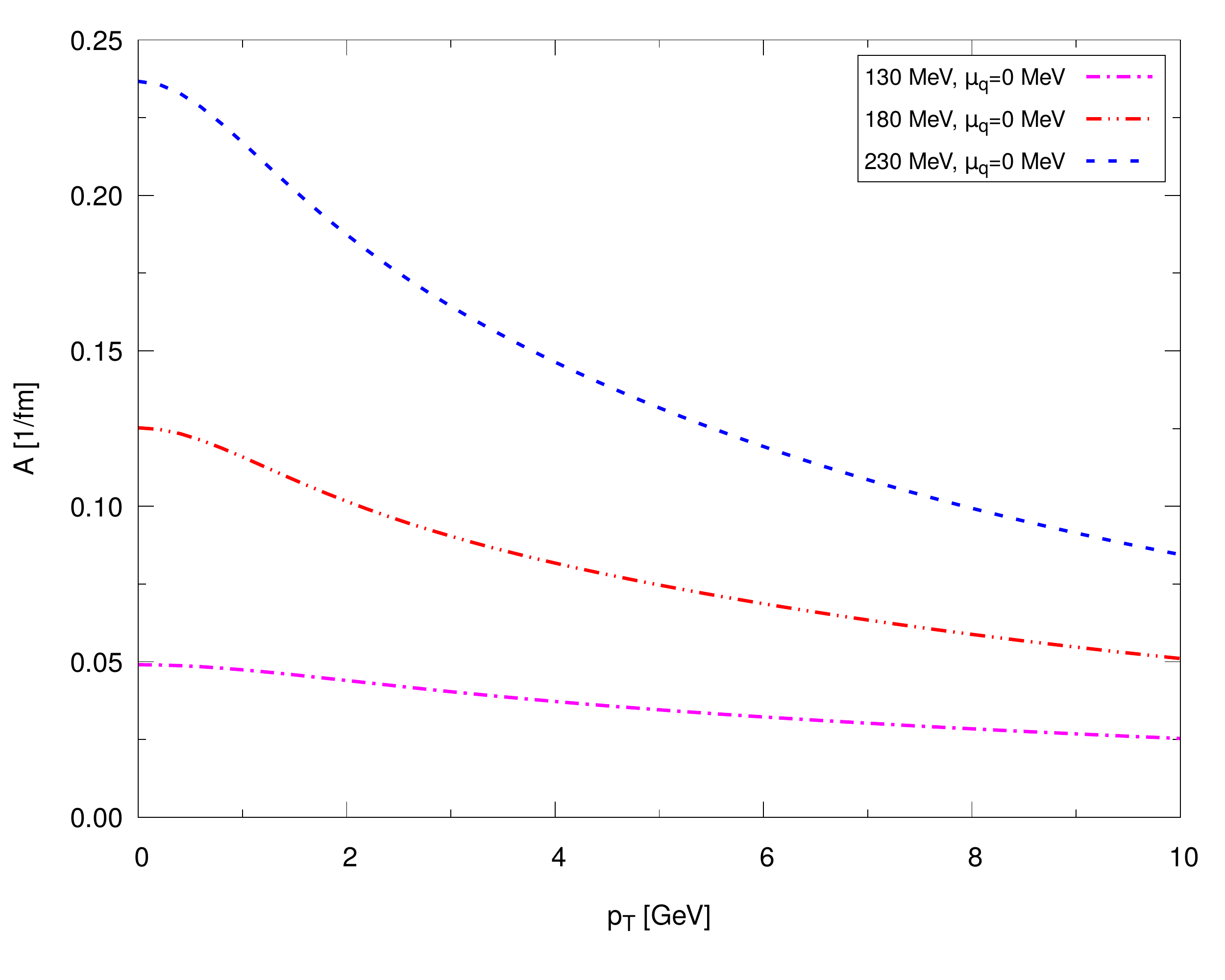}
	\end{minipage} \hspace{3mm}
	\begin{minipage}[b]{0.48\textwidth}
		\includegraphics[width=1\textwidth]{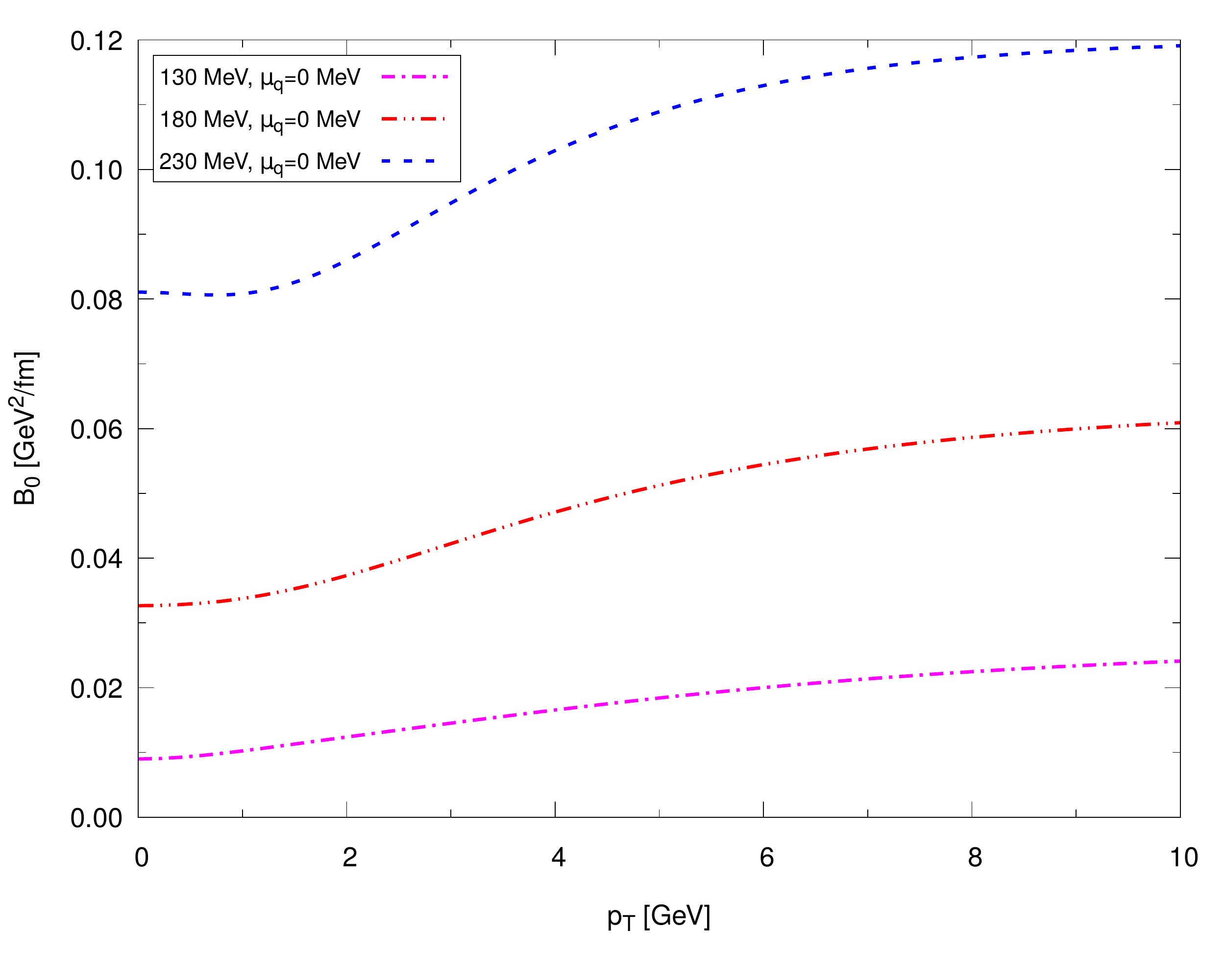}
	\end{minipage}
	\caption{(Color online) Drag (left) and diffusion (right)
          coefficients in the resonance model for charm quarks at
          different temperatures.}
	\label{coeff-quarks}
\end{figure*}

\begin{figure*}[h]
	\begin{minipage}[b]{0.48\textwidth}
		\includegraphics[width=1\textwidth]{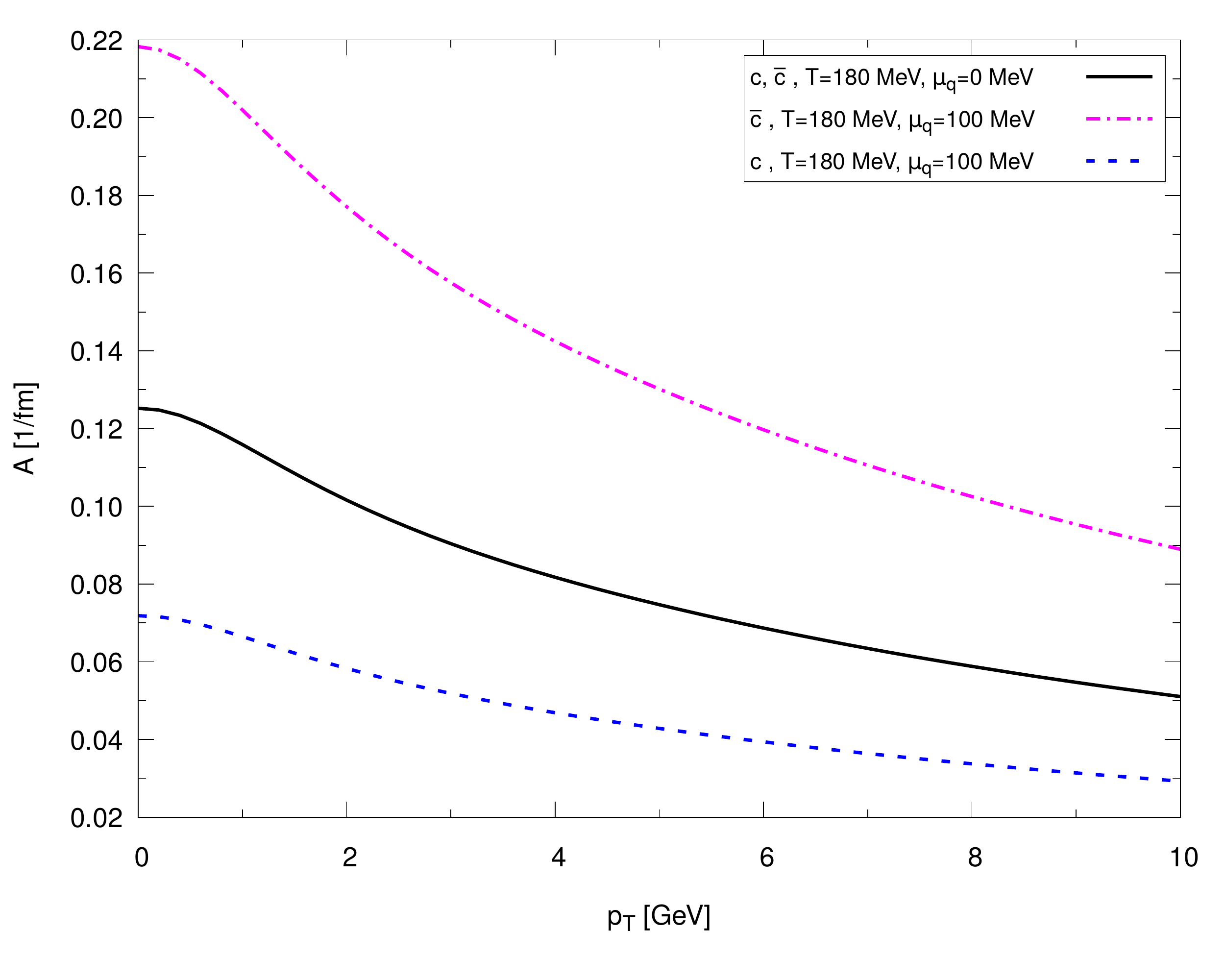}
	\end{minipage} \hspace{3mm}
	\begin{minipage}[b]{0.48\textwidth}
		\includegraphics[width=1\textwidth]{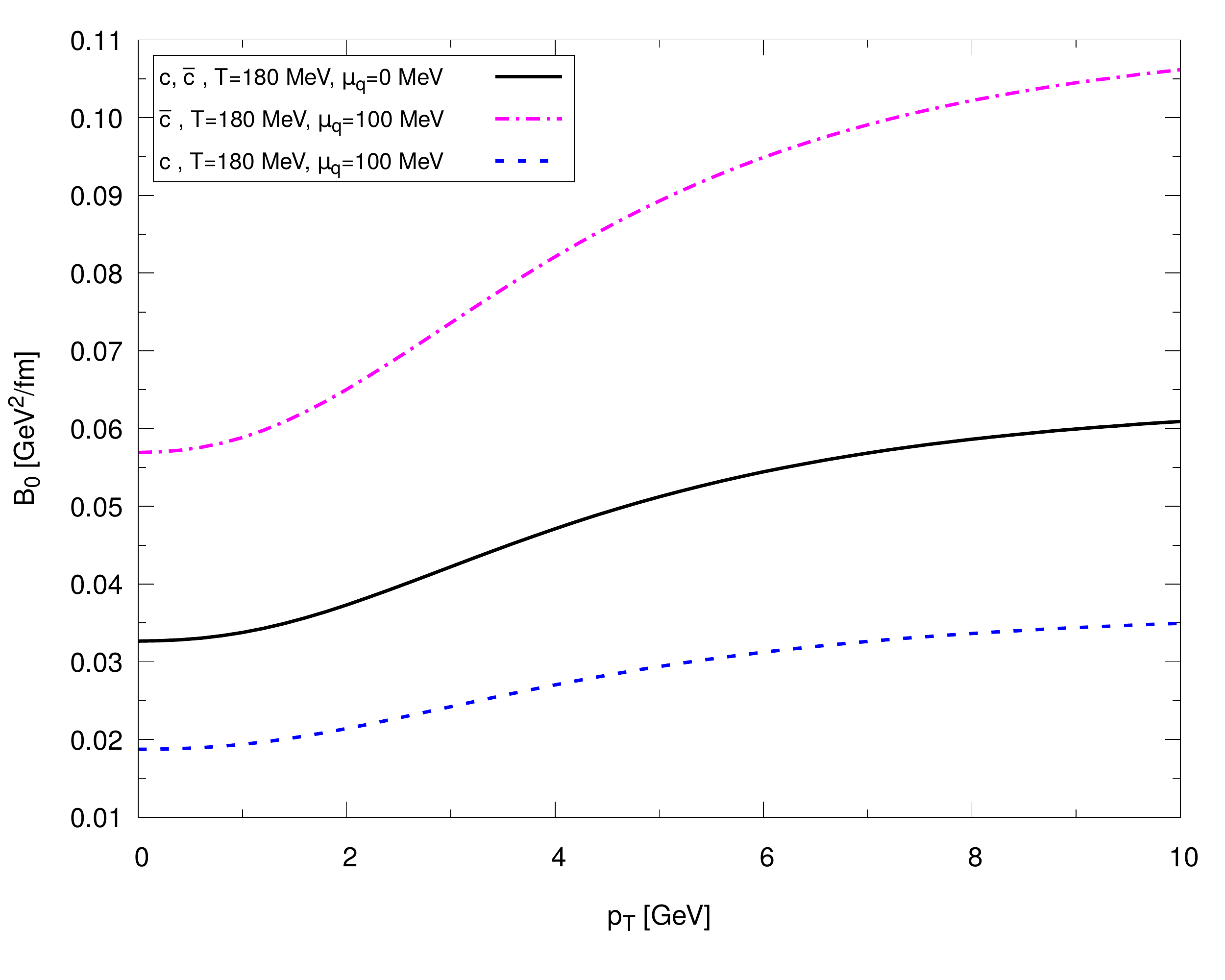}
	\end{minipage}
	\caption{(Color online) Drag (left) and diffusion (right)
          coefficients in the resonance model for charm quarks at
          different temperatures taking into account also a fugacity
          factor $\ee^{(-\mu/T)}$ for charm and $\ee^{(\mu/T)}$ for
          anti-charm quarks.}
	\label{coeff-quarks-fugacity}
\end{figure*} 
In this work the drag and diffusion coefficients to perform the Langevin
propagation of charm quarks are obtained from a resonance model, in
which the existence of $\D$ mesons in the QGP phase is assumed.  The
resonance model is based on heavy-quark effective theory (HQET) and
chiral symmetry in the light-quark sector \cite{vanHees:2004gq}. In this
model we assume the existence of open-heavy-flavor meson
resonances like the $\D$ mesons, an assumption supported by the finding
in lattice-QCD calculations that hadron-like bound states and/or
resonances might survive the phase transition in both the light-quark
sector (e.g., $\rho$ mesons) and heavy quarkonia (e.g.,
$\mathrm{J}/\psi$).

The heavy-light quark resonance model \cite{vanHees:2004gq} is based on
the Lagrangian:
\begin{equation} 
\begin{split}
  \Lag_{Dcq} = &\Lag_D^0 + \Lag_{c,q}^0 - \ii G_S \left( \bar q \Phi_0^*
    \frac{1+\fslash{v}}{2} c - \bar q \gamma^5 \Phi
    \frac{1+\fslash{v}}{2} c + \text{h.c.} \right) \\
  &- G_V \left( \bar q \gamma^{\mu} \Phi_{\mu}^* \frac{1+\fslash{v}}{2}
    c - \bar q \gamma^5 \gamma^{\mu} \Phi_{1\mu} \frac{1+\fslash{v}}{2}
    c + \text{h.c.}  \right),
\label{L_hq-eff}
\end{split}
\end{equation} 
where $v$ is the heavy-quark four-velocity. The free part
of the Lagrangian is given by
\begin{equation} 
\begin{split}
  \Lag_{c,q}^0 &= \bar{c}(\ii \fslash{\partial}-m_c) c+\bar{q} \, \ii
  \fslash{\partial} q, \\ \Lag_D^0 &= (\partial_{\mu}
  \Phi^{\dagger})(\partial^{\mu} \Phi) + (\partial_{\mu}
  {\Phi_0}^{*\dagger})(\partial^{\mu} \Phi_0^*) -m_S^2(\Phi^{\dagger}
  \Phi+\Phi_0^{*\dagger} \Phi_0^*) \\
  & \quad -\frac{1}{2} (\Phi_{\mu \nu}^{*\dagger} \Phi^{*\mu \nu} +
  \Phi_{1 \mu \nu}^{\dagger} \Phi_1^{\mu \nu}) + m_V^2
  (\Phi_{\mu}^{*\dagger} \Phi^{*\mu} + \Phi_{1 \mu}^{\dagger}
  \Phi_1^{\mu}),
\label{L_0-eff}
\end{split}
\end{equation} 
in which $\Phi$ and $\Phi_0^*$ are pseudo-scalar and scalar meson fields
(corresponding to $\D$ and $\D_0^*$ mesons). Because of the chiral
symmetry restoration in the QGP phase, the existence of mass degenerate
chiral-partner states is also assumed. Further from heavy-quark
effective symmetry it is expected to have spin independence for both the
coupling constants, $G_S=G_V$, and the masses, $m_S=m_V$. For the
strange-quark states we consider only the vector and pseudo-scalar
states ($D_s^*$ and $D_s$, respectively).

The $\D$-meson propagators are dressed with the corresponding one-loop self
energy. Assuming charm-quark masses of $m_c=1.5\;\mathrm{GeV}$, we
adjust the masses of the physical $\D$-meson-like resonances to
$m_{\mathrm{D}}=2 \; \GeV$, in approximate agreement with the $T$-matrix models of
heavy-light quark interactions in
\cite{Blaschke:2002ws,Blaschke:2003ji}. The strong-coupling constant is
chosen as $\alpha_s=g^2/(4 \pi)=0.4$, such as to obtain resonance widths
of $\Gamma_{D} =0.75 \; \GeV$.

We use these propagators to compute the elastic $Qq$- and
$Q\overline{q}$-scattering matrix elements, which are then used in
Eq.\ (\ref{2.1.10b}) and (\ref{2.1.10c}) for the evaluation of the
pertinent drag and diffusion coefficients for the heavy quarks. It turns
out that particularly the $s$-channel processes through a $\D$-meson
like resonance provide a large efficiency for heavy-quark diffusion
compared to the pQCD cross sections for the same elastic scattering
processes, resulting in charm-quark equilibration times
$\tau_{\text{eq}}^c =2$-$10 \; \fm/c$.

The relation of elastic heavy-quark-scattering matrix elements with the
drag and diffusion coefficients in the Langevin approach is given by
integrals of the form
\begin{equation}
\begin{split}
\label{2.1.10b} 
\erw{X(\bvec{p}')} = \frac{1}{2 \omega_{\bvec{p}}} &
\tildeint{\bvec{q}} \tildeint{\bvec{p}'}\tildeint{\bvec{q}'}\\
& \times \frac{1}{\gamma_Q} \sum_{g,q} |\mathcal{M}|^2 (2 \pi)^4
\delta^{(4)}(p+q-p'-q') f_{q,g}(\bvec{q}) X(\bvec{p}') \ ,
\end{split}
\end{equation} 
where the invariant scattering-matrix elements are
\begin{equation}
\begin{split} \sum  |\mathcal{M}|^2 = &\frac{64 \pi}{s^2}
(s-m_q^2+m_Q^2)^2(s-m_Q^2-m_q^2)^2 \\ & \times N_f \sum_{a} d_a \left
(|T_{a,l=0}(s)|^2+ 3|T_{a,l=1}(s) \cos \theta_{\text{cm}}|^2 \right).
\end{split}
\end{equation} 
In Eq.\ (\ref{2.1.10b}) the integrations run over the three momenta of
the incoming light quark or gluon and the momenta of the outgoing
particles. The sum over the matrix element is taken over the spin and
color degrees of freedom of both the incoming and outgoing particles;
$\gamma_Q=6$ is the corresponding spin-color degeneracy factor for the
incoming heavy quark, and $f_{q,g}$ stands for the Boltzmann
distribution function for the incoming light quark or gluon. When
adopting this notation, the drag and diffusion coefficients are given by
\begin{equation}
\begin{split}
\label{2.1.10c} A(\bvec{p}) &= \erw{1-\frac{\bvec{p}
\bvec{p}'}{\bvec{p}^2}}, \\ B_{0}(\bvec{p}) &= \frac{1}{4}
\erw{\bvec{p}'{}^2-\frac{(\bvec{p}' \bvec{p})^2}{\bvec{p}^2}} , \\
B_{1}(\bvec{p}) &= \frac{1}{2} \erw{\frac{(\bvec{p'}
\bvec{p})^2}{\bvec{p}^2} - 2 \bvec{p}' \bvec{p} + \bvec{p}^2}.
\end{split}
\end{equation} 
We also include the leading-order perturbative QCD cross sections for
elastic gluon heavy-quark scattering \cite{Combridge:1978kx}, including
a Debye screening mass $m_{\mathrm{D}g}=g T$ in the gluon propagators, thus
controlling the $t$-channel singularities in the matrix elements.

\subsection{Drag and diffusion coefficients for D-mesons}

To account for the combined effect of $\D^+$ and $\D^0$ ($\D^-$ and
$\bar{\D}^0$) mesons we implement the transport coefficients using the
$\D$-meson ($\bar{\D}$-meson) isospin-averaged scattering amplitudes. In
this way we are incorporating possible ``off-diagonal transitions'' in
which the heavy meson can exchange flavor like
$\D^+ \pi^0 \rightarrow \D^0 \pi^+$.

Below the hadronization temperature the $\D$ and $\bar{\D}$ mesons
interact with the hadrons that compose the thermal bath. We assume that
the main contribution to the drag force and diffusion coefficients is
due to their scattering with the most abundant hadronic species. For the
microscopic calculation of transport coefficients we consider the set of
pseudoscalar light mesons $\pi$, K, $\bar{\mathrm{K}}$, $\eta$ and the
baryons N, $\bar{\mathrm{N}}$, $\Delta$, $\bar{\Delta}$.

A detailed presentation of the effective Lagrangian for heavy mesons and
transport coefficients is described in Refs.~\cite{Abreu:2011ic,
Abreu:2012et, Tolos:2013kva, Torres-Rincon:2014ffa}. Here we only review
the basic aspects of the methodology.  We split the discussion between
the interaction of $\D$ mesons with lighter mesons, and with baryons. The
two sectors have in common that the effective Lagrangian follows from
the principles of chiral and heavy-quark spin symmetry (HQSS), and the
final scattering matrix elements satisfy exact unitarity
constraints. Unitarity is assured by the implementation of a
unitarization procedure to the perturbative scattering amplitudes
obtained from the effective theory.
\subsubsection{Interaction with light mesons} 

The effective Lagrangian describing $\D$ mesons and the light
pseudoscalar mesons is described in
Ref.~\cite{Abreu:2011ic,Tolos:2013kva} (and references therein). The
$\D$ meson is incorporated within a $J=0$ isotriplet
$D=(D^0,D^+,D^+_s)$. In addition, the $J=1$ meson field
$D^*_\mu=(D^{*0},D^{*+},D_{s}^{*+})_\mu$ is also introduced in
accordance to HQSS. The set of $\mathrm{SU}(3)_f$ (pseudo-)Goldstone bosons is
introduced via the exponential representation
$U=u^2=\exp \left(\frac{\sqrt{2} \ii \Phi}{f} \right)$, where the matrix
\begin{equation} \Phi= \left(
\begin{array}{ccc} \frac{1}{\sqrt{2}} \pi^0 + \frac{1}{\sqrt{6}} \eta &
\pi^+ & K^+ \\ \pi^- & - \frac{1}{\sqrt{2}} \pi^0 + \frac{1}{\sqrt{6}}
\eta & K^0 \\ K^- & {\bar K}^0 & - \frac{2}{\sqrt{6}} \eta
\end{array} \right) \ ,
\end{equation} 
and $f$ is the pion decay constant in the chiral limit. The
leading-order (LO) Lagrangian is fixed by chiral symmetry and HQSS. It
incorporates the standard LO chiral perturbation theory for the
Goldstone bosons, and
\begin{equation}
\begin{split} 
  \Lag_{LO} = & \langle \nabla^\mu D \nabla_\mu D^\dag \rangle -
  m_{\D}^2 \langle DD^{\dag} \rangle - \langle \nabla^\mu
  D^{*\nu} \nabla_\mu D^{*\dag}_{\nu} \rangle \\
  &+ m_{\D}^2 \langle D^{*\mu} D_\mu^{* \dag} \rangle + ig \langle D^{*
    \mu} u_\mu D^\dag - D u^\mu D_\mu^{*\dag} \rangle \\
  &+ \frac{g}{2m_{\D}} \langle D^*_\mu u_\alpha \nabla_\beta D_\nu^{*\dag}
  - \nabla_\beta D^*_\mu u_\alpha D_\nu^{*\dag}\rangle \epsilon^{\mu \nu
    \alpha\beta}\ ,
\end{split}
\end{equation} 
where $m_{\D}$ is the tree-level heavy-meson mass, the bracket denotes
trace in flavor space, and
\begin{alignat}{2} 
  u_\mu &= \ii \left( u^\dag \partial_\mu u -
    u \partial_\mu u^\dag \right) \ , \\
  \nabla_\mu &= \partial_\mu - \frac{1}{2} \left( u^\dag \partial_\mu u
    + u \partial_\mu u^\dag \right) \ ,
\end{alignat} 
are the auxiliary axial vector field, and the covariant derivative,
respectively. The coupling $g$ connects heavy and light mesons and can
be fixed such that the decay width of the process
$\D^* \rightarrow \D+\pi$ is reproduced. The Lagrangian is further
expanded up to next-to-leading order (NLO) in chiral counting.  This
order (not reproduced here) is needed to account for the light-meson
masses and additional interactions between heavy and light sectors.  The
expression of the perturbative potential at NLO is
\begin{equation} 
\begin{split}
  \label{eq:mespot} 
V^{\text{meson}}_{ij} =& \frac{C_{0,ij}}{2 f^2} (p_1 \cdot
  p_2 - p_1 \cdot p_4) + \frac{2C_{1,ij} h_1}{f^2} \\
  &+ \frac{2C_{2,ij}}{f^2} h_3 (p_2 \cdot p_4) + \frac{2C_{3,ij}}{f^2}
  h_5 \left[(p_1 \cdot p_2) (p_3 \cdot p_4) \right. + \left. (p_1 \cdot p_4) (p_2 \cdot p_3) \right] \ ,
\end{split} 
\end{equation}
where $i,j$ denote the incoming and outgoing scattering channels
($1,2 \rightarrow 3,4$), $C_{n,ij}$ are numerical coefficients depending
on the isospin, spin, strangeness and charm quantum numbers, and $h_n$
are the low-energy coefficients, appearing at NLO and not fixed by
symmetry arguments alone, but by matching physical observables to
experimental data~\cite{Abreu:2011ic}.  Equation~(\ref{eq:mespot})
provides the NLO scattering amplitudes for meson-meson (elastic and
inelastic) scattering. The interactions of $\bar{\D}$ mesons are
obtained by appropriate charge conjugations.

To increase the validity to moderate energies we impose exact unitarity
on these amplitudes. This is achieved by the solution of the
Bethe-Salpeter equation, or $T$-matrix approach similar to the one used
for the partonic case.  We use $V$ as the kernel for the $T-$matrix
equation, in a full coupled-channel basis. The integral equation is
simplified within the ``on-shell'' approximation~\cite{Abreu:2011ic} and
transformed into an algebraic equation $T=V+V\tilde{G}T$, which is
readily solved by
\begin{equation} \label{eq:Tamp} T_{ij} = [1-V\tilde{G}]_{ik}^{-1}
V_{kj} \ ,
\end{equation} 
where $\tilde{G}$ is the so-called loop function (integral over the
internal momentum of the two-particle propagator).

In addition to the exact unitarity satisfied by $T$, the unitarization
method produces a set of resonance and bound states in some of the
scattering channels, appearing as poles in the complex-energy plane of
$T$. The identification of these poles with experimental states, helps
us to fix the unknown parameters of the effective approach (low-energy
constants and the regularization parameters of $\tilde{G}$).  In
particular, we obtain the $D_0^*(2400)$ in the $(I,J^P)=(1/2,0^+)$
channel, and the bound state $D^*_{s0}(2317)$ in the
$(I,J^P)=(0,0^+)$.
\begin{figure*}[ht]
	\begin{minipage}[b]{0.48\textwidth}
		\includegraphics[width=1\textwidth]{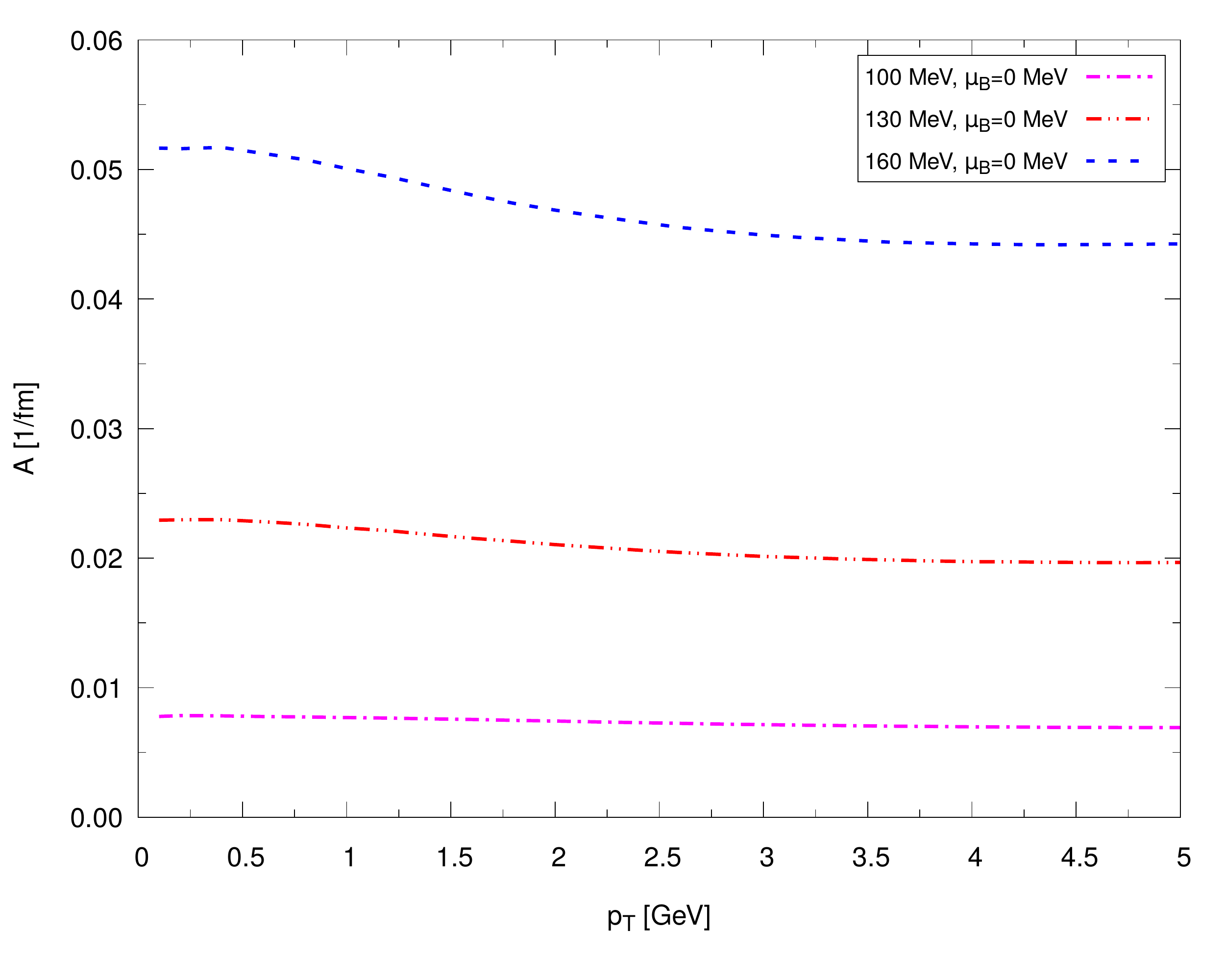}
	\end{minipage} \hspace{3mm}
	\begin{minipage}[b]{0.48\textwidth}
		\includegraphics[width=1\textwidth]{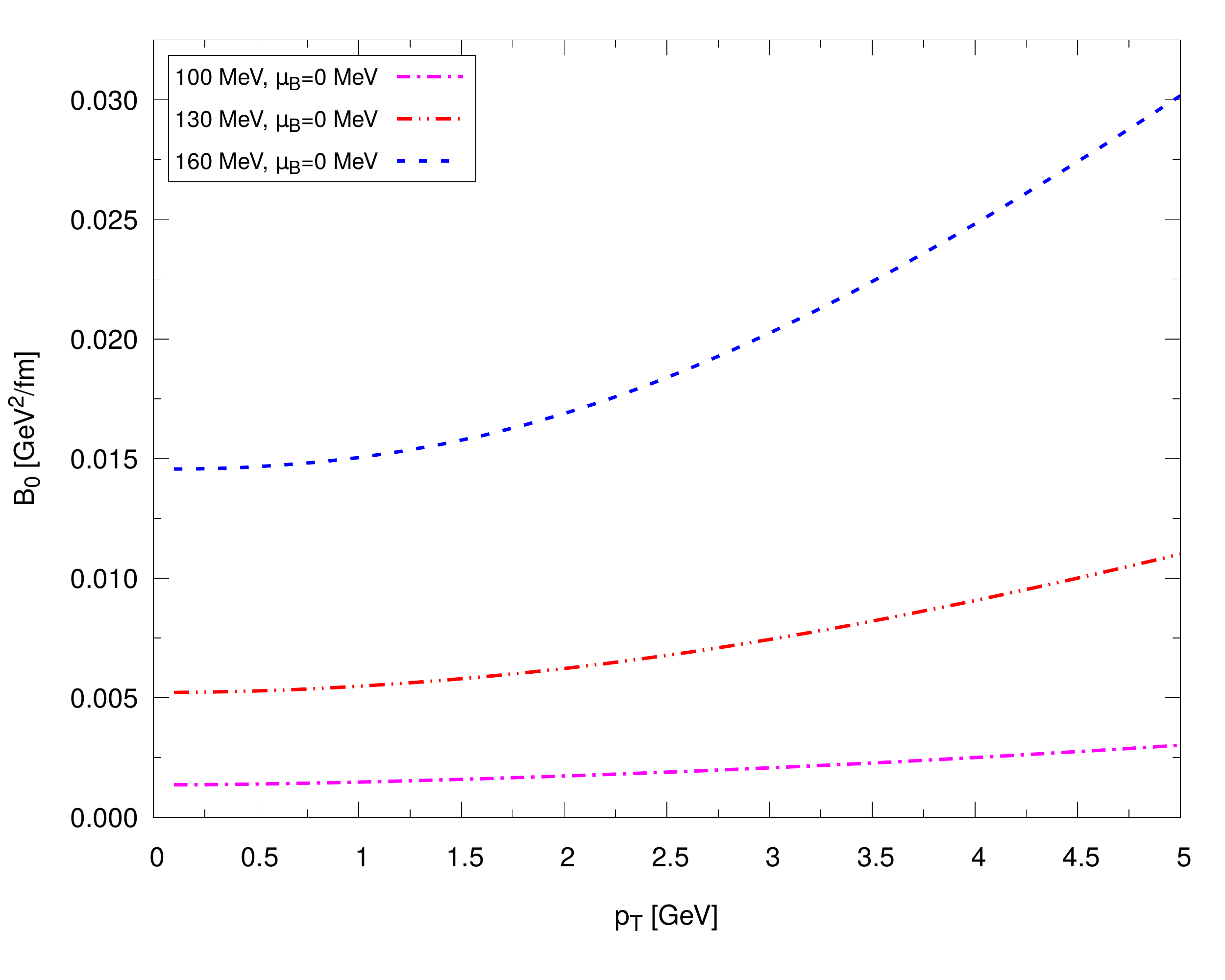}
	\end{minipage}
	\caption{(Color online) Drag (left) and diffusion (right)
coefficients at different temperatures for $\D$ mesons interacting with
the pseudoscalar meson octet $\pi$, $K$, $\bar{K}$, $\eta$.}
	\label{coeff-mesons}
\end{figure*} 
In Fig.~\ref{coeff-mesons} we present the drag force (left panel) and
diffusion coefficient (right panel) of $\D$ mesons interacting with
light mesons as functions of momentum for several temperatures at
$\mu_{\text{B}}=0$. For large momentum---beyond the natural application of the
effective Lagrangian---the interactions are taken assuming constant
cross sections. Although the qualitative behavior of the transport
coefficients is similar to the case for $c$ quarks, notice that the
numerical values are one order of magnitude smaller.

\subsubsection{Interaction with baryons} 

The interaction of $\D$ mesons with baryons follows a parallel
methodology using an effective Lagrangian based on chiral and HQSS
symmetries.  In this case the formalism is taken from
Refs.~\cite{GarciaRecio:2008dp,Gamermann:2011mq,Romanets:2012hm,Garcia-Recio:2013gaa,Torres-Rincon:2014ffa}.
The Lagrangian is considered at LO in chiral expansion, and is further
reduced to a Weinberg-Tomozawa interaction when the Goldstone bosons
participate in the interaction. Then, the $\text{SU}(3)_f$ chiral
symmetry is enlarged to $\text{SU}(6)$ symmetry (spin times flavor).
From the degrees of freedom introduced in the effective description, we
focus on those involved in the interaction of the $\D$ meson with
$N,{\bar N}, \Delta$ and ${\bar \Delta}$ baryons.

The tree-level meson-baryon scattering amplitudes have the structure
\begin{equation} 
  V_{ij}^{\text{baryon}} = \frac{D_{ij}}{4\,f_i f_j}
  (2\sqrt{s}-M_i-M_j) \sqrt{\frac{M_i+E_i}{2M_i}}
  \sqrt{\frac{M_j+E_j}{2M_j}} \ ,
\end{equation} 
where $M_i$, $E_i$ and $f_i$ denote respectively the baryon mass,
C.M. energy, and the meson decay constant participating in the $i$
channel. The $D_{ij}$ are numerical coefficients depending on the
quantum numbers of the scattering channel.

As in the meson sector, these amplitudes are used as kernels in a
coupled-channel $T$-matrix approach. It is again solved in the
``on-shell'' approximation to obtain the solution given of
Eq.\~(\ref{eq:Tamp}), which satisfies exact unitarity. A large set of
resonant and bound states are dynamically generated by the unitarization
procedure. The most prominent ones being the $\Lambda_c (2595)$ in the
$(I,J^P)=(0,1/2^-)$ channel and the $\Sigma_c (2550)$ in the
$(I,J^P)=(1,3/2^-)$ channel.

Once the scattering amplitudes are fixed, the $\D$-meson transport
coefficients are computed---like in the partonic case---within the
Fokker-Planck approximation. The drag force and the diffusion
coefficients are calculated using the same equations as
in~(\ref{2.1.10b},\ref{2.1.10c}), but implementing quantum statistics
instead. Pertinent isospin-spin degeneracy factors are used for each
degree of freedom.

The dependence of the transport coefficients on the chemical potential
has been addressed in Ref.~\cite{Tolos:2013kva}.  To an excellent
approximation the fugacity ($z=\ee^{\mu_B/T}$) factorizes out of the
expression of the meson-baryon transport coefficients (and $z^{-1}$
factorizes out for the antibaryon case). In this respect, the transport
coefficients of the $\D$ meson can be constructed by a linear
combination of the transport coefficients of mesons, baryon and
antibaryon at $\mu_B=0$, with respective coefficients $1$, $z$, $z^{-1}$
(for the $\bar{\D}$ meson, baryon and antibaryon coefficients should be
reversed).

\begin{figure*}[h]
	\begin{minipage}[b]{0.48\textwidth}
		\includegraphics[width=1\textwidth]{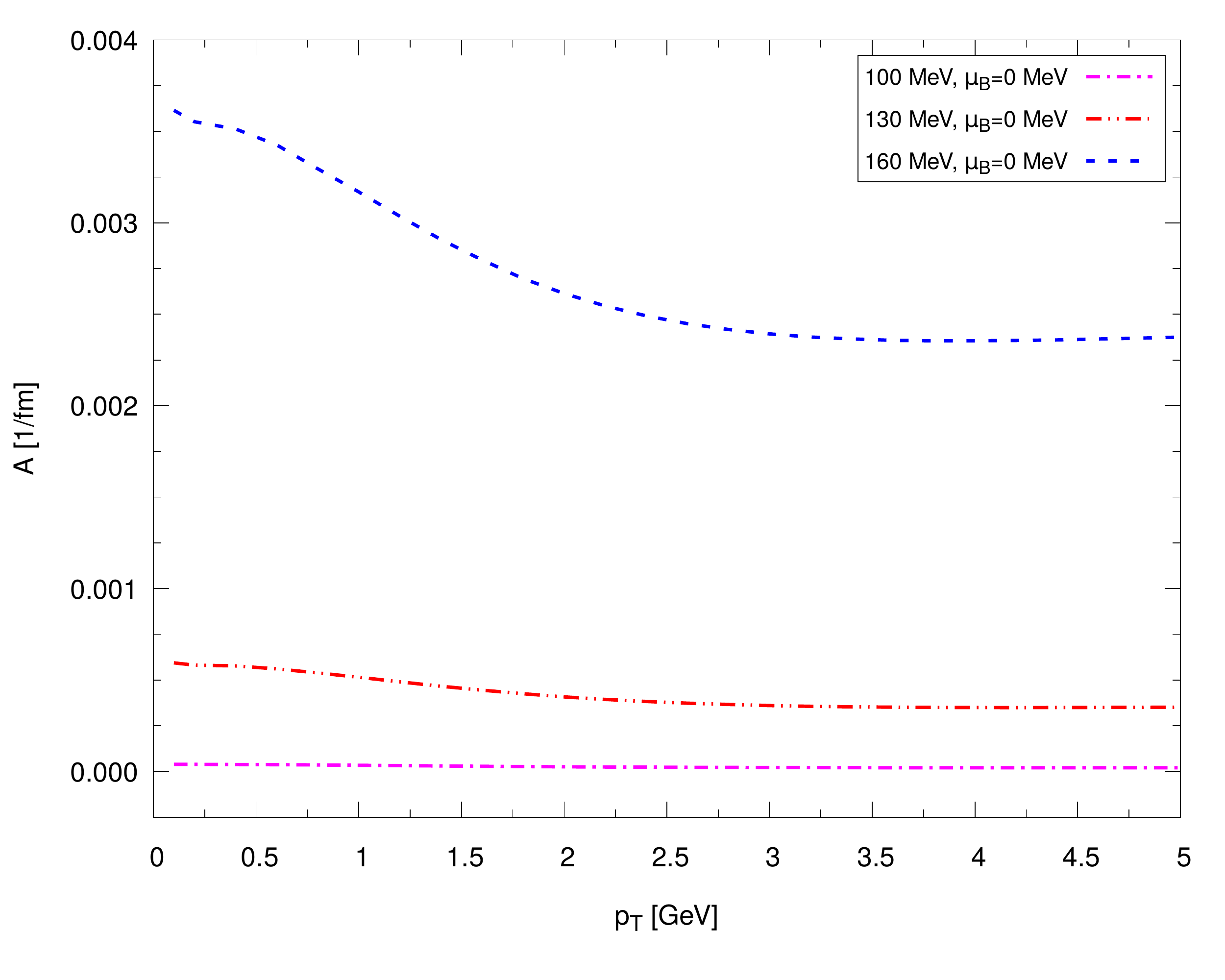}
	\end{minipage} \hspace{3mm}
	\begin{minipage}[b]{0.48\textwidth}
		\includegraphics[width=1\textwidth]{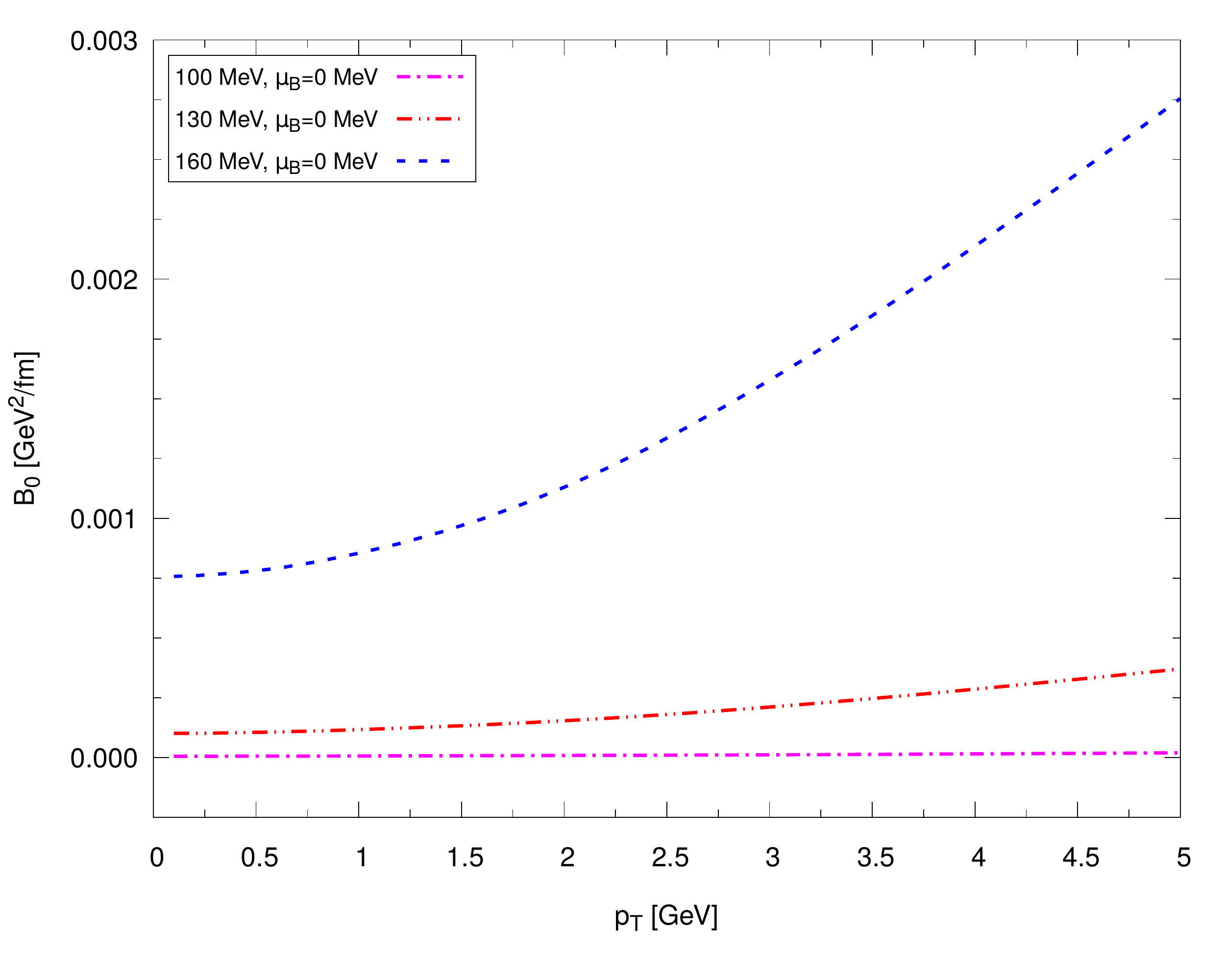}
	\end{minipage}
	\caption{(Color online) Drag (left) and diffusion (right)
coefficients at different temperatures for $\D$ mesons interacting with
baryons $N$ and $\Delta$.}
	\label{coeff-baryons}
\end{figure*}

In Fig.~\ref{coeff-baryons} we show the transport coefficients for the
$\D$ mesons interacting with baryons (alternatively, $\bar{\D}$ with
antibaryons). Due to the Boltzmann suppression of baryons, the transport
coefficients are considerably suppressed with respect to those for
mesons.

\begin{figure*}[h]
	\begin{minipage}[b]{0.48\textwidth}
		\includegraphics[width=1\textwidth]{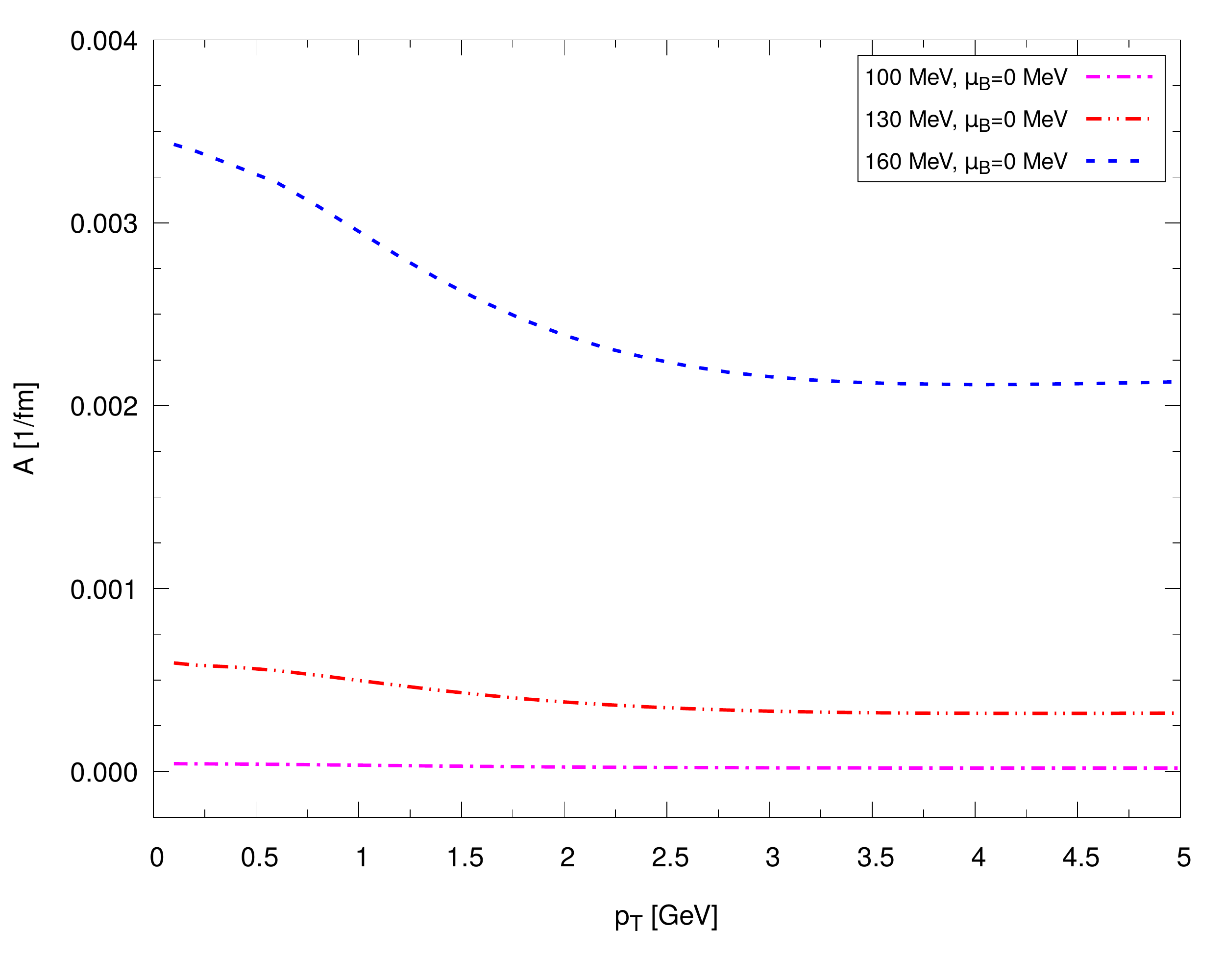}
	\end{minipage} \hspace{3mm}
	\begin{minipage}[b]{0.48\textwidth}
		\includegraphics[width=1\textwidth]{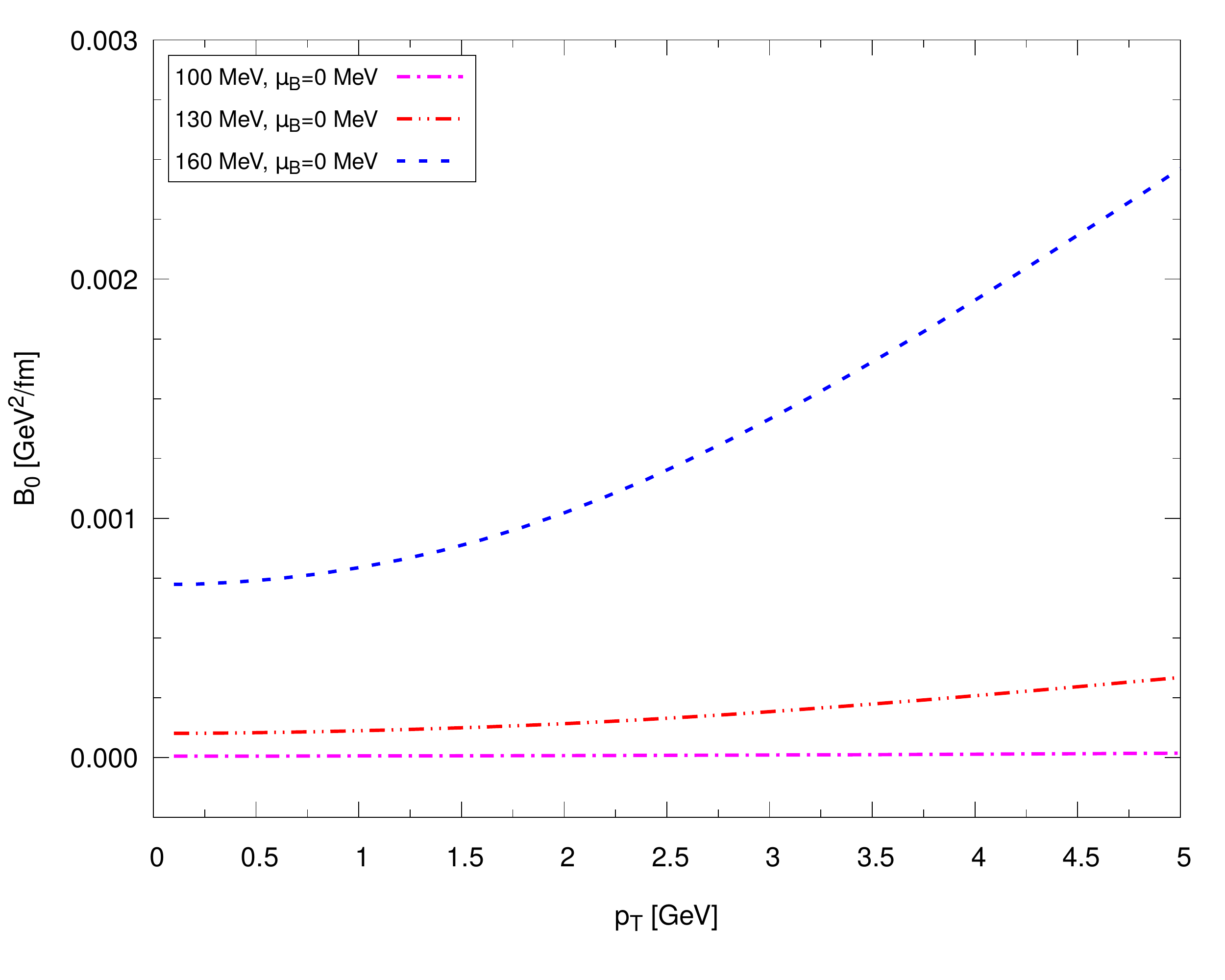}
	\end{minipage}
	\caption{(Color online) Drag (left) and diffusion (right)
          coefficients at different temperatures for $\bar{\D}$ mesons
          interacting with baryons $N$ and $\Delta$.}
	\label{coeff-antibaryons}
\end{figure*}

In Fig.~\ref{coeff-antibaryons} we present a similar plot of the
coefficients for the $\bar{\D}$ mesons interacting with baryons
(equivalently, $\D$ mesons with antibaryons). Let us note that for the
rather different cross sections as compared to the previous case, the
transport coefficients are very similar. The reason is that the
transport coefficients are not very sensitive to the details of the
scattering amplitude (resonance peaks, channel openings...), but only to
the thermal average of it, which is similar in both cases. However, we
note that the $\D$-meson-baryon interaction is stronger, with more
resonances contributing to the total cross section. This is reflected in
slightly larger coefficients.

\begin{figure*}[h]
	\begin{minipage}[b]{0.48\textwidth}
		\includegraphics[width=1\textwidth]{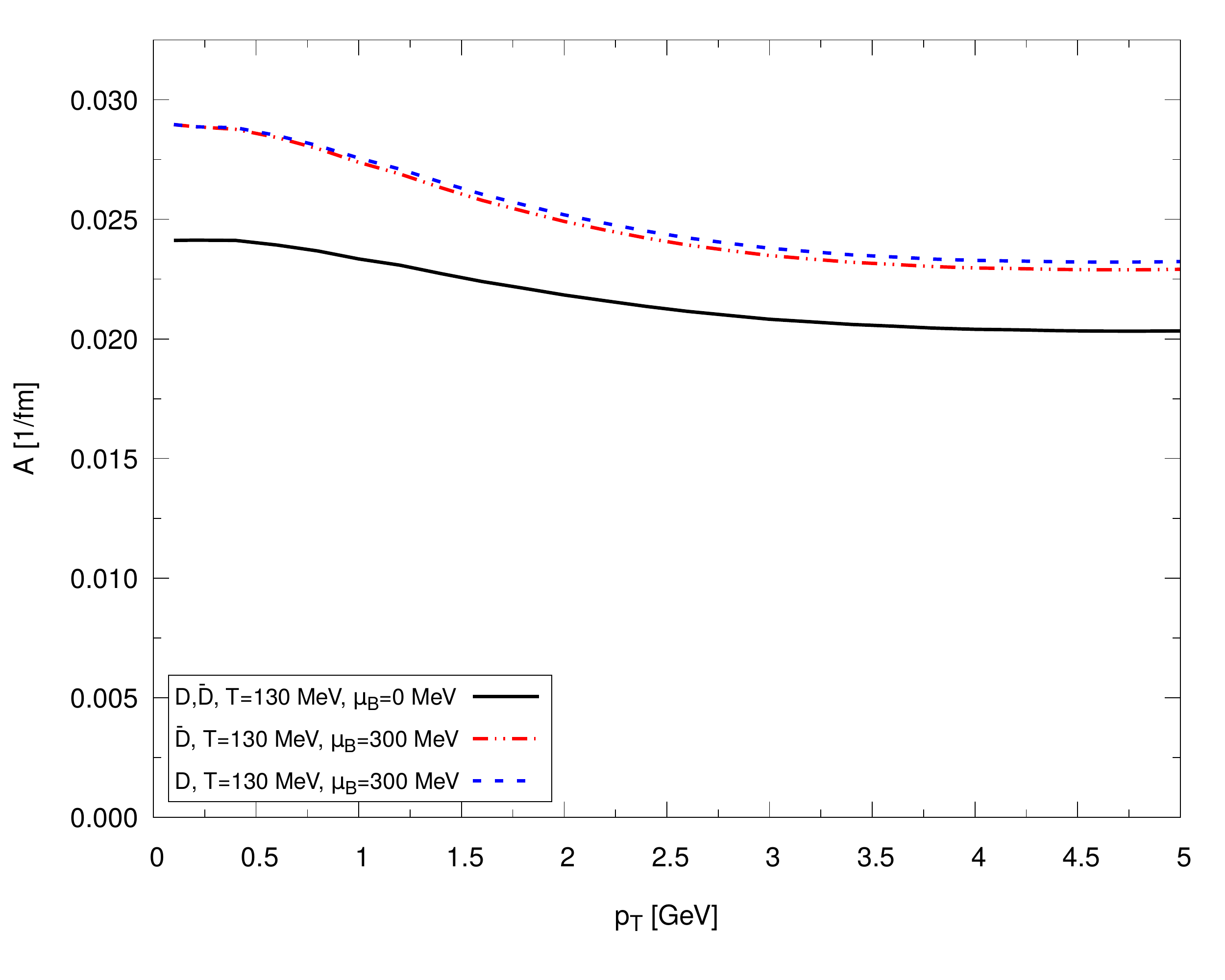}
	\end{minipage} \hspace{3mm}
	\begin{minipage}[b]{0.48\textwidth}
		\includegraphics[width=1\textwidth]{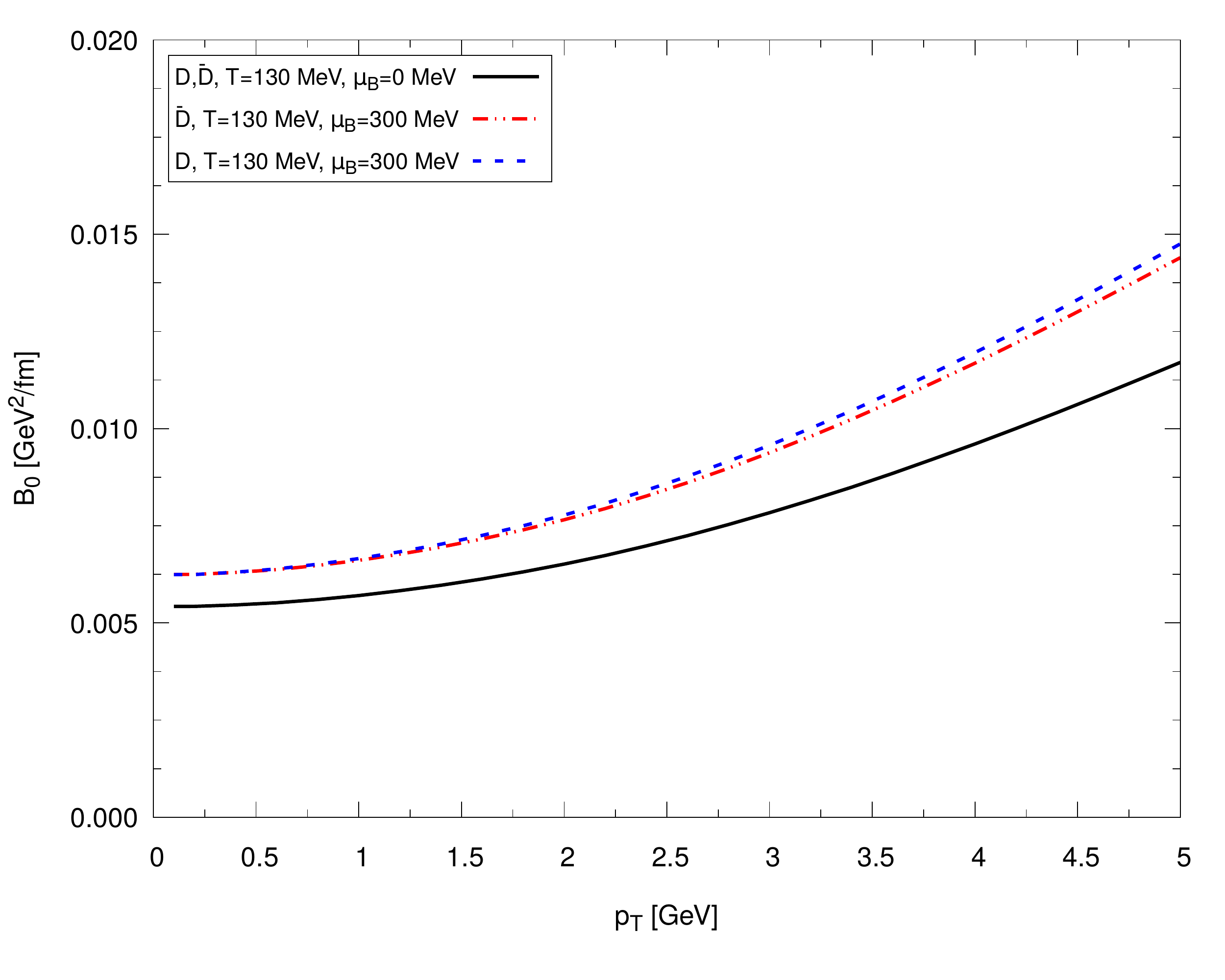}
	\end{minipage}
	\caption{(Color online) Drag (left) and diffusion (right)
          coefficients at different temperatures for $\D$ and $\bar{\D}$
          mesons, taking into account a fugacity factor.}
	\label{coeff-D-AB}
\end{figure*}

\begin{figure}[h]
	\includegraphics[width=0.48\textwidth]{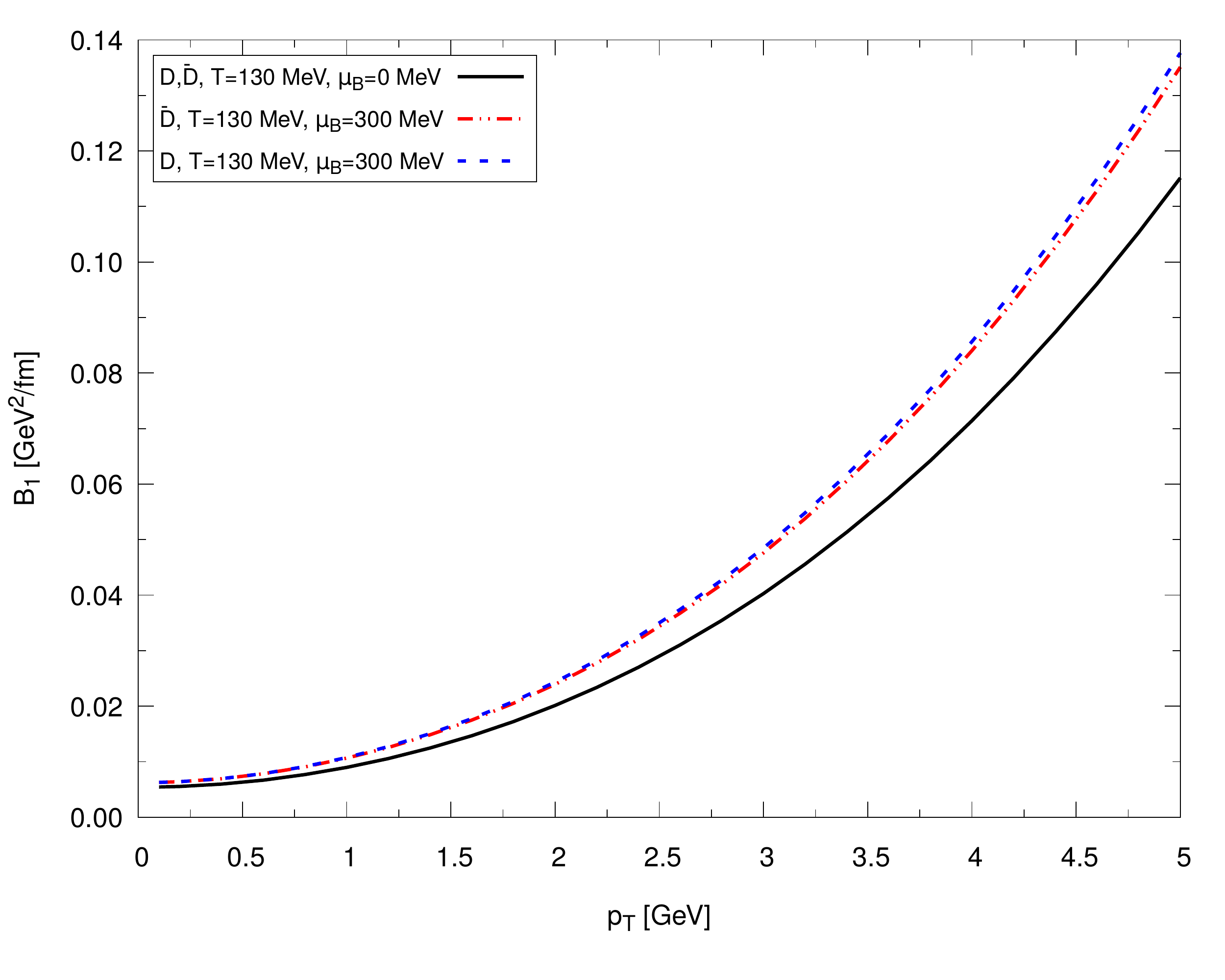}
	\caption{(Color online) $B_1$ coefficients at different
          temperatures for $\D$ and $\bar{\D}$ mesons, taking into
          account a fugacity factor.}
	\label{coeff-D-B1}
\end{figure}

Finally, in Figs.~\ref{coeff-D-AB} and~\ref{coeff-D-B1} we show the
effect of the baryochemical potential. A sizable increase of the drag
and diffusion coefficients is obtained for moderate values of the
chemical potential, entirely due to the baryon and antibaryon
contributions. For higher $\mu_{\mathrm{B}}$ this important increase of the
coefficients produces a large energy loss and momentum diffusion of $\D$
mesons in dense matter.

\section{Implementation of the numerical simulations} \label{implementation}

Notice: except in case of explicit distinctions, in this section we will
use the term $c$, \emph{charm} quark and $\D$ mesons both for particles
and anti-particles. More precisely, we will consider $D^+$ and $D^-$
only, excluding all other open charm mesons.
\begin{figure*}[h]
	\begin{minipage}[b]{0.48\textwidth}
		\includegraphics[width=1\textwidth]{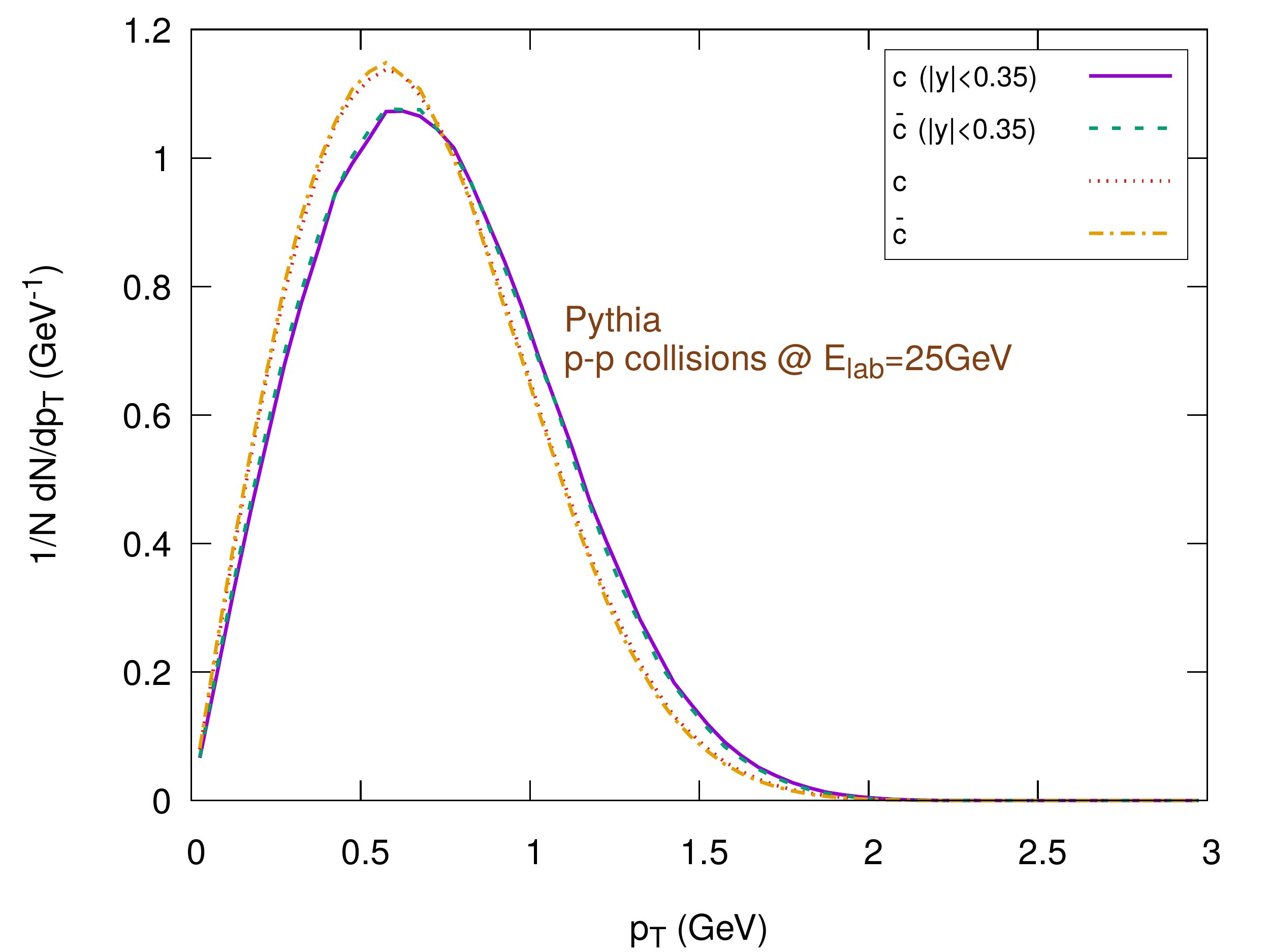}
	\end{minipage} \hspace{3mm}
	\begin{minipage}[b]{0.48\textwidth}
		\includegraphics[width=1\textwidth]{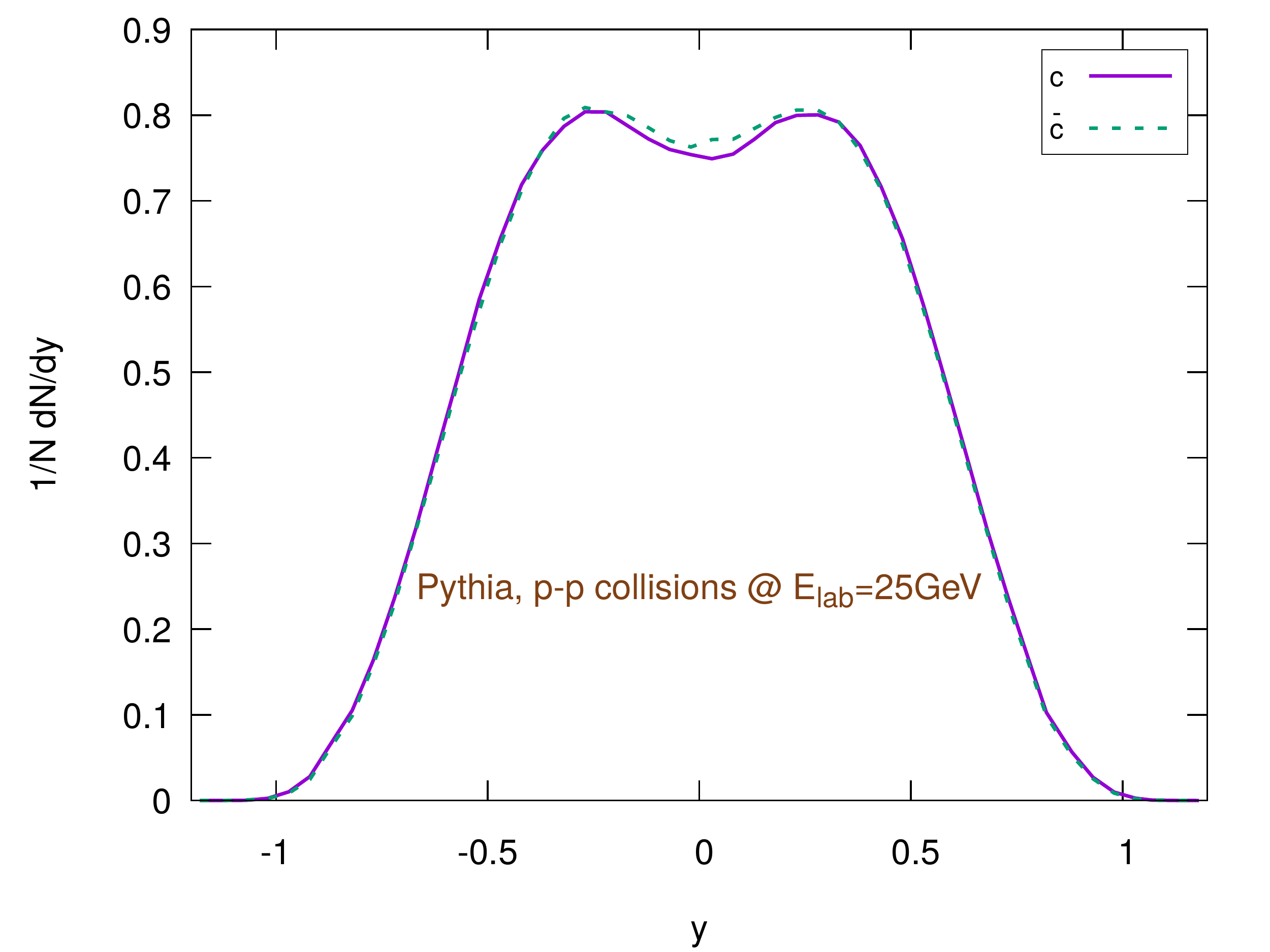}
	\end{minipage}
	\caption{(Color online) Spectra of initial charms and
anti-charms as sampled with Pythia. The left figure shows the
(normalized) $1/N \dd N/\dd p_{\text{T}}$ distribution (in the rapidity range
$|y|<0.35$), the right figure shows the (normalized) $1/N \dd N/ \dd y$
distribution.}
	\label{initial_charms}
\end{figure*}

\begin{figure*}[h]
	\begin{minipage}[b]{0.48\textwidth}
		\includegraphics[width=1\textwidth]{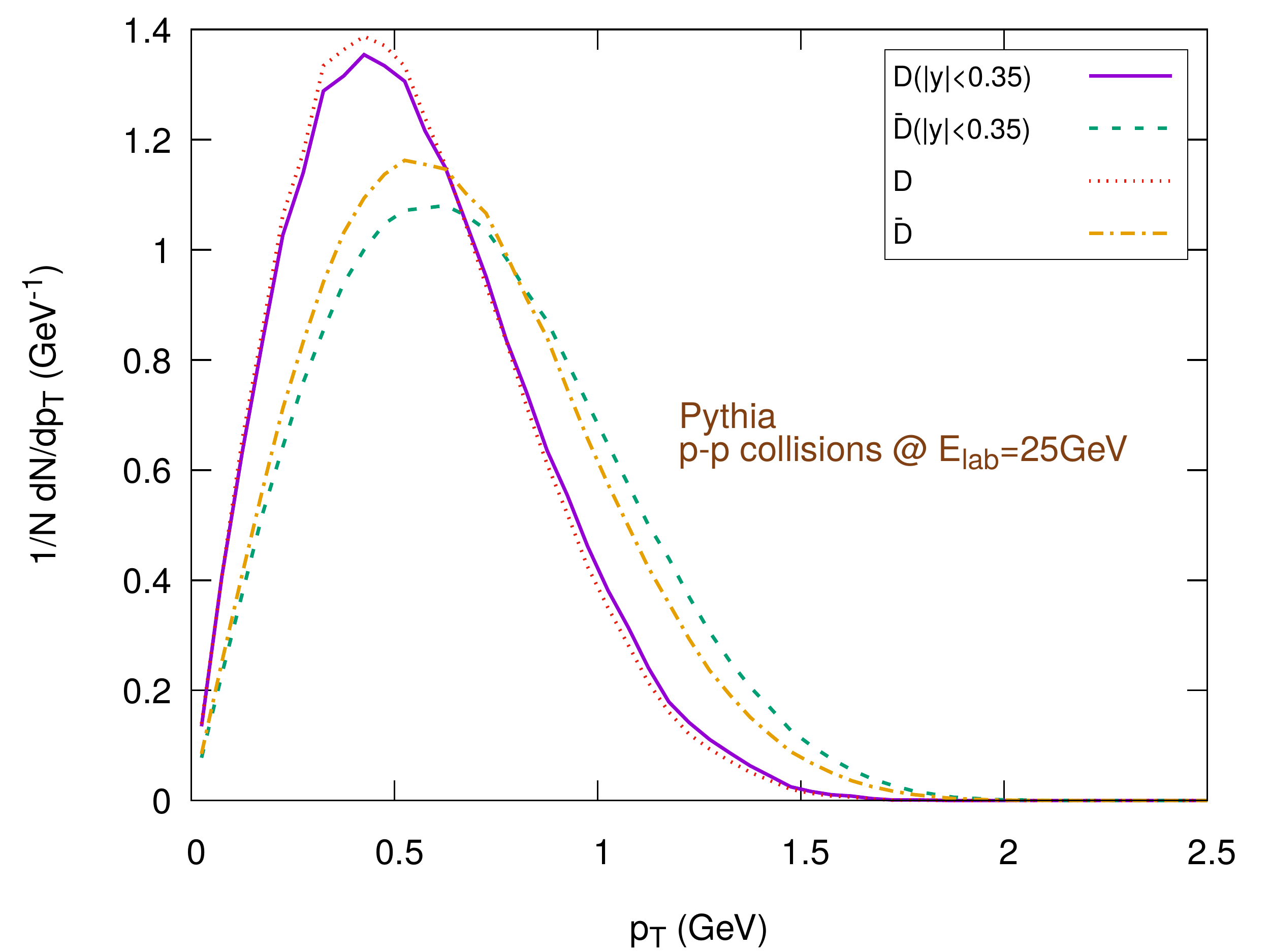}
	\end{minipage} \hspace{3mm}
	\begin{minipage}[b]{0.48\textwidth}
		\includegraphics[width=1\textwidth]{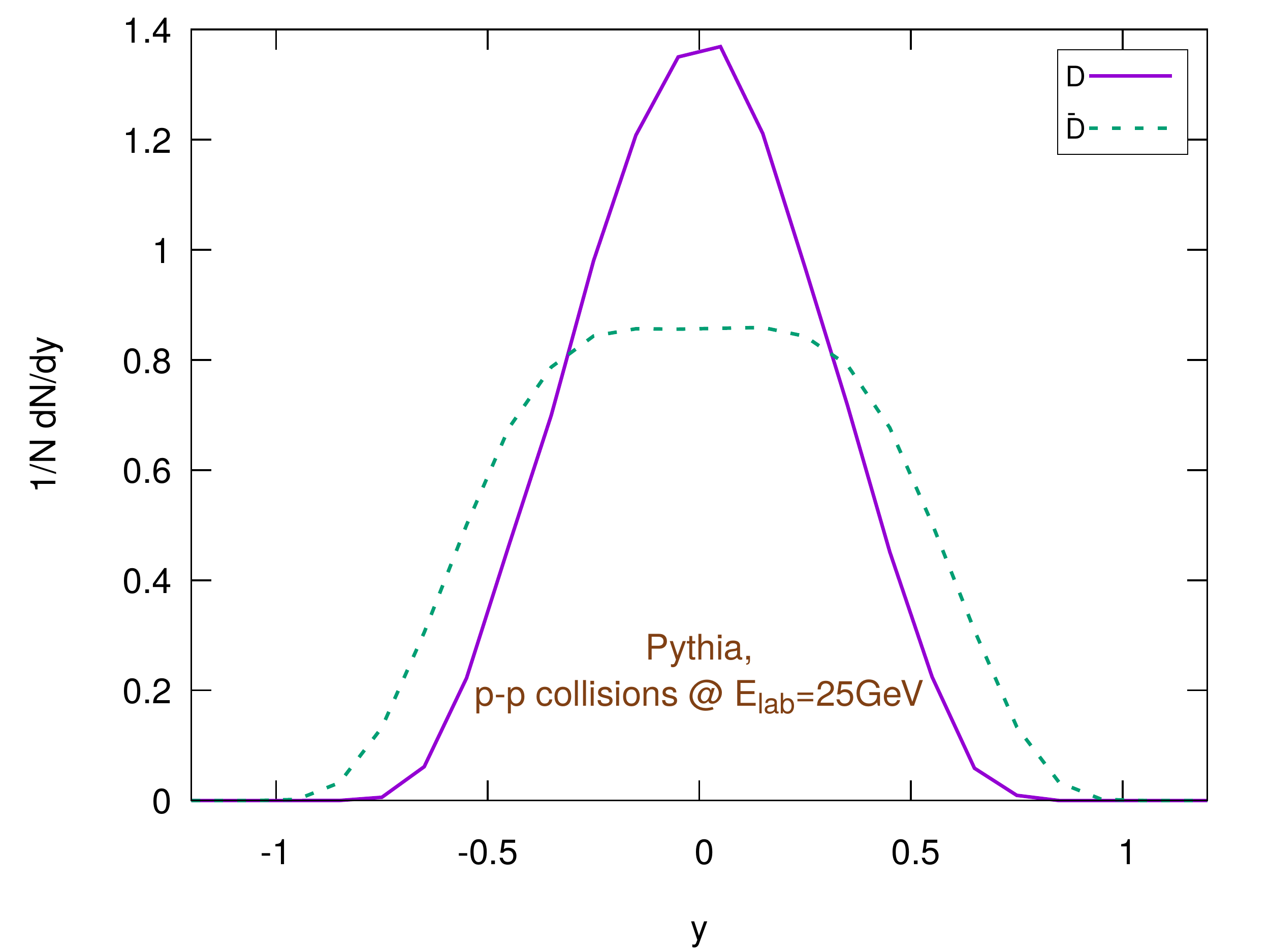}
	\end{minipage}
	\caption{(Color online) Spectra of $\D/\bar{\D}$-mesons in p-p
		collisions sampled with Pythia. The left figure shows the
		(normalized) $1/N \dd N/\dd p_{\text{T}}$ distribution (in the rapidity
		range $|y|<0.35$), the right figure shows the (normalized)
		$1/N \dd N/\dd y$ distribution.}
	\label{dmesons_in_pp}
\end{figure*}

We use Pythia 8.2 \cite{Sjostrand:2006za,Sjostrand:2014zea} to obtain a
set of $10^6$ charm-anti-charm quark pairs by performing p+p collisions
at $E_{\text{lab}}=25\,\GeV$, enabling the \emph{SoftQCD}
mode\footnote{For technical reasons in Pythia the simulations are
  performed in a fixed target set-up and the four-momenta are boosted
  back to the center of mass frame.}. The initial charm and anti-charm
distributions versus transverse momentum and rapidity are shown in
Fig.(\ref{initial_charms}). Pythia is also used to compute the $\D/\bar{\D}$ mesons momentum distribution in  $p-p$ collisions, shown in 
Fig.\ \ref{dmesons_in_pp} with respect to the transverse momentum
(left) and rapidity (right). Here one observes
(Fig.~(\ref{dmesons_in_pp}), right) that the different production
channels $\text{pp} \rightarrow \D\bar{\D}+X$ and
$\text{pp} \rightarrow \bar{\D}\Lambda_c+X$ lead to different initial
rapidity distributions for the charm and anti-charm channels.

After this preliminary step, we perform the Langevin propagation of the
charm quarks in the background medium, first modeling it with the UrQMD
hydrid model\cite{Petersen:2008dd} and then with the UrQMD
coarse-graining approach\cite{Endres:2014zua}.

To obtain the space-time points of the production of the charm quarks,
we perform an UrQMD run with elastic zero degree scatterings between the
colliding nuclei (Monte Carlo Glauber initial conditions), saving the
space-time coordinates of the points where collisions between the
nucleons happened. In the subsequent full UrQMD runs, for each event we
distribute over these collision points around 140000 $c$-$\bar{c}$ pairs
previously created with Pythia. The (anti-)charm quarks propagate along
straight lines without interacting with any particle until the onset of
the hydrodynamical phase, i.e. after the two nuclei have completely
passed through each other at
$t=(2R_{\text{nucl}})/(\sqrt{\gamma^2_{\text{CM}}-1})\approxeq3.5\,\fm$.
The timestep for the Langevin propagation is
$\dd t_{\text{Langevin}}=0.01\, \dd t_{\text{hydro}}$ for each hydro
timestep. We have checked that this accuracy is sufficient to obtain
stable results. At each Langevin iteration step we use the values of the
fluid temperature $T$ and fluid velocity components $v_i$ to perform a
bilinear interpolation of the transport coefficients (which depend on
the momentum $p$ and the temperature $T$). The finite baryon chemical
potential is taken into account by multiplying the drag and diffusion
coefficients of the charm quarks by a fugacity factor $\ee^{\mu_q/T}$ for
$\bar{c}$ quarks $\ee^{-\mu_q/T}$ for $c$ quarks ($\mu_q=\mu_B/3$). For
the $\D$ mesons we use
$K_{\D}(T,\mu_B,p)=K_{\text{mesons}}^{\D}(T,p)+\ee^{\mu_B/T}K^{\D}_{\text{baryons}}(T,p)+\ee^{-\mu_B/T}K^{\D}_{\text{antibar}}(T,p)$
and
$K_{\bar{\D}}(T,\mu_B,p)=K^{\D}_{\text{mesons}}(T,p)+\ee^{-\mu_B/T}K^{\D}_{\text{baryons}}(T,p)+\ee^{\mu_B/T}K^{\D}_{\text{antibar}}(T,p)$,
where the $K$ is any of the transport coefficients $A$, $B_\perp$,
$B_\parallel$ and $K^{\D}_{\text{mesons}}$, $K^{\D}_{\text{baryons}}$,
$K^{\D}_{\text{antib}}$ are the contributions coming from the
interactions of $\D$ mesons with other mesons, baryons and anti-baryons,
respectively. In our model, we assume that the medium affects the
propagation of the heavy quarks, but the medium itself is not affected
by the heavy quarks that we inject. There is also no interaction between
the injected charm quarks. This approximation allows us to use a large
number of charm quarks per event, thus reducing considerably the number
of events needed to reach a sufficient statistics.

We assume to have instantaneous hadronization and decoupling processes
which happen at the same temperature $T_c$, that means that the
$c(\bar{c})$ quarks immediately become $\D(\bar{\D})$-mesons as soon
as they are found to be in a fluid cell with a temperature $T<T_c$ and,
on the contrary, $\D(\bar{\D})$-mesons become $c(\bar{c})$ quarks if they
are in a cell with $T>T_c$. 

We consider hadronization either through coalescence or Peterson
fragmentation. We assume a constituent quark rest mass for up and down
quarks of $m_{u,d}=369\,\MeV$, a charm quark mass $m_c=1.5\,\GeV$ and a
$\D$-meson mass $m_{\D}=1.869\,\GeV$ (we neglect the $5\,\MeV$ mass
difference between $\D^+/\D^-$ and $\D^0/\bar{\D}^0$). The velocity
components $v_x, v_y, v_z$ of the light quarks are taken as equal to the
fluid velocity, i.e. thermal smearing is omitted. The probability of
hadronization by coalescence, $P_{\text{coa}}$\cite{Greco:2003vf}, in terms of
the four momentum components $p^{\mu}$ of the light quarks is given by:
\begin{equation} 
P_{\text{coa}}=\exp \left \{ \left [(\Delta
p^0)^2-\sum\limits_{i=1}^3(\Delta
p^i)^2-(\Delta_m)^2 \right]\sigma^2\right \}.
\label{prob_to_coalescence}
\end{equation} 
Here the $\Delta p$ are the differences between the
four-momentum components of the heavy and the light quark,
$\Delta_m=m_{c} - m_{u,d}$, $\sigma=\sqrt{\frac{8}{3(\hbar c)^2} r^2_{\D_{(\mathrm{rms})}}}$
and $r^2_{\D_{(\mathrm{rms})}}$ is the mean squared radius of the
$\D$-meson.

In case of coalescence, the four-momentum of the newly formed
$\D(\bar{\D})$-meson is given by the sum of the four-momenta of the
constituent quarks, while, in case of Peterson fragmentation, the
$\D(\bar{\D})$-meson obtains a fraction of momentum of the charm quark
according to the distribution\cite{Peterson:1982ak}:
\begin{equation} D(z)=\frac{H}{z[1-(1/z)-\epsilon_p/(1-z)]^2}.
\label{peterson_frag_eq}
\end{equation} 
Here $H$ is a normalization constant, $z$ the momentum fraction obtained
in the fragmentation and $\epsilon_p$ a parameter. Peterson
fragmentation is the only process allowed for heavy quarks hadronizing
in the void, a condition that may occur in the coarse graining
approach. On average, roughly $80\%$ of the times the hadronization channel is Peterson fragmentation, more than it is commonly expected at low collision energies, especially if we consider that in Eq. (\ref{prob_to_coalescence}) we removed the dependence on the spatial distance between the heavy and the light quark in the probability distribution, which is present, for example, in the original Ref. \cite{Greco:2003vf} or in Ref. \cite{Song:2015sfa}. At the moment, we do not have a clear explanation for this issue.\\
In the case of the charm quarks originating from $\D$ mesons
entering into cells with $T>T_c$, we maintain the
four-ve\-lo\-ci\-ty. We evolve the UrQMD hydro simulations until the
energy density over the grid is below $0.3\varepsilon_0$
($\varepsilon_0=146.5\,\MeV/\fm^3$), then, using the phase-space data
(position and velocities) of the charm quarks at the beginning of the
hydro phase, we repeat each series using the coarse-graining approach.

We maintain the same time step for the Langevin propagation process that
we use in hydro, i.e. $8\cdot10^{-4}\,\fm$, so, since the time
resolution of the coarse graining data is $0.2\,\fm$ for reactions with impact parameter $b=3\fm$ and $0.4\,\fm$ for reactions with impact parameter $b=3\fm$, we perform 250 and 500 iterations per coarse-graining time step, respectively. As before for the hydro case,
we check that the choice of the time step has no effect on the final
results. We start the simulations in the coarse-graining approach from
$3.6\,\fm$, propagating again the charm quarks along straight lines from
the hydro starting time until this time. The method is the same as in
the hydrid approach. However, in addition we can now follow the bulk
evolution of the system until $t=75\,\fm$. To avoid spurious effects in
the coarse-graining simulations due to a few cells with low statistics
and therefore unrealistic momentum transfers, we limit the fugacity
factors to lie in the range [0.01-100], after a comparison with the
hydro case.\\
The $\D$ mesons decay weakly into non charmed hadrons before reaching the detectors, however they are relatively long-lived, with proper mean decay lengths of order $100\,\mu\textrm{m}$\cite{Tanabashi:2018oca}, therefore their decay products are not affected by hadronic rescattering and the decay vertices can be accurately reconstructed. This is the reason why we did not consider important to simulate also their decay into directly observed hadrons. However, probably we will reconsider open heavy flavor meson decays in future studies, when including also excited states\cite{Acharya:2017jgo}, or when working at higher collision energies and interested in distinguishing the prompt $\D$ mesons signal from the feed-down of heavier particles.\\
For each combination of the parameters, we run 2000 events.

\section{Results}

\begin{figure*}[h]
	\begin{minipage}[b]{0.48\textwidth}
		\includegraphics[width=1\textwidth]{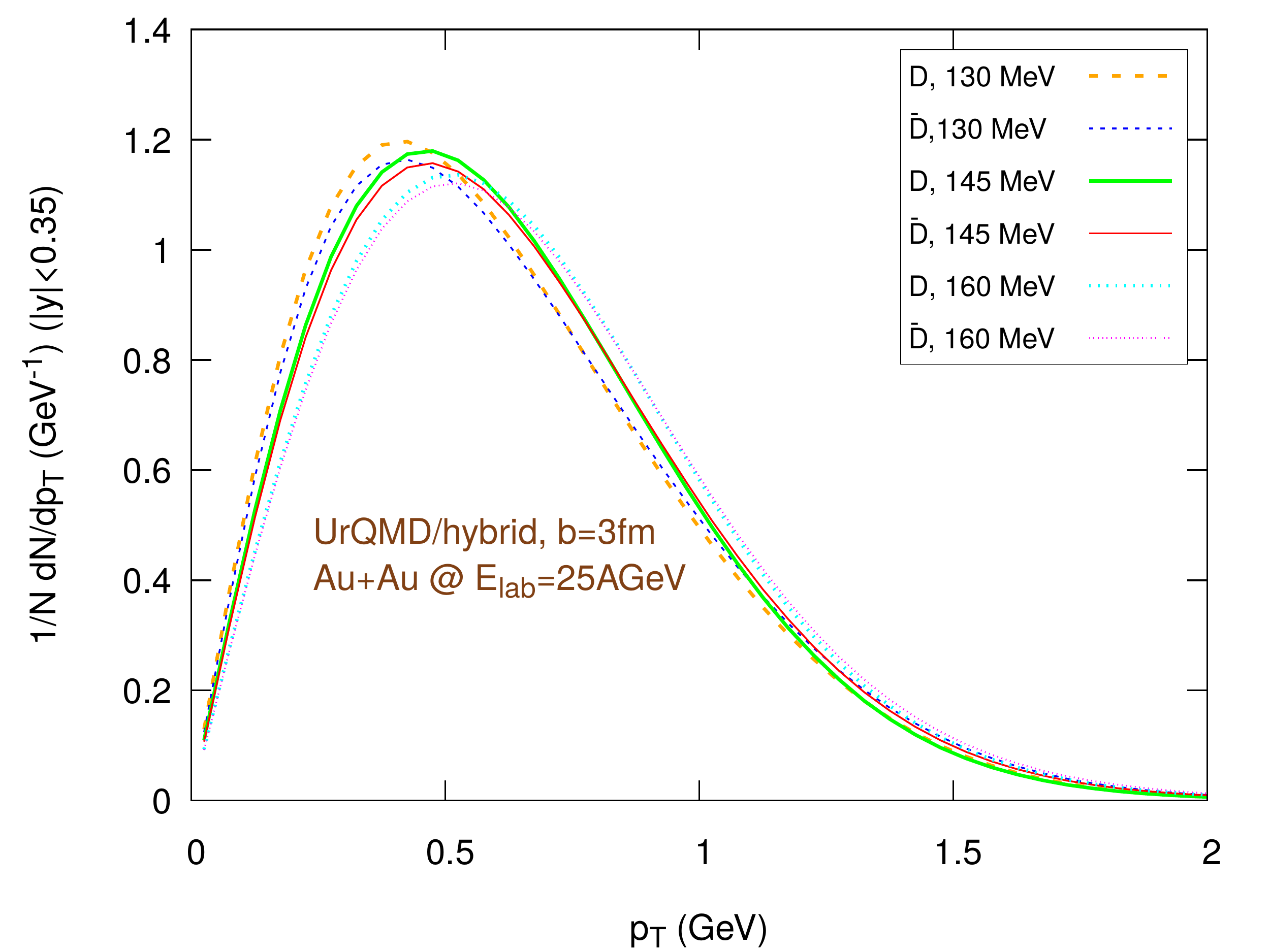}
	\end{minipage} \hspace{3mm}
	\begin{minipage}[b]{0.48\textwidth}
		\includegraphics[width=1\textwidth]{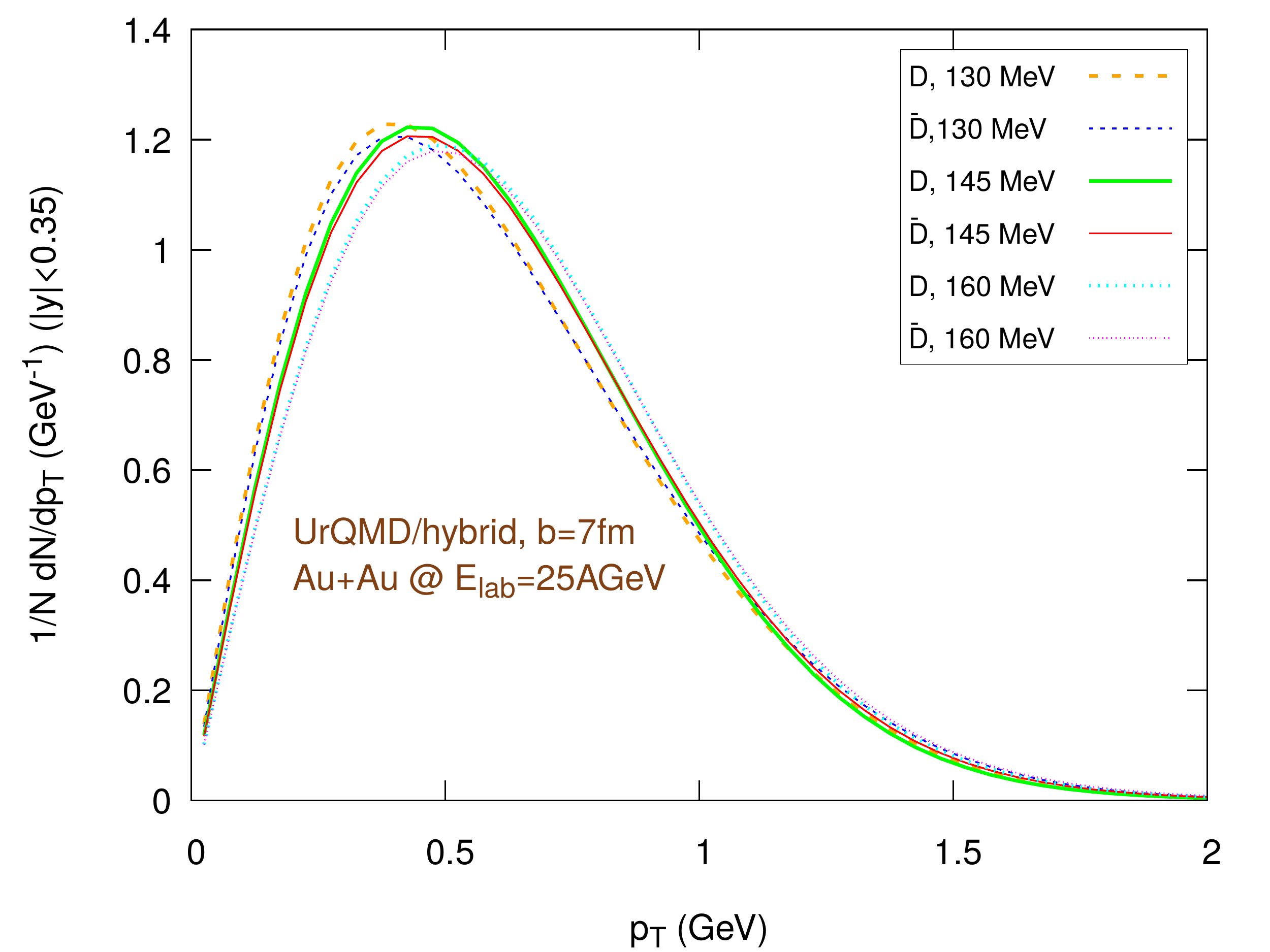}
	\end{minipage}
	\caption{(Color online) Normalized
		$1/N \dd N/\dd p_{\text{T}}$ distribution of the final $\D/\bar{\D}$ mesons,
		in the rapidity range $|y|<0.35$, for Au+Au collisions at
          $E_{\text{lab}}=25\,\AG$, using the UrQMD/hybrid
          model. The hadronization parameters are $\epsilon_p=0.05$ and
          $\erw{r_{\D_{\mathrm{rms}}}}=0.6\,\textrm{fm}$. Left: $b=3\,\fm$,
          right: $b=7\,\fm$.}
	\label{dNdpT_hy}
\end{figure*}

\begin{figure*}[h]
	\begin{minipage}[b]{0.48\textwidth}
		\includegraphics[width=1\textwidth]{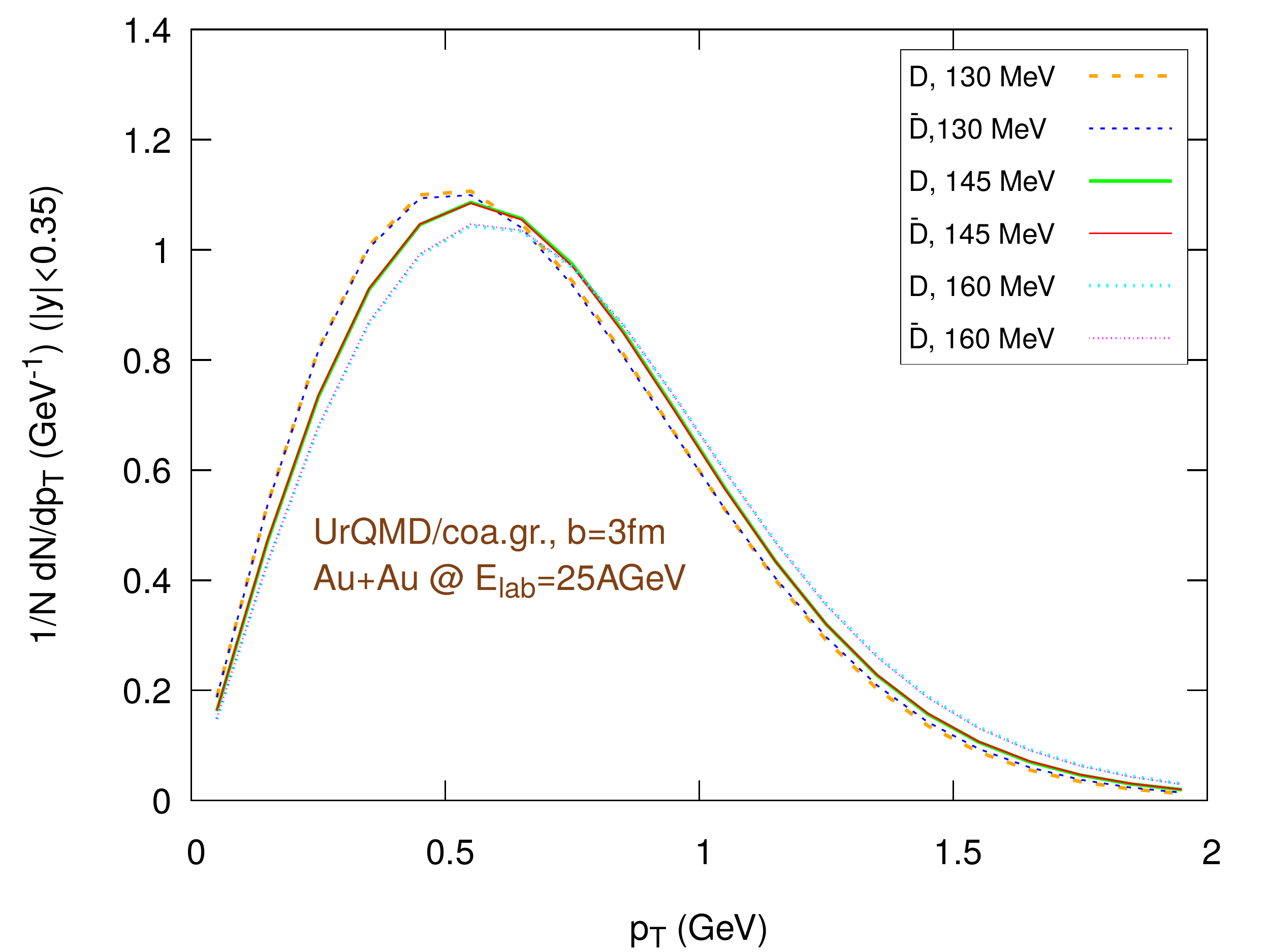}
	\end{minipage} \hspace{3mm}
	\begin{minipage}[b]{0.48\textwidth}
		\includegraphics[width=1\textwidth]{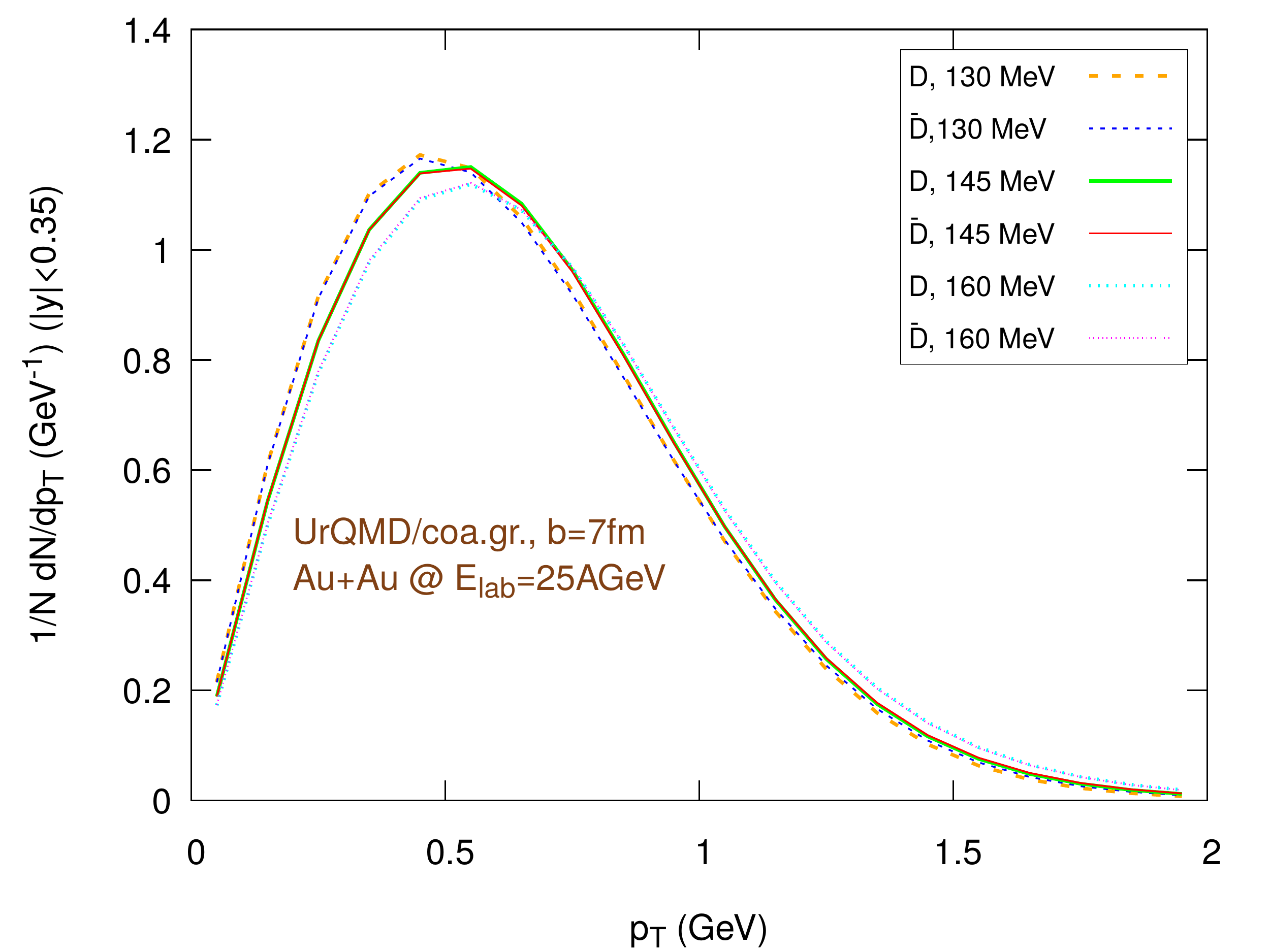}
	\end{minipage}
	\caption{(Color online) Normalized
		$1/N \dd N/\dd p_{\text{T}}$ distribution of the final $\D/\bar{\D}$ mesons,
		in the rapidity range $|y|<0.35$, for Au+Au collisions at
		$E_{\text{lab}}=25\,\AG$, using the UrQMD/coarse-graining
		approach. The hadronization parameters are $\epsilon_p=0.05$ and
		$\erw{r_{\D_{\mathrm{rms}}}}=0.6\,\textrm{fm}$. Left: $b=3\,\fm$,
		right: $b=7\,\fm$.}
	\label{dNdpT_cg}
\end{figure*}

\begin{figure*}[h]
	\begin{minipage}[b]{0.48\textwidth}
		\includegraphics[width=1\textwidth]{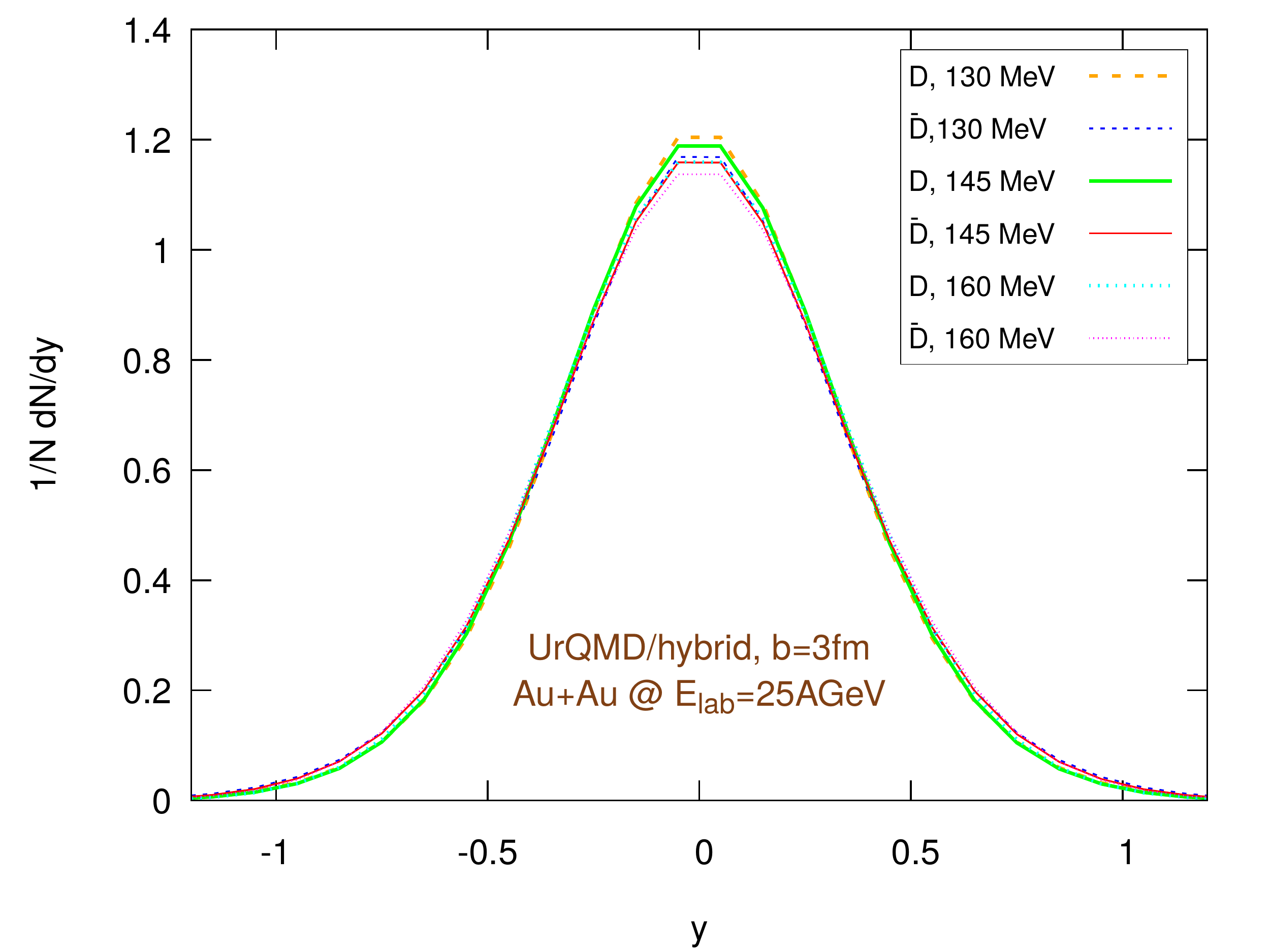}
	\end{minipage} \hspace{3mm}
	\begin{minipage}[b]{0.48\textwidth}
		\includegraphics[width=1\textwidth]{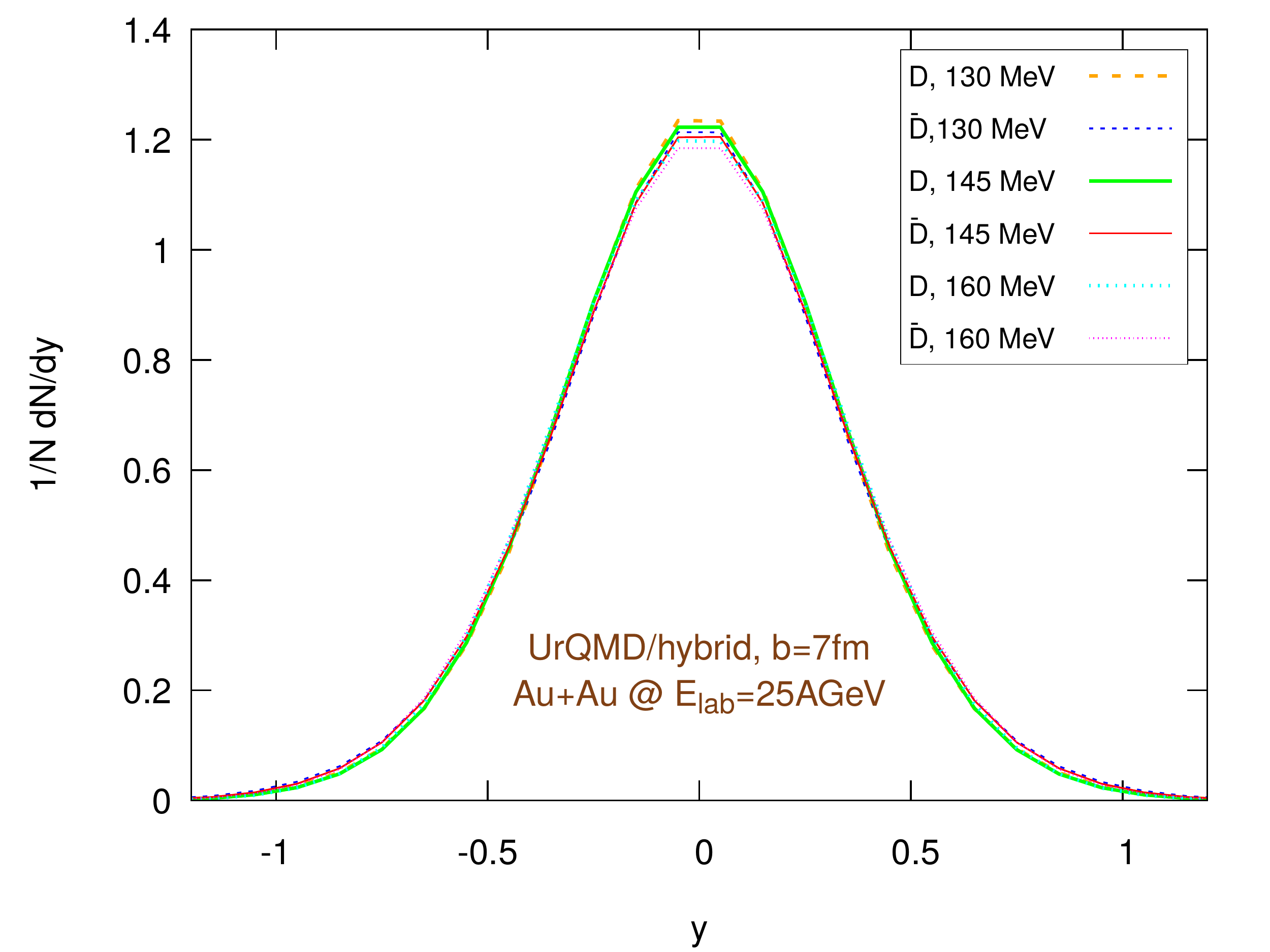}
	\end{minipage}
	\caption{(Color online) Final $1/N \dd N/\dd y$
		distribution using the UrQMD/hybrid model for Au+Au
		collisions at 25 GeV per nucleon in the lab frame, assuming
		different hadronization temperatures, with fixed parameters:
		$\epsilon_p=0.05$ and $\erw{r_{\D_{\mathrm{rms}}}}=0.6\,\textrm{fm}$.
		Left: $b=3\,\fm$, right: $b=7\,\fm$.}
	\label{dndy_hy}
\end{figure*}

\begin{figure*}[h]
	\begin{minipage}[b]{0.48\textwidth}
		\includegraphics[width=1\textwidth]{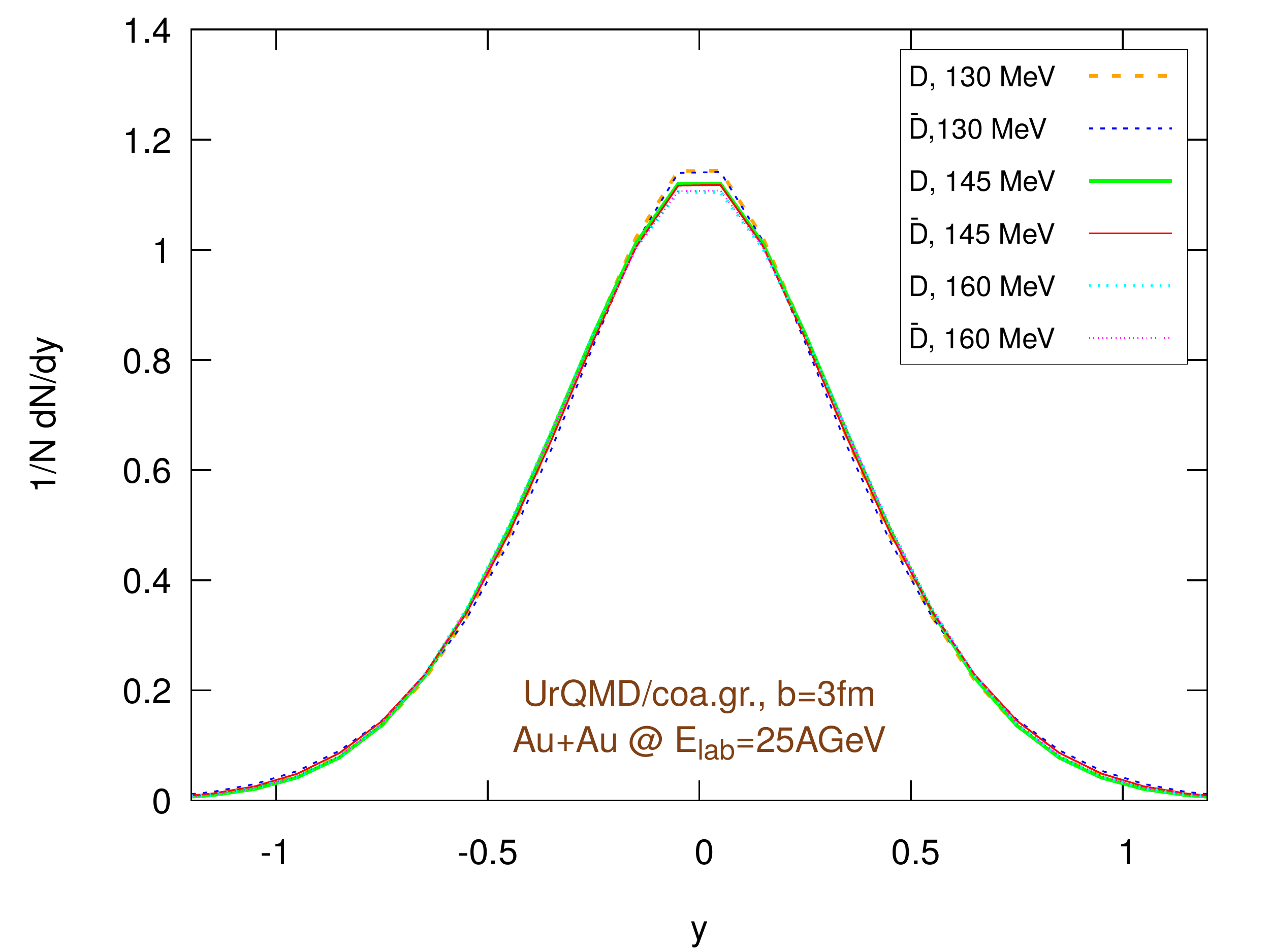}
	\end{minipage} \hspace{3mm}
	\begin{minipage}[b]{0.48\textwidth}
		\includegraphics[width=1\textwidth]{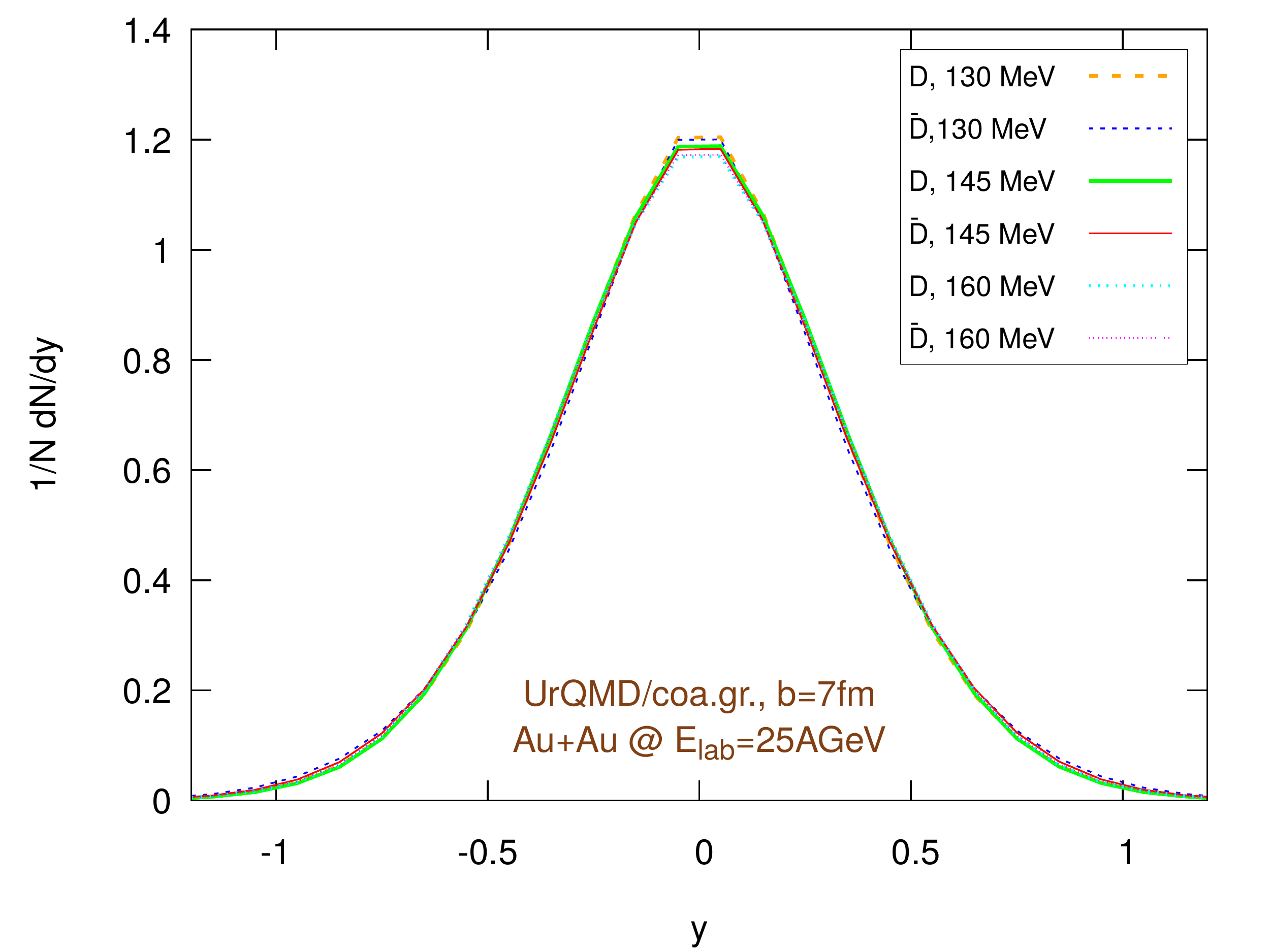}
	\end{minipage}
	\caption{(Color online) Final $1/N \dd N/\dd y$ distribution
		using the UrQMD/coarse-graining approach for Au+Au collisions
		at 25 GeV per nucleon in the lab frame, assuming different
		hadronization temperatures, with fixed parameters:
		$\epsilon_p=0.05$ and
		$\erw{r_{\D_{\mathrm{rms}}}}=0.6\,\textrm{fm}$. Left: $b=3\,\fm$,
		right: $b=7\,\fm$.}
	\label{dndy_cg}
\end{figure*}

\begin{figure*}[h]
	\begin{minipage}[b]{0.48\textwidth}
		\includegraphics[width=1\textwidth]{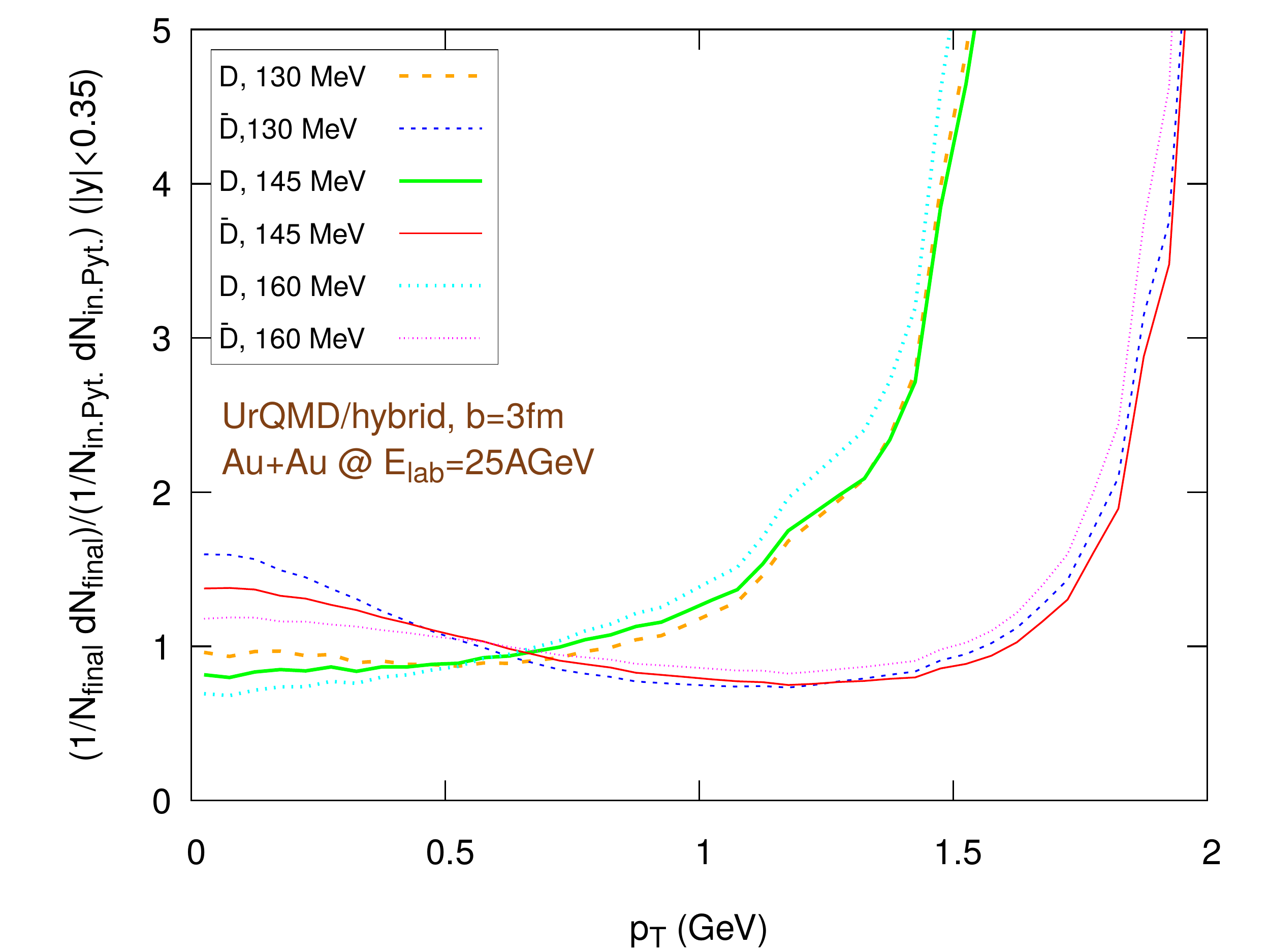}
	\end{minipage} \hspace{3mm}
	\begin{minipage}[b]{0.48\textwidth}
		\includegraphics[width=1\textwidth]{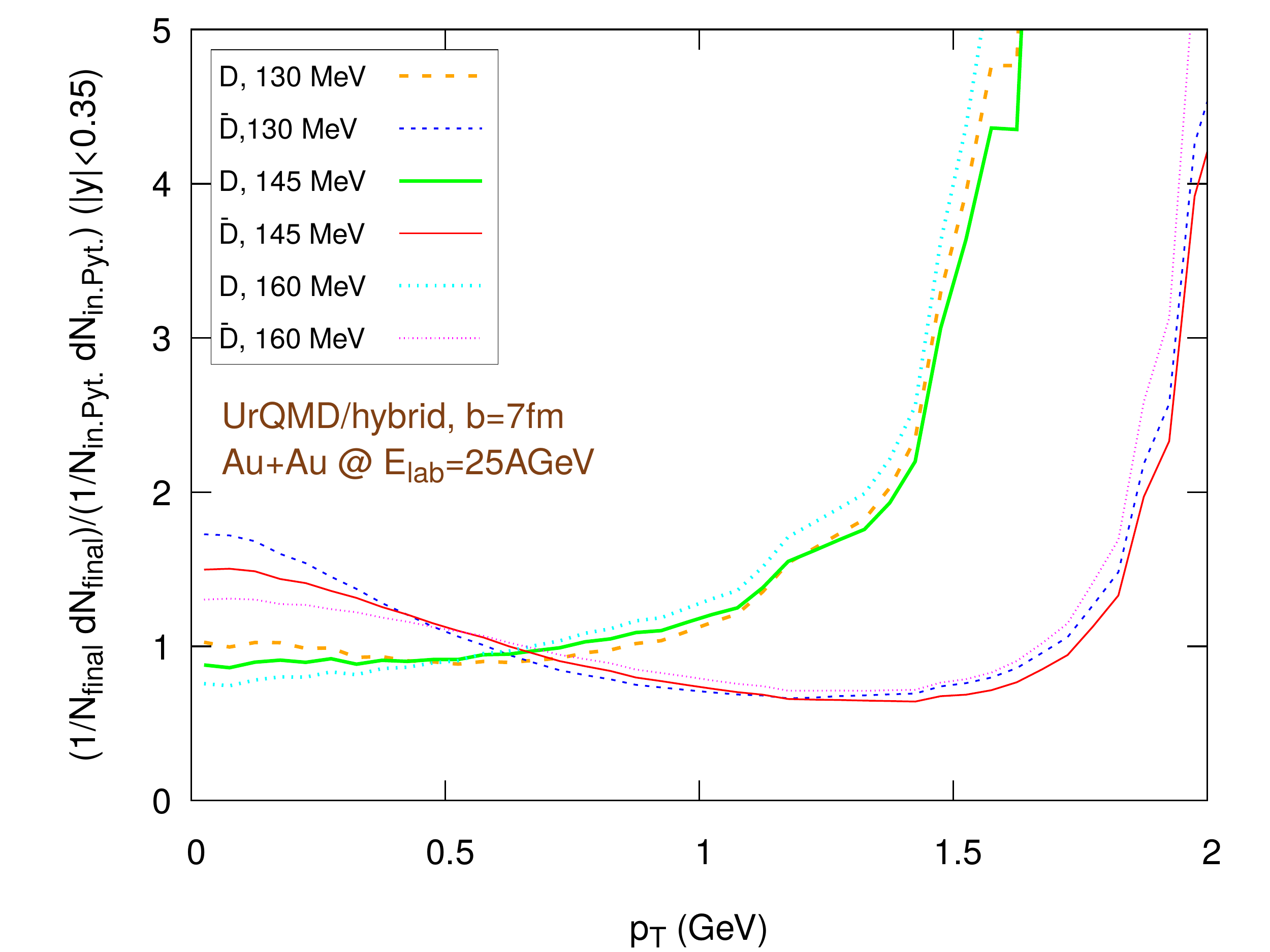}
	\end{minipage}
	\caption{(Color online)  $\tilde{R}_{AA}$, i.e. the ratio of the
		individually normalized distributions
		$1/N_{\text{final}} \dd N_{\text{final}}/\dd p_{\text{T}}$ in
		Au+Au collisions and
		$1/N_{\text{in. Pyt.}}  \dd N_{\text{in. Pyt.}}/\dd p_{\text{T}}$ in pp
		collisions (simulated with Pythia), in the rapidity range
		$|y|<0.35$, for Au+Au collisions at
          $E_{\text{lab}}=25\,\AG$, using the UrQMD/hybrid
          model. The hadronization parameters are $\epsilon_p=0.05$ and
          $\erw{r_{\D_{\mathrm{rms}}}}=0.6\,\fm$. Left: $b=3\,\fm$,
          right: $b=7\,\fm$.} 
	\label{raa_hy}
\end{figure*}

\begin{figure*}[h]
	\begin{minipage}[b]{0.48\textwidth}
		\includegraphics[width=1\textwidth]{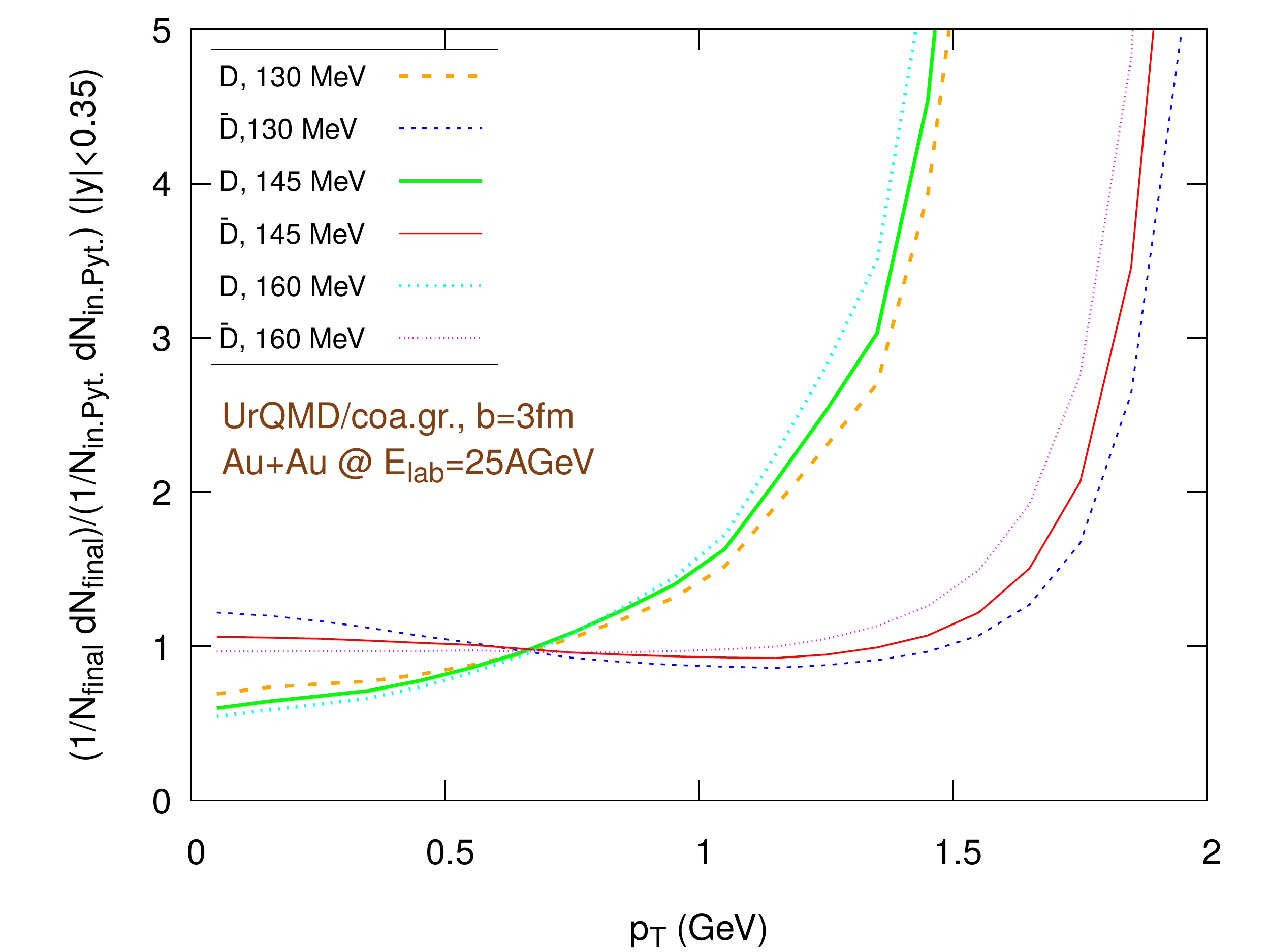}
	\end{minipage} \hspace{3mm}
	\begin{minipage}[b]{0.48\textwidth}
		\includegraphics[width=1\textwidth]{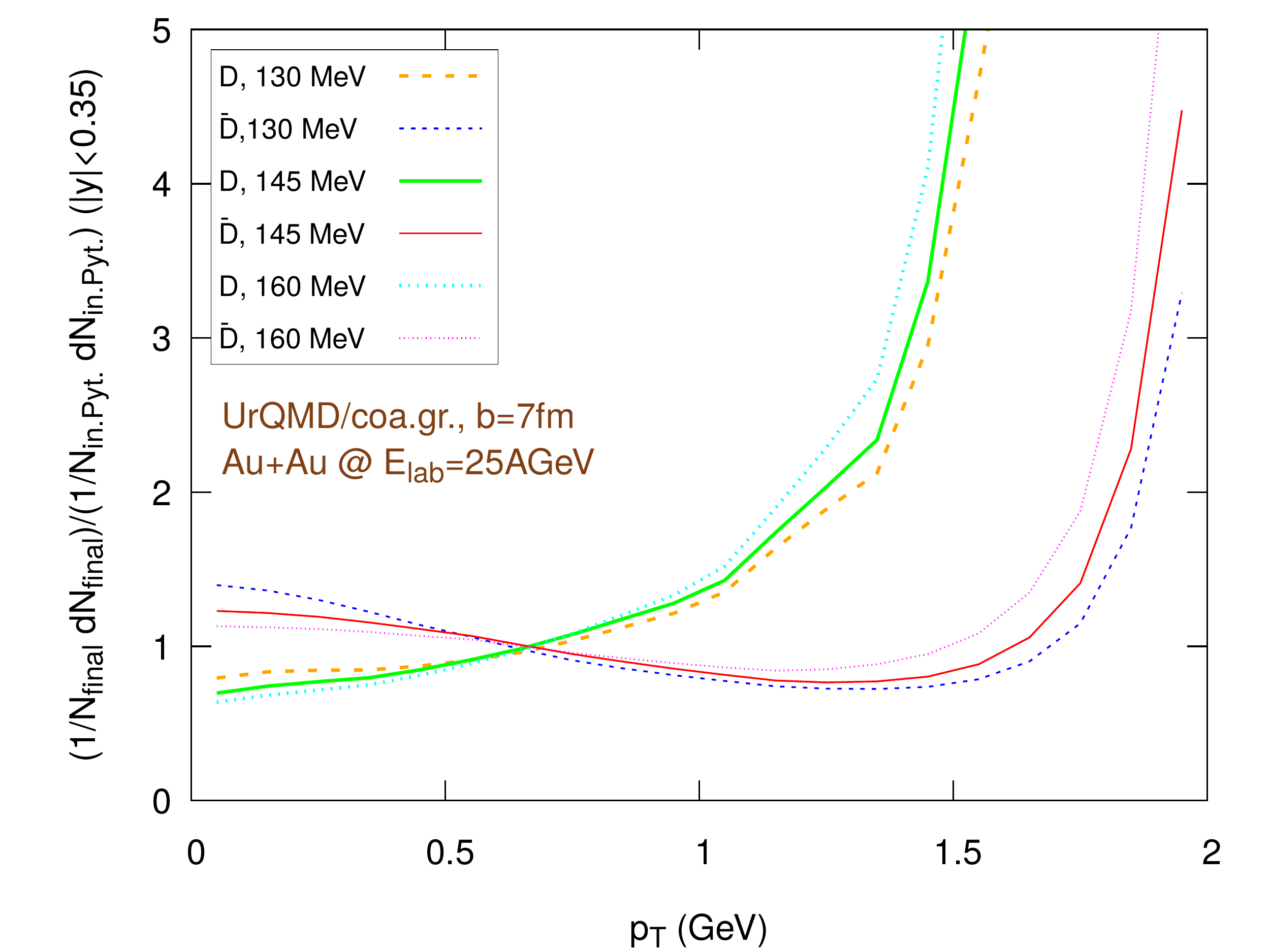}
	\end{minipage}
	\caption{(Color online) $\tilde{R}_{AA}$, i.e. the ratio of the
		individually normalized distributions
		$1/N_{\text{final}} \dd N_{\text{final}}/\dd p_{\text{T}}$ in
		Au+Au collisions and
		$1/N_{\text{in. Pyt.}}  \dd N_{\text{in. Pyt.}}/\dd p_{\text{T}}$ in pp
		collisions (simulated with Pythia), in the rapidity range
		$|y|<0.35$, for Au+Au collisions at
		$E_{\text{lab}}=25\,\AG$, using the UrQMD/coarse-graining approach. The hadronization parameters are $\epsilon_p=0.05$ and
		$\erw{r_{\D_{\mathrm{rms}}}}=0.6\,\fm$. Left: $b=3\,\fm$,
		right: $b=7\,\fm$.} 
	\label{raa_cg}
\end{figure*}

\begin{figure*}[h]
	\begin{minipage}[b]{0.48\textwidth}
		\includegraphics[width=1\textwidth]{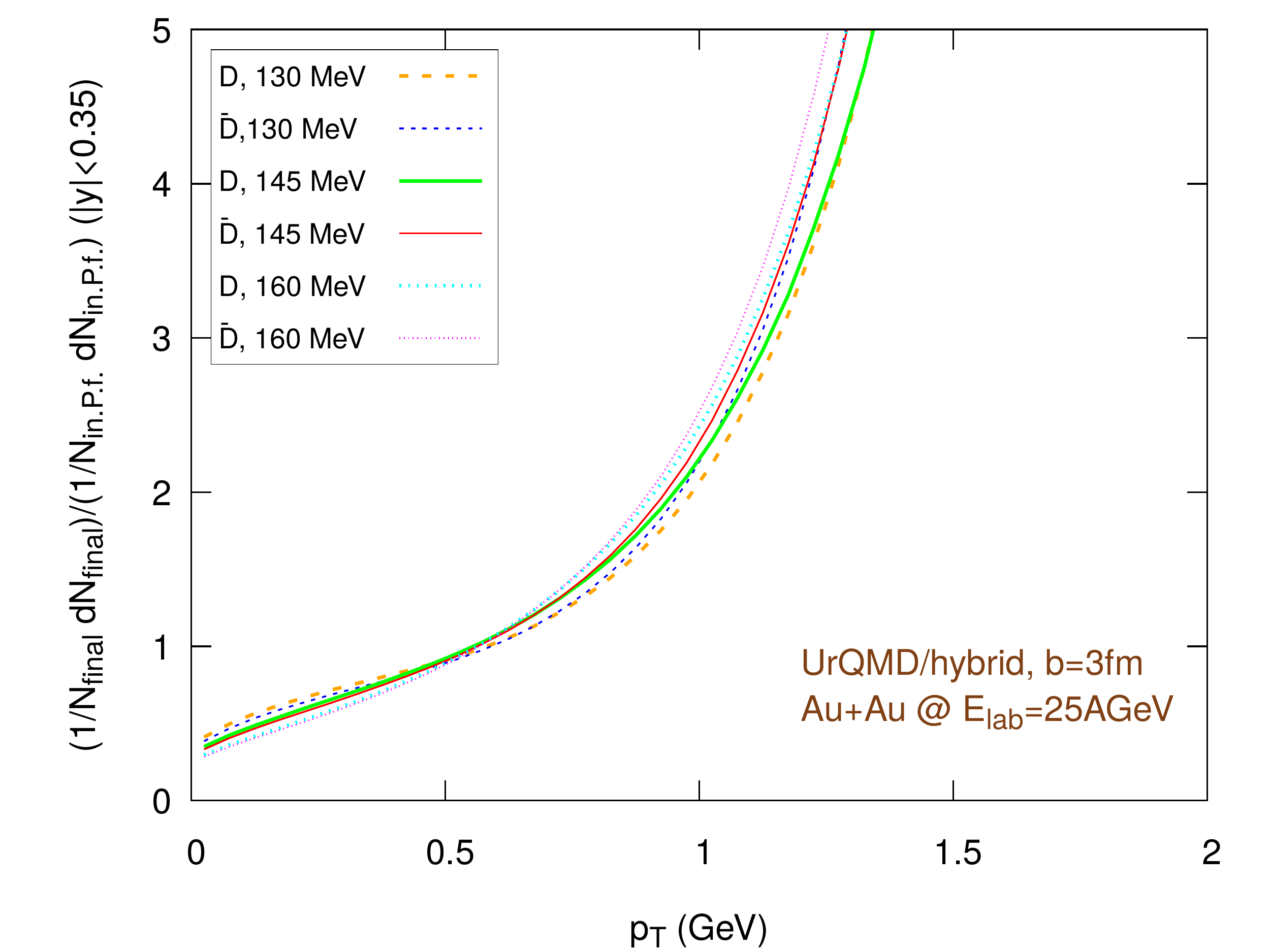}
	\end{minipage} \hspace{3mm}
	\begin{minipage}[b]{0.48\textwidth}
		\includegraphics[width=1\textwidth]{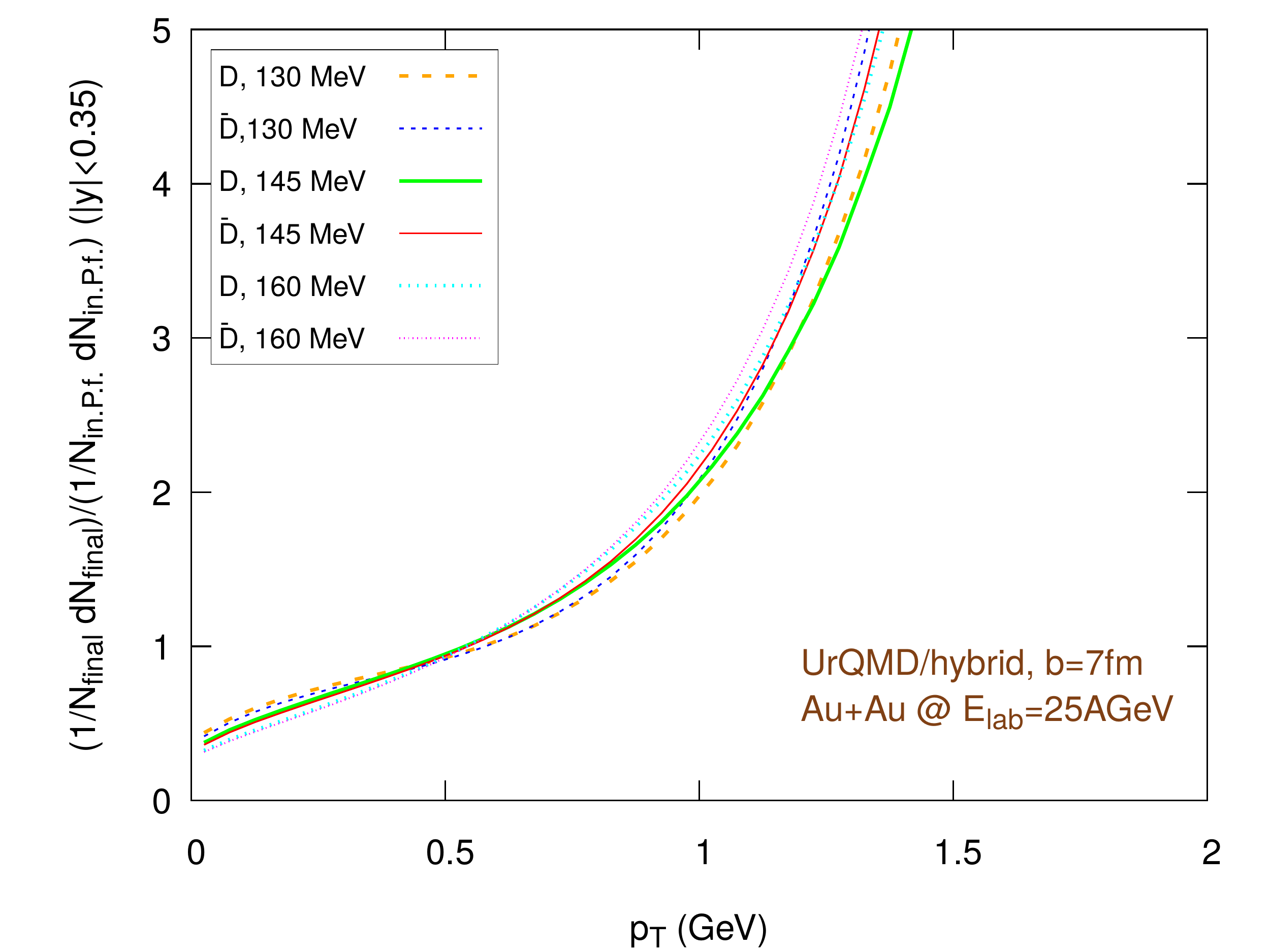}
	\end{minipage}
	\caption{(Color online) $\tilde{R}_{AA}$, i.e. the ratio of the
		individually normalized distributions
		$1/N_{\text{final}} \dd N_{\text{final}}/\dd p_{\text{T}}$ in
		Au+Au collisions and
		$1/N_{\text{in. Pyt.}}  \dd N_{\text{in. P. f.}}/\dd p_{\text{T}}$ in pp
		collisions (Pythia + Peterson fragmentation), in the rapidity range
		$|y|<0.35$, for Au+Au collisions at
		$E_{\text{lab}}=25\,\AG$, using the UrQMD/hybrid model. The hadronization parameters are $\epsilon_p=0.05$ and
		$\erw{r_{\D_{\mathrm{rms}}}}=0.6\,\fm$. Left: $b=3\,\fm$,
		right: $b=7\,\fm$.}
	\label{raa_sim_hy}
\end{figure*}

\begin{figure*}[h]
	\begin{minipage}[b]{0.48\textwidth}
		\includegraphics[width=1\textwidth]{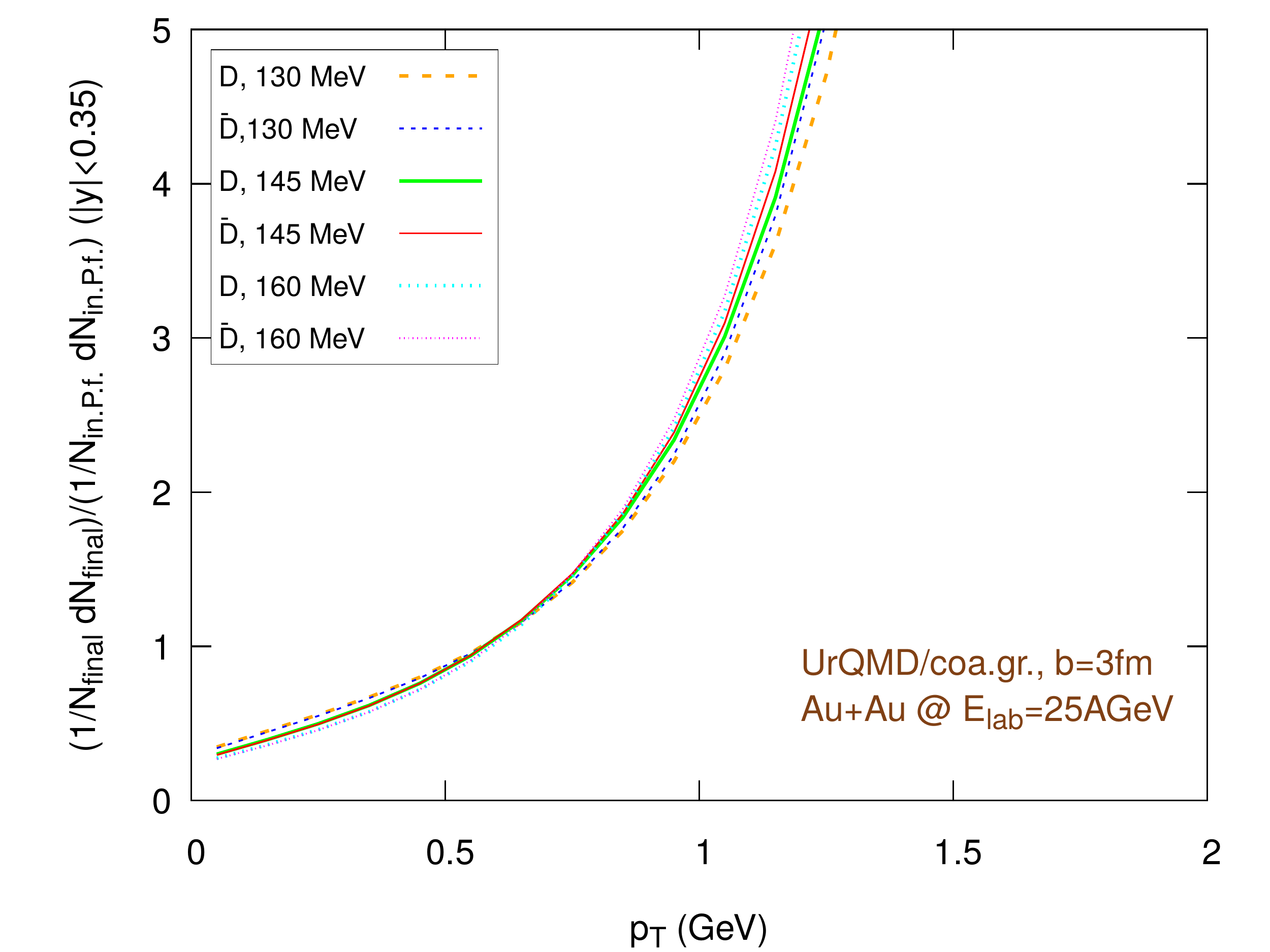}
	\end{minipage} \hspace{3mm}
	\begin{minipage}[b]{0.48\textwidth}
		\includegraphics[width=1\textwidth]{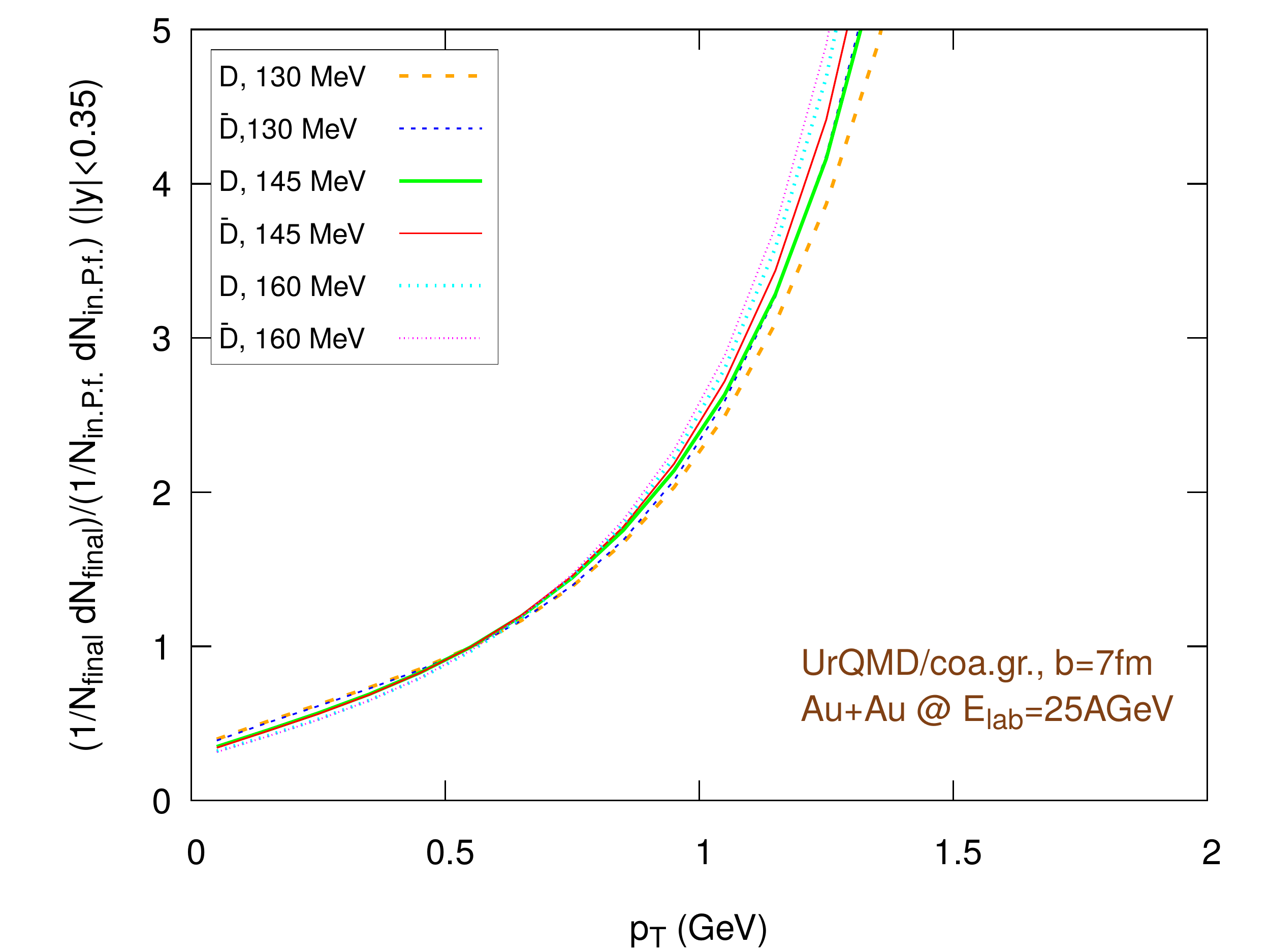}
	\end{minipage}
	\caption{(Color online) $\tilde{R}_{AA}$, i.e. the ratio of the
		individually normalized distributions
		$1/N_{\text{final}} \dd N_{\text{final}}/\dd p_{\text{T}}$ in
		Au+Au collisions and
		$1/N_{\text{in. Pyt.}}  \dd N_{\text{in. P. f.}}/\dd p_{\text{T}}$ in pp
		collisions (Pythia + Peterson fragmentation), in the rapidity range
		$|y|<0.35$, for Au+Au collisions at
		$E_{\text{lab}}=25\,\AG$, using the UrQMD/coarse-graining approach. The hadronization parameters are $\epsilon_p=0.05$ and
		$\erw{r_{\D_{\mathrm{rms}}}}=0.6\,\fm$. Left: $b=3\,\fm$,
		right: $b=7\,\fm$.}
	\label{raa_sim_cg}
\end{figure*}

\begin{figure*}[h]
	\begin{minipage}[b]{0.48\textwidth}
		\includegraphics[width=1\textwidth]{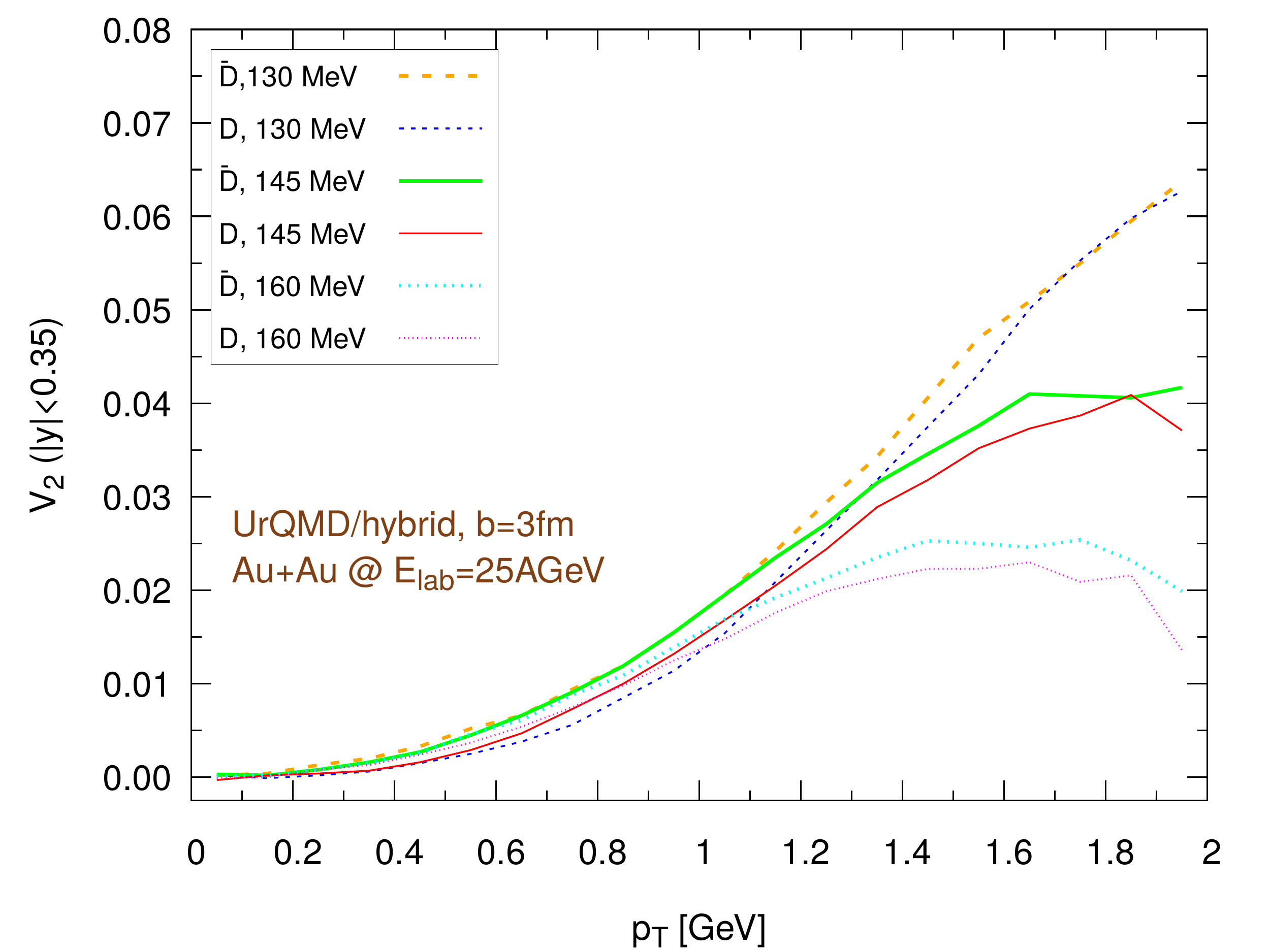}
	\end{minipage} \hspace{3mm}
	\begin{minipage}[b]{0.48\textwidth}
		\includegraphics[width=1\textwidth]{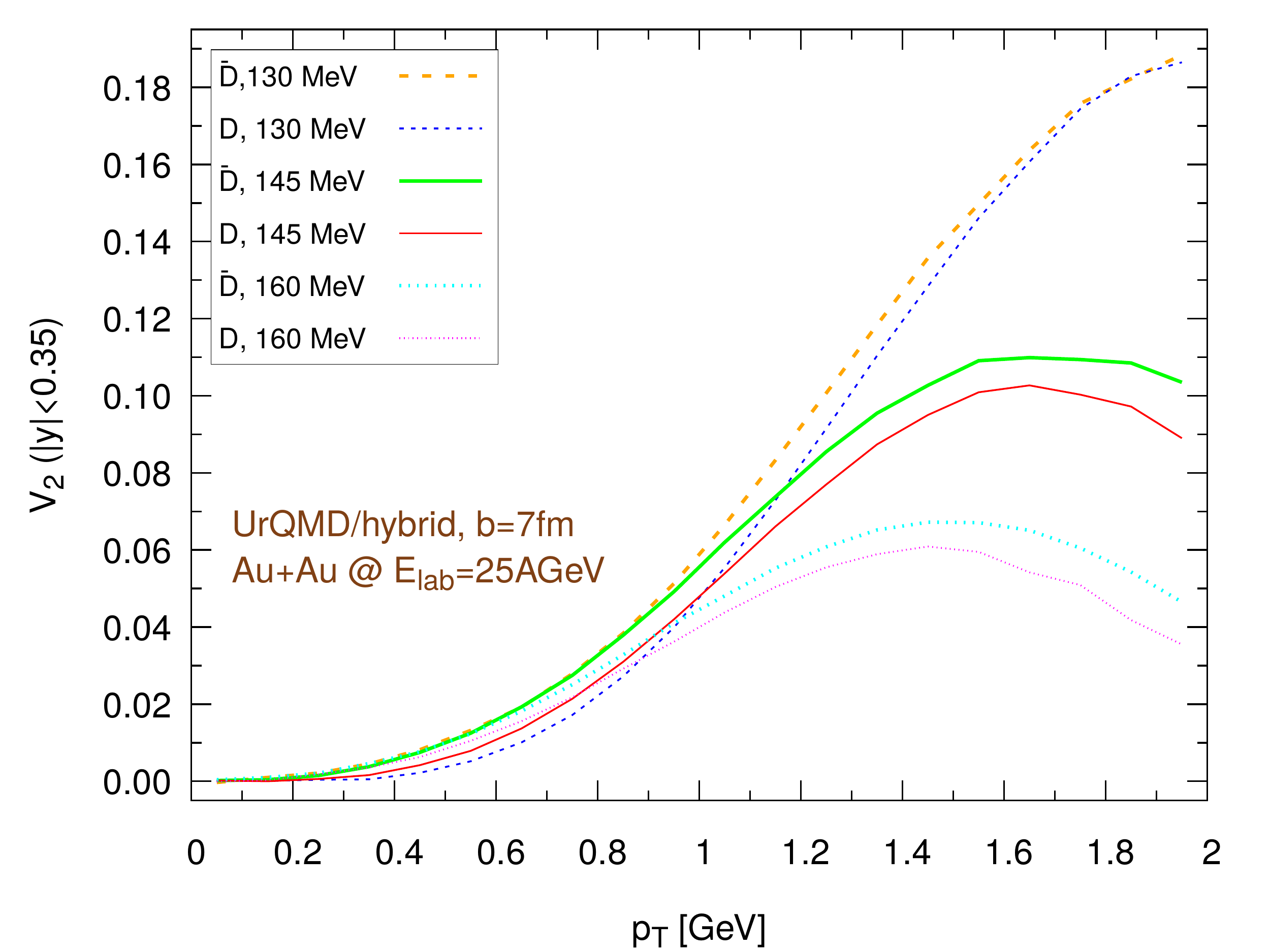}
	\end{minipage}
	\caption{(Color online) Elliptic flow of $\D/\bar{\D}$ mesons
		($|y|<0.35$) within the UrQMD/hybrid approach in Au+Au
		collisions at $E_{\text{lab}}=25$ AGeV. We show different
		hadronization temperatures, with fixed parameters:
		$\epsilon_p=0.05$ and $\erw{r_{\D_{\mathrm{rms}}}}=0.6\,\fm$.  Left: $b=3\,\fm$,
		right: $b=7\,\fm$. (Note the
		different scales on the ordinate.)}
	\label{v2-hydro-full}
\end{figure*}

\begin{figure*}[h]
	\begin{minipage}[b]{0.48\textwidth}
		\includegraphics[width=1\textwidth]{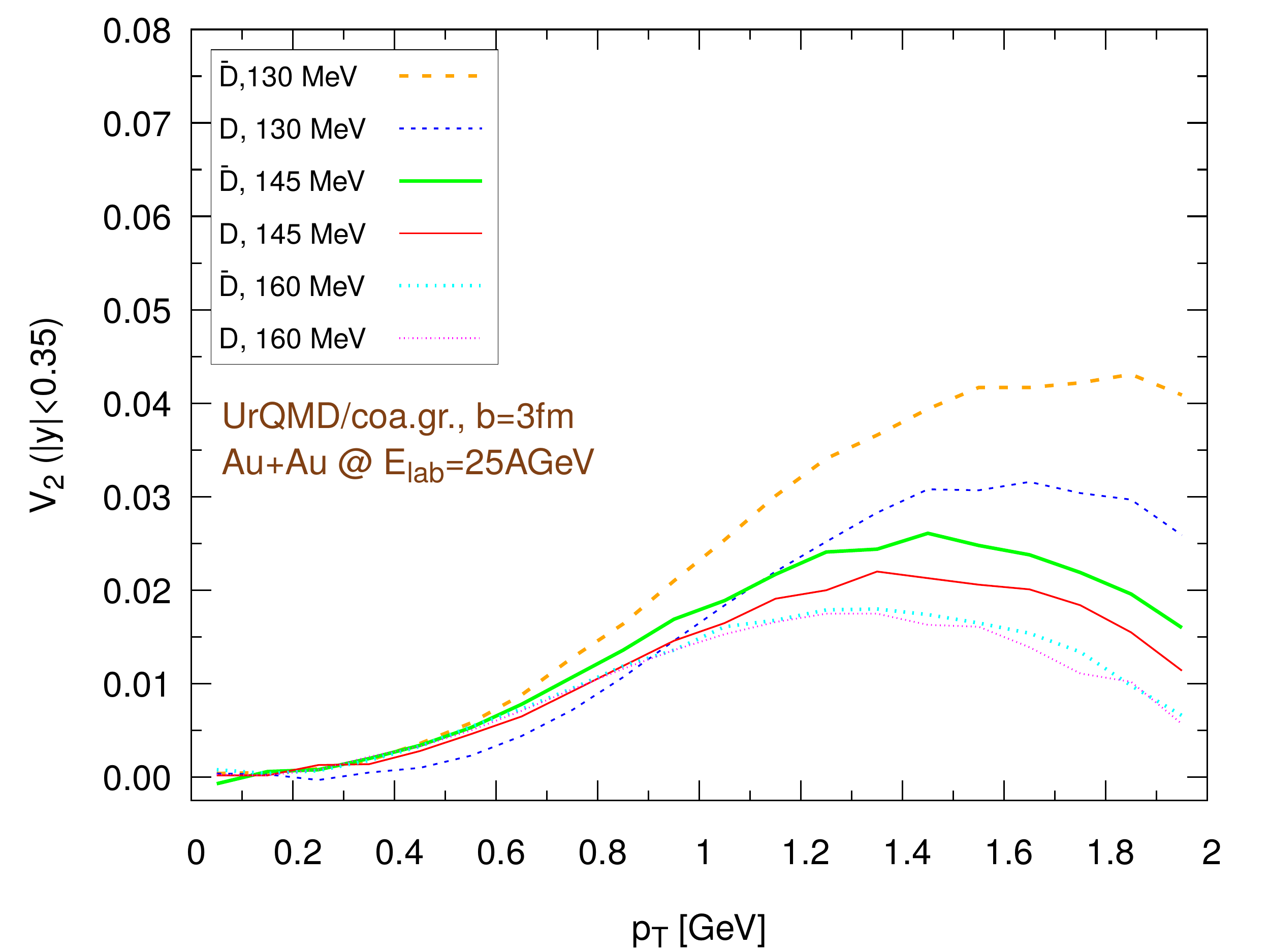}
	\end{minipage} \hspace{3mm}
	\begin{minipage}[b]{0.48\textwidth}
		\includegraphics[width=1\textwidth]{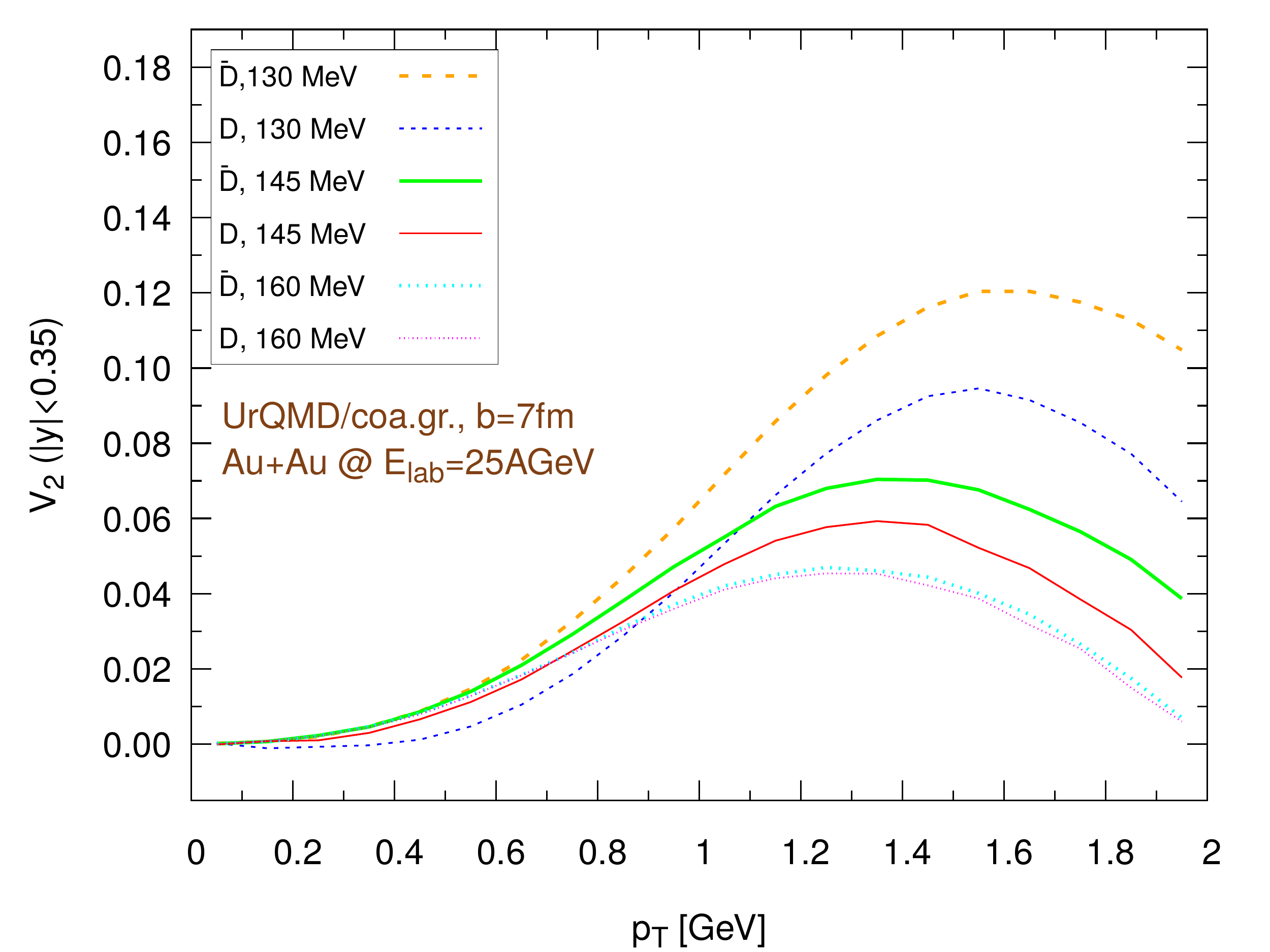}
	\end{minipage}
	\caption{(Color online) Elliptic flow of $\D/\bar{\D}$ mesons
		($|y|<0.35$) within the UrQMD/coarse-graining approach in
		Au+Au collisions at $E_{\text{lab}}=25$ AGeV. We show different
		hadronization temperatures, with fixed parameters:
		$\epsilon_p=0.05$ and $\erw{r_{\D_{\mathrm{rms}}}}=0.6\,\fm$.  Left: $b=3\,\fm$,
		right: $b=7\,\fm$.}
	\label{v2-cg-full}
\end{figure*}

\begin{figure*}[h]
	\begin{minipage}[b]{0.48\textwidth}
		\includegraphics[width=1\textwidth]{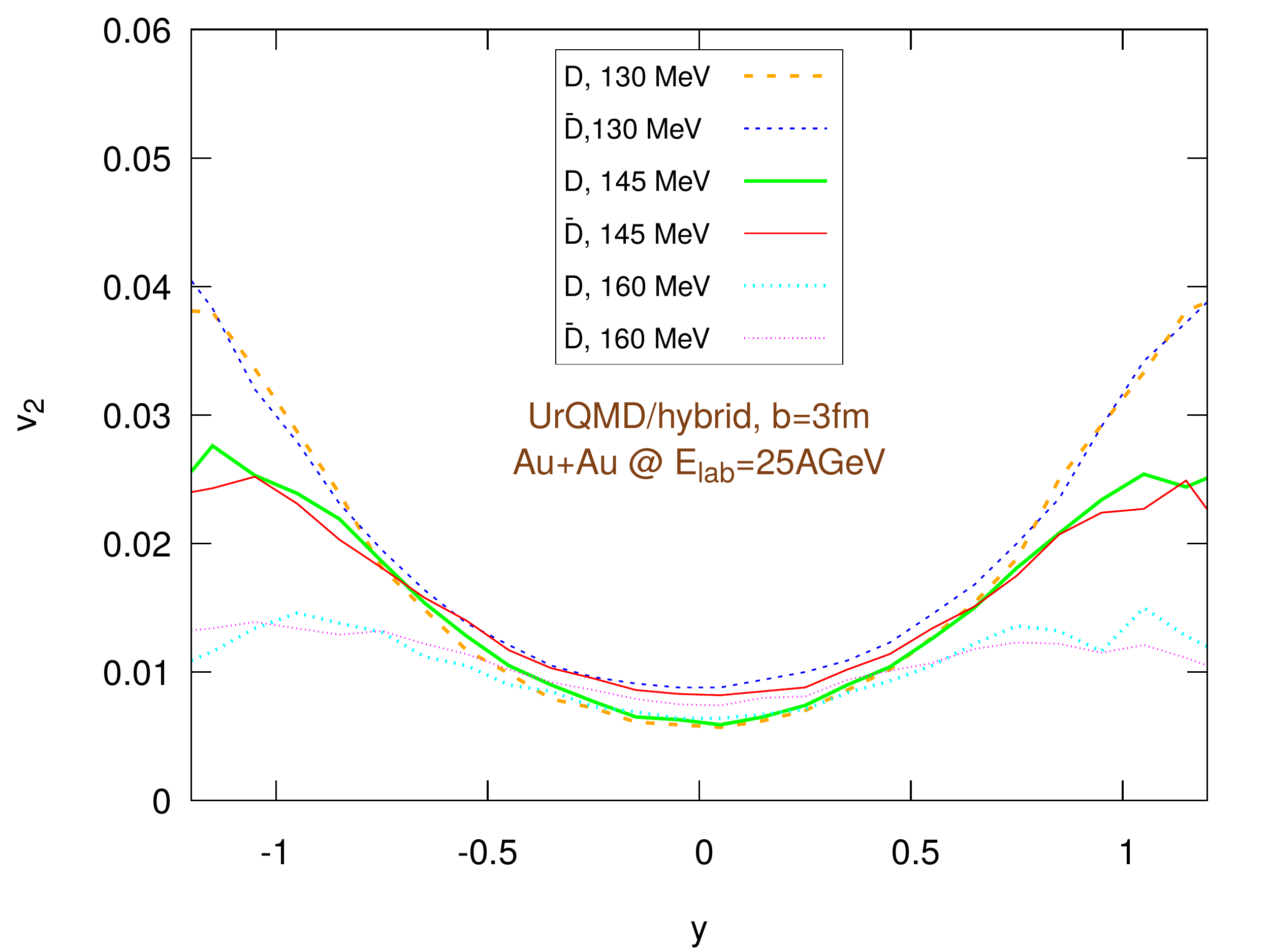}
	\end{minipage} \hspace{3mm}
	\begin{minipage}[b]{0.48\textwidth}
		\includegraphics[width=1\textwidth]{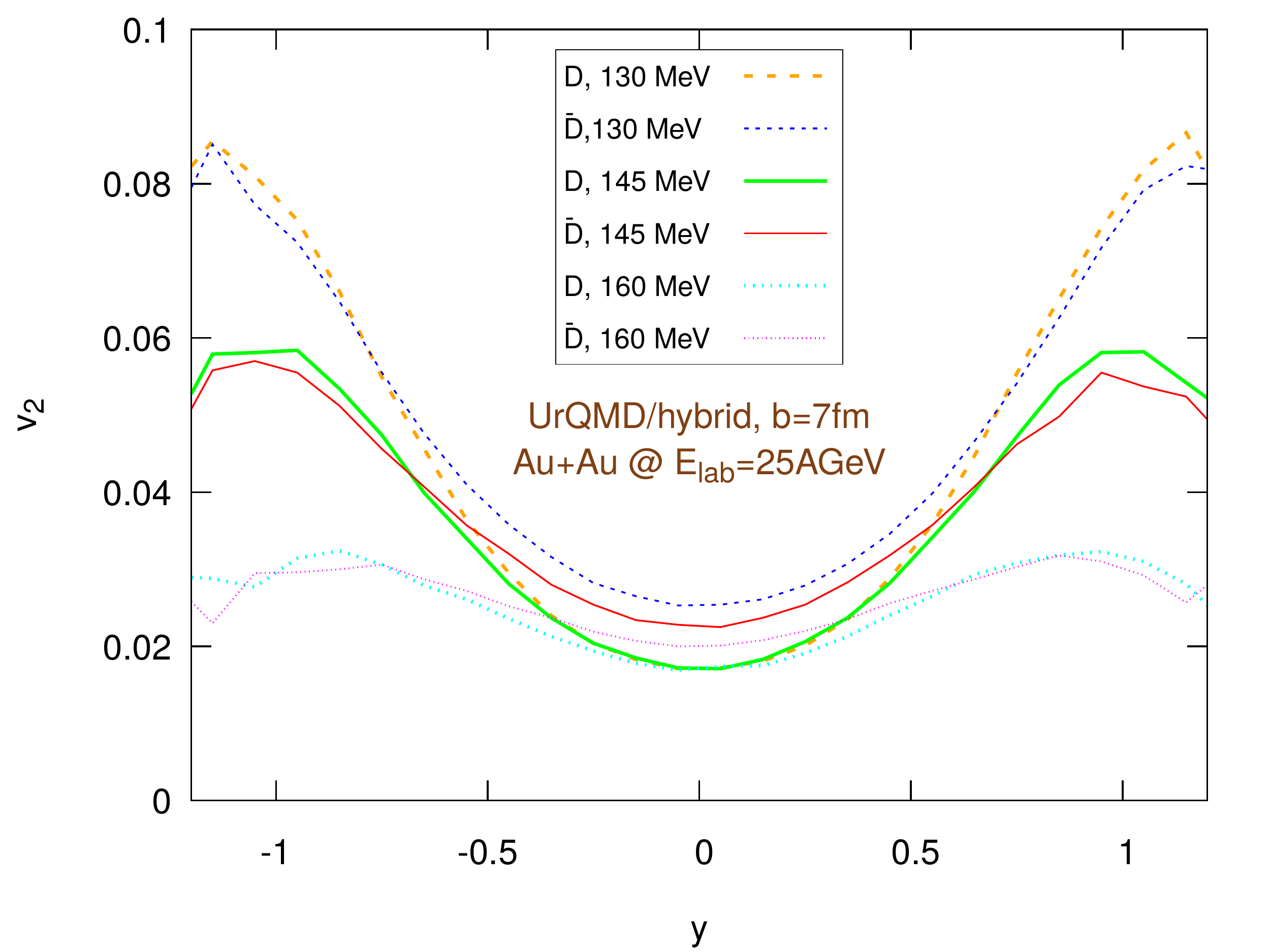}
	\end{minipage}
	\caption{(Color online) Elliptic flow of $\D/\bar{\D}$ mesons
		with respect to rapidity within the UrQMD/hybrid model in
		Au+Au collisions at $E_{\text{lab}}=25$ AGeV. We show different
		hadronization temperatures, with fixed parameters:
		$\epsilon_p=0.05$ and $\erw{r_{\D_{\mathrm{rms}}}}=0.6\,\fm$. Left: $b=3\,\fm$,
		right: $b=7\,\fm$.}
	\label{v2y_hy}
\end{figure*}

\begin{figure*}[h]
	\begin{minipage}[b]{0.48\textwidth}
		\includegraphics[width=1\textwidth]{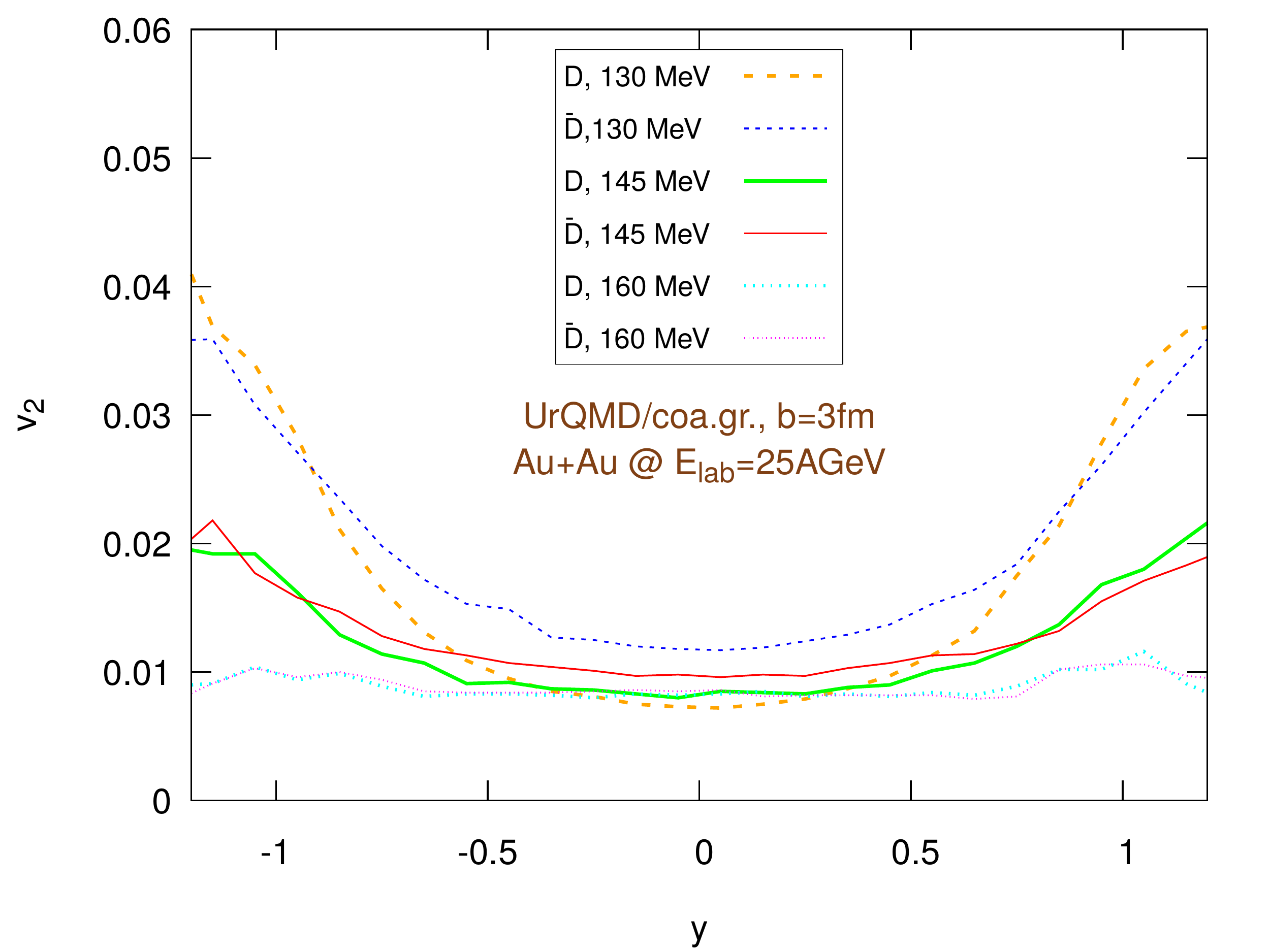}
	\end{minipage} \hspace{3mm}
	\begin{minipage}[b]{0.48\textwidth}
		\includegraphics[width=1\textwidth]{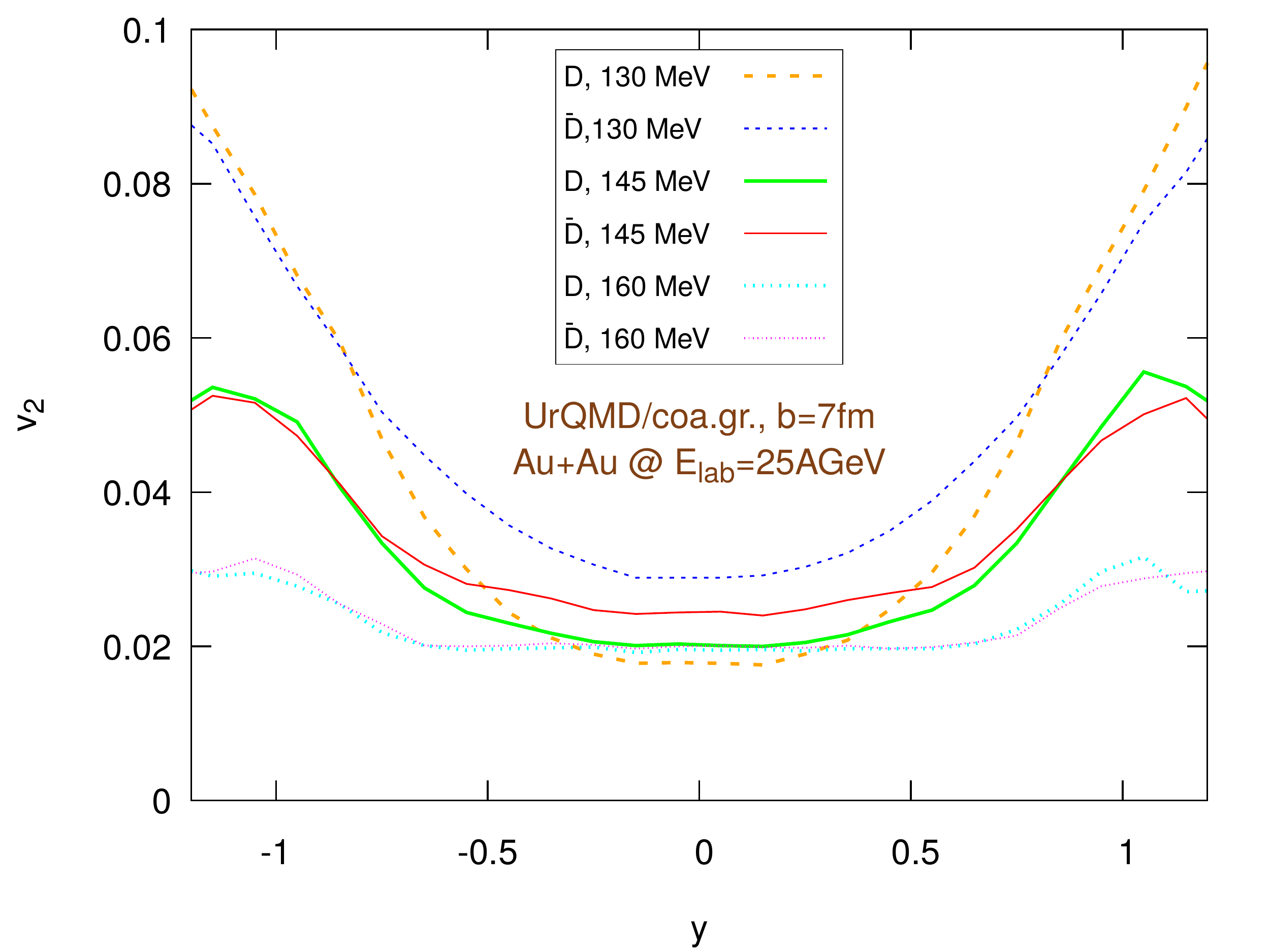}
	\end{minipage}
	\caption{(Color online) Elliptic flow of $\D/\bar{\D}$ mesons
		    with respect to rapidity within the UrQMD/coarse-graining approach in
			Au+Au collisions at $E_{\text{lab}}=25$ AGeV. We show different
			hadronization temperatures, with fixed parameters:
			$\epsilon_p=0.05$ and $\erw{r_{\D_{\mathrm{rms}}}}=0.6\,\fm$.  Left: $b=3\,\fm$,
			right: $b=7\,\fm$.}
	\label{v2y_cg}
\end{figure*}

We simulate Au+Au reactions at $E_{\text{lab}}=25\,A\GeV$ at fixed
impact parameters $b=3\,\fm$ and $b=7\,\fm$. The simulations are
performed both for the hybrid set-up and the coarse-graining
approach. The elliptic flow at mid-rapidity is calculated in the
reaction plane as:
\begin{equation}
v_2(p_{\text{T}})=\left\langle\dfrac{p_x^2-p_y^2}{p_x^2+p_y^2}\right\rangle,\quad
|y|<0.35,
\label{v2-def}
\end{equation} 
where $E$, $p_x$, $p_y$ and $p_z$ are the four-momentum components,
$p_{\text{T}}=\sqrt{p_x^2+p_y^2}$ is the transverse momentum,
$y=1/2\ln[(E+p_z)/(E-p_z)]$ is the rapidity and the averages are made
over all charm particles produced in all the events of a series at a
given impact parameter.
 
\subsection{Dependence on the hadronization temperature} 

To explore the sensitivity of the $\D/\bar{\D}$ elliptic flow and momentum distribution on the
lifetime of the partonic phase, we evaluate the effect of three
different hadronization temperatures: $160\,\MeV$, $145\,\MeV$ and
$130\,\MeV$. In all cases we perform the Langevin propagation until the
local temperature of the computational cell is above $60\,\MeV$. In
(Eq.\ \ref{prob_to_coalescence}), which gives the probability to
hadronize by coalescence, we set $\erw{r_{\D_{\mathrm{rms}}}}=0.6\,\fm$, while
in the fragmentation function (Eq.\ \ref{peterson_frag_eq}) we set
$\epsilon_p=0.05$.

 Figures \ref{dNdpT_hy} and \ref{dNdpT_cg} show the transverse momentum distribution for the final $\D/\bar{\D}$ mesons in the UrQMD/hybrid model and in the UrQMD/coarse-graining approach, respectively, while figures \ref{dndy_hy} and \ref{dndy_cg} show their rapidity distributions.
 
 Figures \ref{raa_hy} and \ref{raa_cg} show $\tilde{R}_{AA}$,
 i.e.
 $\tilde{R}_{AA}=\dfrac{1/N_{AA}\dd N/\dd p_{\text{T}} |_{AA}}{1/N_{\text{pp}}\dd
 	N/\dd p_{\text{T}} |_{\text{pp}}}$,  where the distribution in $\text{pp}$ is taken from
 Fig.~(\ref{dmesons_in_pp}, left). In particular, Fig. \ref{raa_hy} refers to the UrQMD/hybrid model, while Fig. \ref{raa_cg} refers to the coarse graining approach. The left and right sides of the figures refer to reactions at fixed impact parameter $b=3\,\fm$ and $b=7\,\fm$, respectively. A general trend observed in both
 scenarios and for both impact parameters is the strong increase of
 $\tilde{R}_{AA}$ with increasing transverse momentum. This effect is due to
 energy conservation, which limits the maximum $p_{\mathrm{T}}$ available
 in pp reactions to
 $p^{\text{max}}_T=(\sqrt{s_{\text{pp}}}-2m_p)/2\simeq2.5\,\GeV$. Therefore,
 we expect and observe this in the $R_{AA}$ as a strong increase.
 
 To explore more in depth the uncertainties of the initial state, Figures
 \ref{raa_sim_hy} and \ref{raa_sim_cg} show the same $\tilde{R}_{AA}$
 distributions as before, however now with a different pp
 baseline. Instead of $\D$-mesons from Pythia, we extract the charm
 quarks from Pythia in pp and hadronize them according to the Peterson
 fragmentation. As in the previous case, we observe a good consistency between the results coming from the UrQMD/hydro and the UrQMD/coarse-graining models. However, although essential features like the rise of $\tilde{R}_{AA}$ at ``high'' $p_T$ do not change when
 switching between the pp baselines, from a quantitative perspective there are noticeable differences. In particular, in Figures \ref{raa_sim_hy} and \ref{raa_sim_cg} we miss the strong distinction between the $\tilde{R}_{AA}$ of particles and anti-particles visible in Figures \ref{raa_hy} and \ref{raa_cg}, due to the internal Pythia non-perturbative machinery and the inclusion of additional hadronization channels, already mentioned at the beginning of Sect. (\ref{implementation}), which introduces a sharp difference in the spectra of $D$ and $\bar{D}$ mesons, clearly shown in Fig. (\ref{dmesons_in_pp}). On the other hand, it is well known that Pythia focuses on high-energy collisions and results at low energies obtained with non-tuned default program parameters should be taken with care. Anyway, the differences in the $\tilde{R}_{AA}$ depending on the chosen pp baseline suggest that Peterson fragmentation might tend to overlook important details of the hadronization process and they call for the development and/or the adoption of more sophisticated models. 
Regarding the normalized momentum distribution with respect to the rapidity, again we observe a good agreement between the UrQMD/hybrid model (Fig.~\ref{dndy_hy}) and the UrQMD/coarse-graining approach (Fig.~\ref{dndy_cg}). In both cases, for more central collisions we can observe a slightly more evident distinction between particles and anti-particles, in particular for lower hadronization temperature, associated with a small broadening of the distributions. These small effects are consistent with the expected larger interaction with the medium for $b=3\,\fm$.\\
  
The results for the elliptic flow with respect to the transverse momentum are shown in Fig.\ \ref{v2-hydro-full}, in the case of the UrQMD/hybrid model, and in Fig.\ \ref{v2-cg-full},  in the case of the UrQMD/coarse-graining approach. In all cases we observe that the elliptic flow
of $\bar{\D}$ is larger than the elliptic flow of $\D$. As expected this
is because of the fugacity factor which, in the partonic phase, enhances
the transport coefficients for $\bar{\D}$ and suppresses the transport
coefficients for $\D$. We also observe that the elliptic flow is higher
for lower hadronization temperatures. With a larger time spent in the
partonic phase, the larger magnitude of the transport coefficients in
this phase compared to the hadronic phase leads to a stronger elliptic
flow. By comparing $b=3\,\fm$ and the $b=7\,\fm$ collisions in figures
(\ref{v2-hydro-full}) and (\ref{v2-cg-full}) we notice that the $v_2$
for collisions having an impact parameter $b=7\,\fm$ is larger than the
$v_2$ for collisions with $b=3\,\fm$. This behavior is consistent with
the more anisotropic initial energy density spatial distribution in more
peripheral collisions. By comparing Fig.\ \ref{v2-hydro-full} with
Fig.\ \ref{v2-cg-full}, we observe that the $v_2$ in the case of the
UrQMD/hybrid approach is larger than the $v_2$ in the case of the
UrQMD/coarse-graining approach, showing the effects of the different
viscosities in the two different modelings of the medium. In the
UrQMD/coarse-graining approach the enhancement of the elliptic flow when
switching from $b=3\,\fm$ to $b=7\,\fm$ is weaker than in the UrQMD/hybrid
approach. This also indicates that partial thermalization might play a
role.\\
Figures \ref{v2y_hy} and \ref{v2y_cg} show the dependence of the elliptic flow with respect to the rapidity. We observe that the $\bar{\D}$ mesons have a significantly larger elliptic flow than the $\D$ mesons only in the central rapidity region and for lower hadronization temperatures, in particular for peripheral collisions. Moreover, as a general trend, $v_2$ exhibits a minimum for $y=0$, nevertheless in the UrQMD/coarse-graining case the growth of $v_2$ moving away from the central rapidity region becomes important only for $|y|\gtrapprox 0.5$.

\begin{figure*}[h]
	\begin{minipage}[b]{0.48\textwidth}
		\includegraphics[width=1\textwidth]{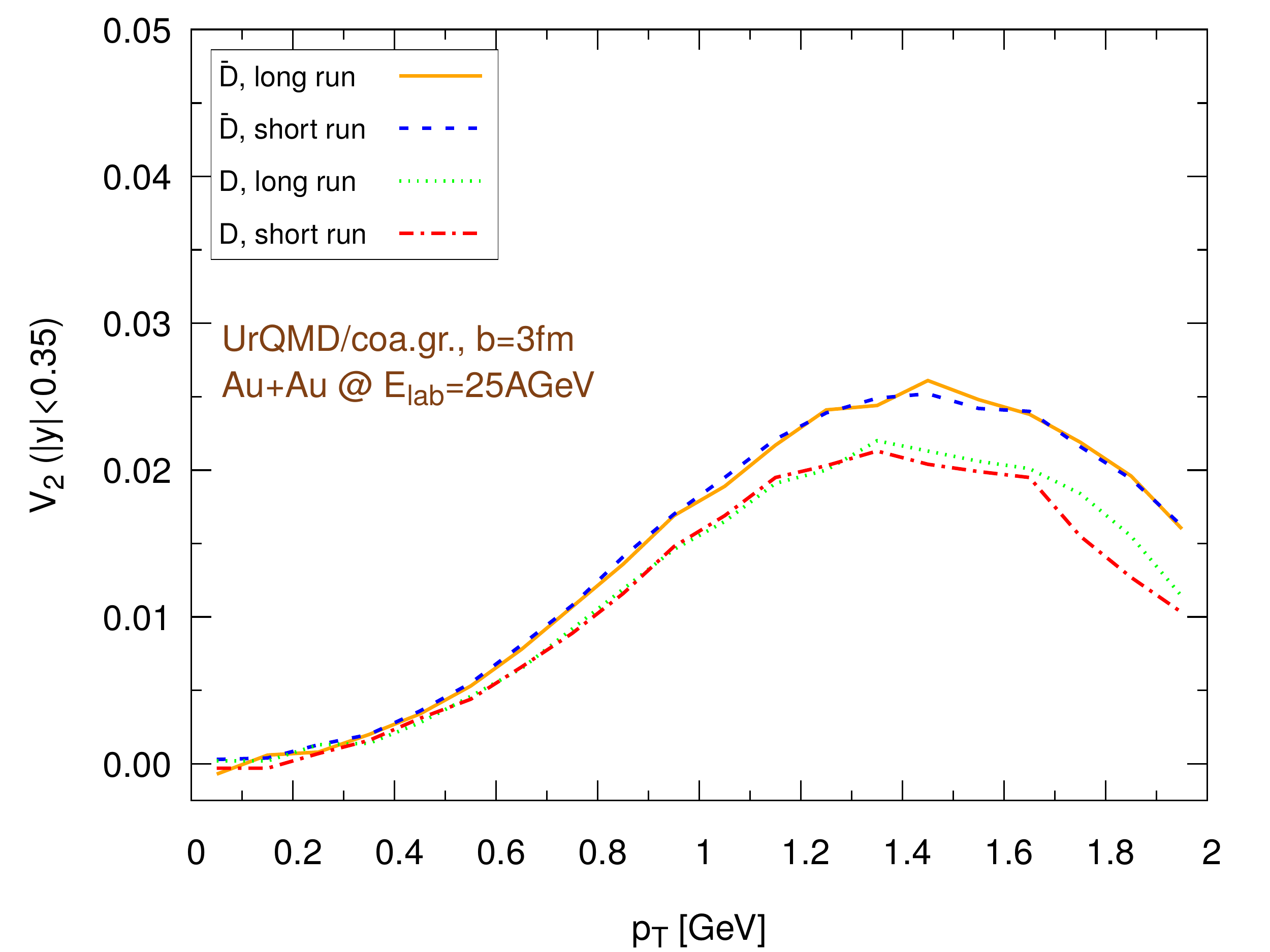}
	\end{minipage} \hspace{3mm}
	\begin{minipage}[b]{0.48\textwidth}
		\includegraphics[width=1\textwidth]{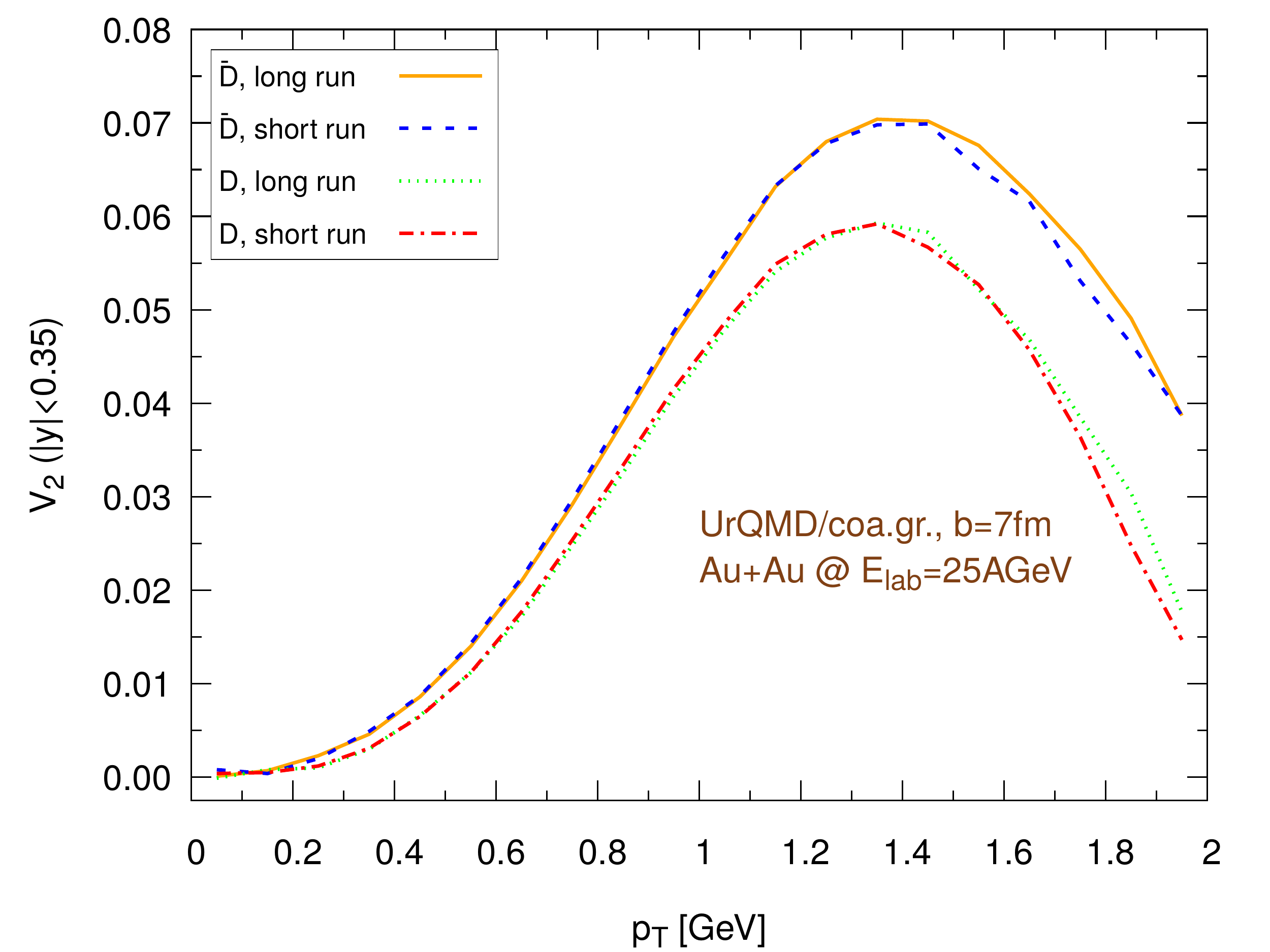}
	\end{minipage}
	\caption{(Color online) Au+Au collisions at
          $E_{\text{lab}}=25\,\AG$ in the UrQMD/hybrid model. The hadronization
          parameters are $\epsilon_p=0.05$ and
          $\erw{r_{\D_{\mathrm{rms}}}}=0.6\,\textrm{fm}$. Left:
          $b=3\,\fm$, right: $b=7\,\fm$. Comparison between the elliptic
          flow of $\D$ mesons ($|y|<0.35$) within the coarse-graining
          approach at two different final times: $75\,\fm$ (\emph{long
            run}) and $22\,\fm$ (\emph{short run}, $b=3\,\fm$ ) or
          $19\,\fm$ (\emph{short run}, $b=7\,\fm$).}
	\label{v2-cg-ts}
\end{figure*}
\subsection{The influence of the late hadronic phase} 

We recall that the final times in hybrid and coarse-graining approach
are different: the condition to stop hydrodynamics (at maximum energy
density of $0.3\varepsilon_0\approx44\,\MeV/\fm^3$) is reached at
$\approx22\,\fm$ for $b=3\,\fm$ collisions and at $\approx19\,\fm$ for
$b=7\,\fm$ collisions, while the coarse-graining approach ends at
$75\,\fm$.  It is important to stress that, since the hydro stopping temperature corresponding to $44\,\MeV/\fm^3$ is lower than $T_c$, the UrQMD/hybrid model always includes a hadronic phase, yet this is considerably shorter than in the UrQMD/coarse-graining approach. To evaluate the impact of this prolongated hadronic phase in the latter case, we
repeat the $T_c=145\,\MeV$ coarse-graining simulations at $E_{lab}=25\,\AG$, with hadronization parameters $\epsilon_p=0.05$ and
$\erw{r_{\D_{\mathrm{rms}}}}=0.6\,\textrm{fm}$, stopping
them at the time of the average hydro ending time, i.e. $22\,\fm$ for
$b=3\,\fm$ collisions and $19\,\fm$ for $b=7\,\fm$ collisions. We evaluate the elliptic flow of $\D$ and $\bar{\D}$ mesons at mid-rapidity, plotted in Fig.~(\ref{v2-cg-ts}) both for $b=3\,\fm$ (left) and for $b=3\,\fm$ (right). In Fig.~(\ref{v2-cg-ts}) the \emph{long run} labels refer to simulations until $t=75\,\fm$, while the \emph{short} label refer to simulations terminated at $22\,\fm$ (left) or $19\,\fm$ (right).
We can notice how the elliptic flow remains basically the same, in both centrality classes and both for $\D$ and $\bar{\D}$ mesons, except for small statistical fluctuations for $p_\mathrm{T}\gsim1.3\,\GeV$. This means that the late hadronic phase does not alter the $\D/\bar{\D}$ distributions. This outcome confirms the expectations, because the transport coefficients for $\D$
mesons are very small at low temperature, which in turn means that the
$\D$ mesons approach free streaming.
 
\subsection{The impact of the hadronization procedure}\label{hadro_procedure}

\begin{figure*}[h]
	\begin{minipage}[b]{0.48\textwidth}
		\includegraphics[width=1\textwidth]{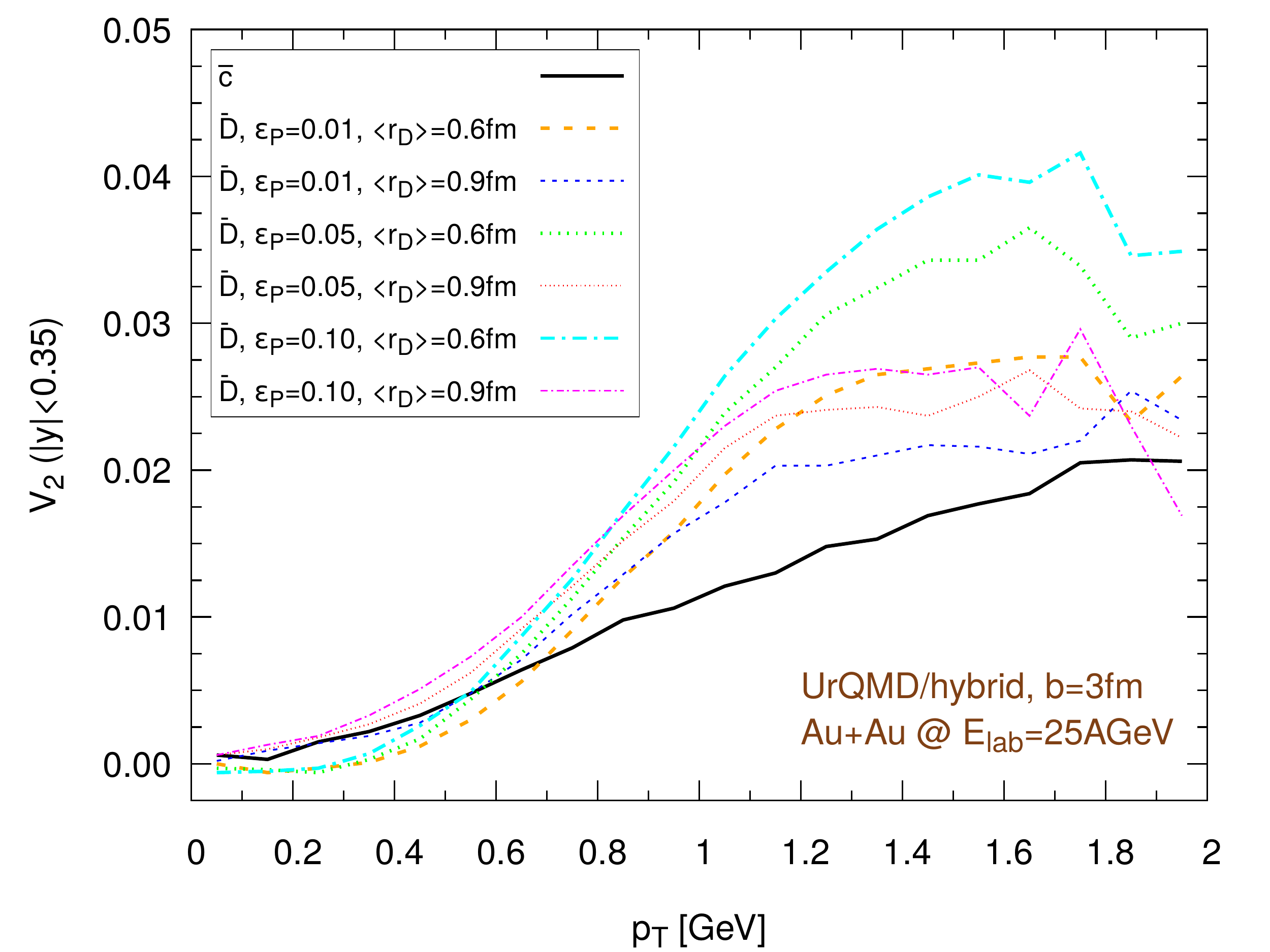}
	\end{minipage} \hspace{3mm}
	\begin{minipage}[b]{0.48\textwidth}
		\includegraphics[width=1\textwidth]{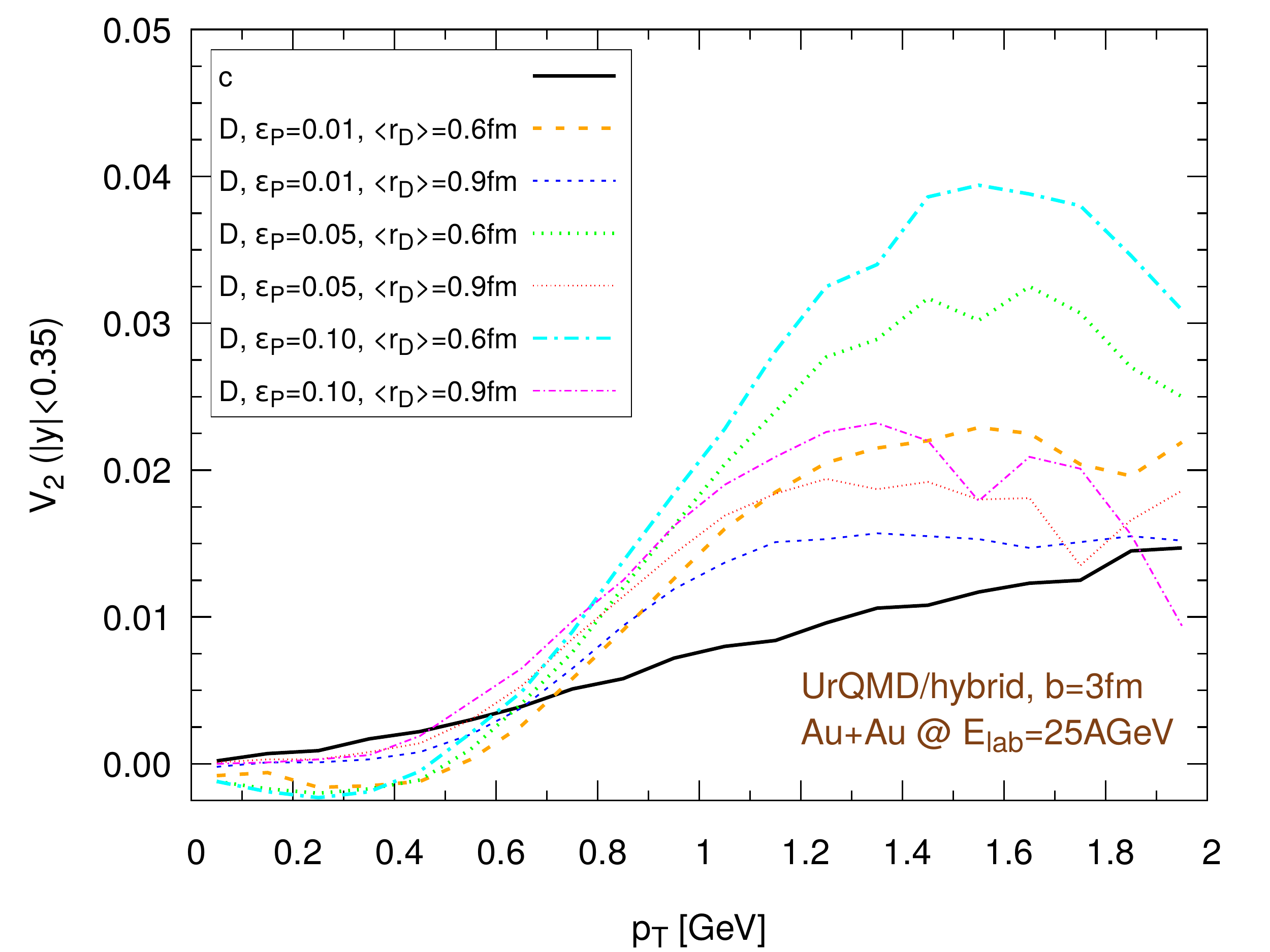}
	\end{minipage}
	\caption{(Color online) Au+Au collisions at
          $E_{\text{lab}}=25\,\AG$, $b=3\,\fm$ in the UrQMD/hybrid
          model. Elliptic flow of charm quarks and D-mesons
          ($|y|<0.35$).  We explore the effect of different choices of
          the hadronization parameters, by performing a single
          hadronization process, without further hadronic propagation in
          the medium. Left: $\bar{c}$ quarks and $\bar{\D}$ mesons,
          right: $c$ quarks and $\D$ mesons.}
	\label{hadro-hy-b3}
\end{figure*}

\begin{figure*}[h]
	\begin{minipage}[b]{0.48\textwidth}
		\includegraphics[width=1\textwidth]{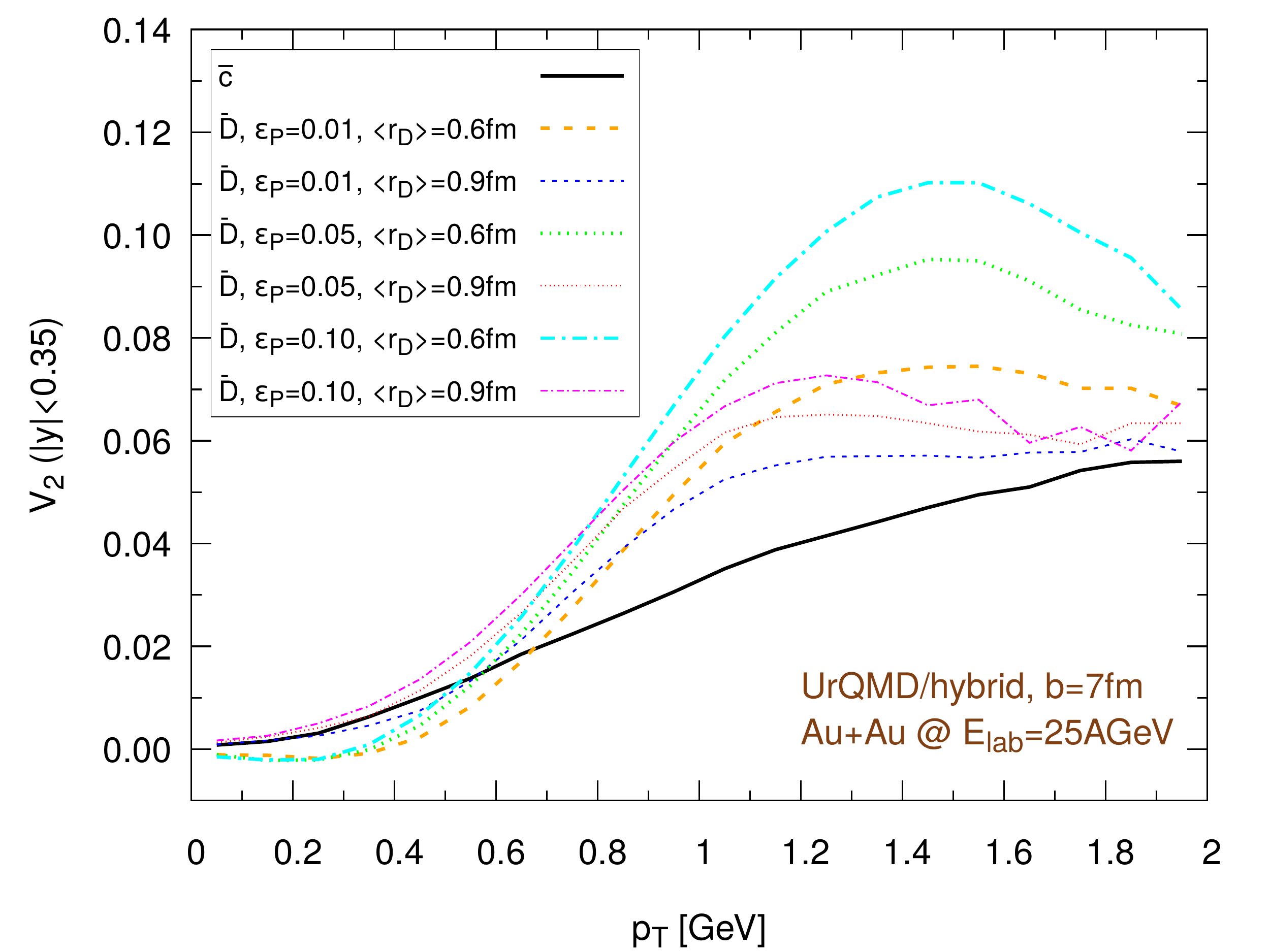}
	\end{minipage} \hspace{3mm}
	\begin{minipage}[b]{0.48\textwidth}
		\includegraphics[width=1\textwidth]{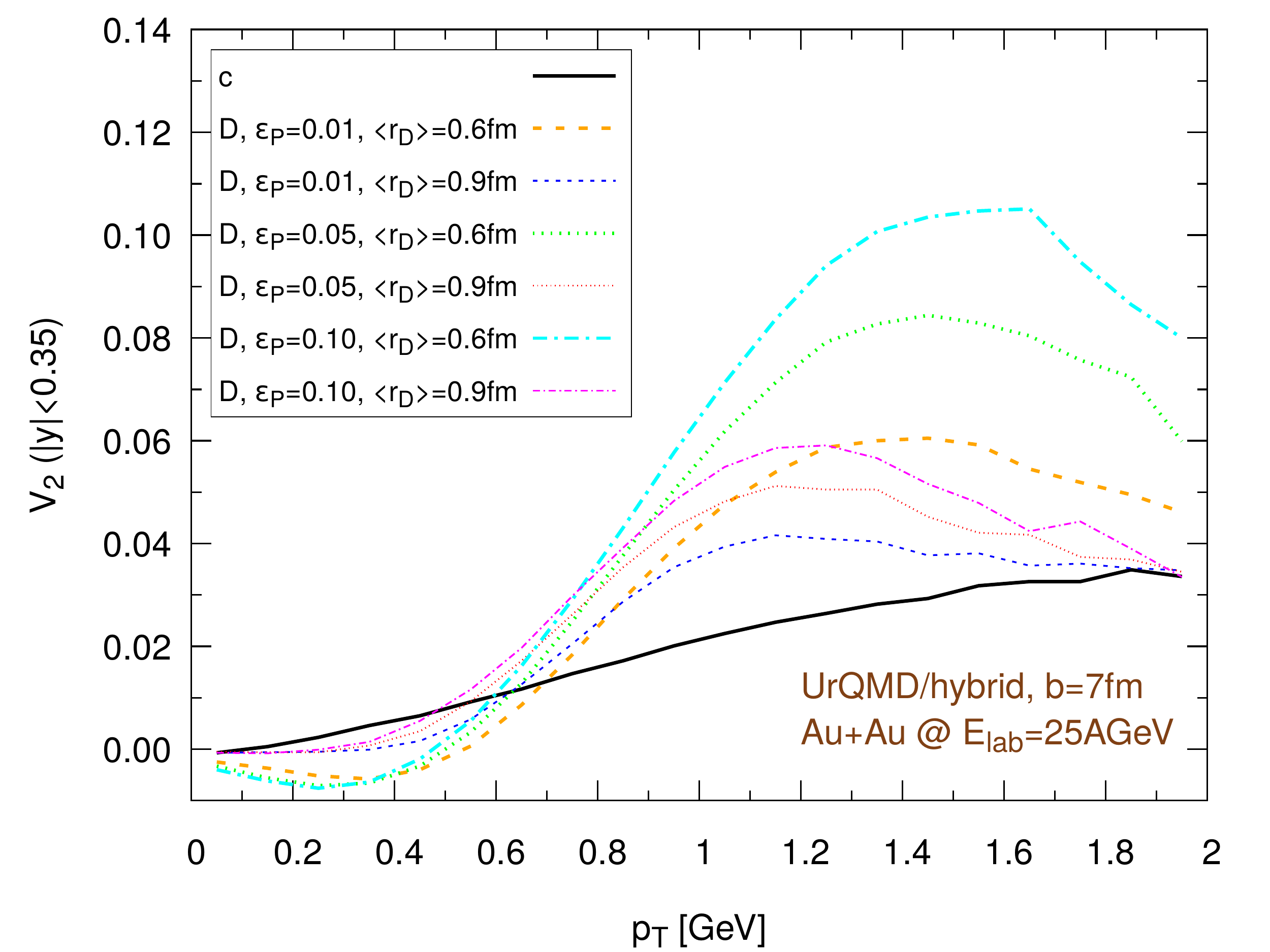}
	\end{minipage}
	\caption{(Color online) Au+Au collisions at $E_{\text{lab}}=25\,\AG$,
$b=7\,\fm$ in the UrQMD/hybrid model. Elliptic flow of charm quarks and
D-mesons ($|y|<0.35$).  We explore the effect of different choices of
the hadronization parameters, by performing a single hadronization
process, without further hadronic propagation in the medium. Left:
$\bar{c}$ quarks and $\bar{\D}$ mesons, right: $c$ quarks and $\D$
mesons.}
	\label{hadro-hy-b7}
\end{figure*}

\begin{figure*}[h]
	\begin{minipage}[b]{0.48\textwidth}
		\includegraphics[width=1\textwidth]{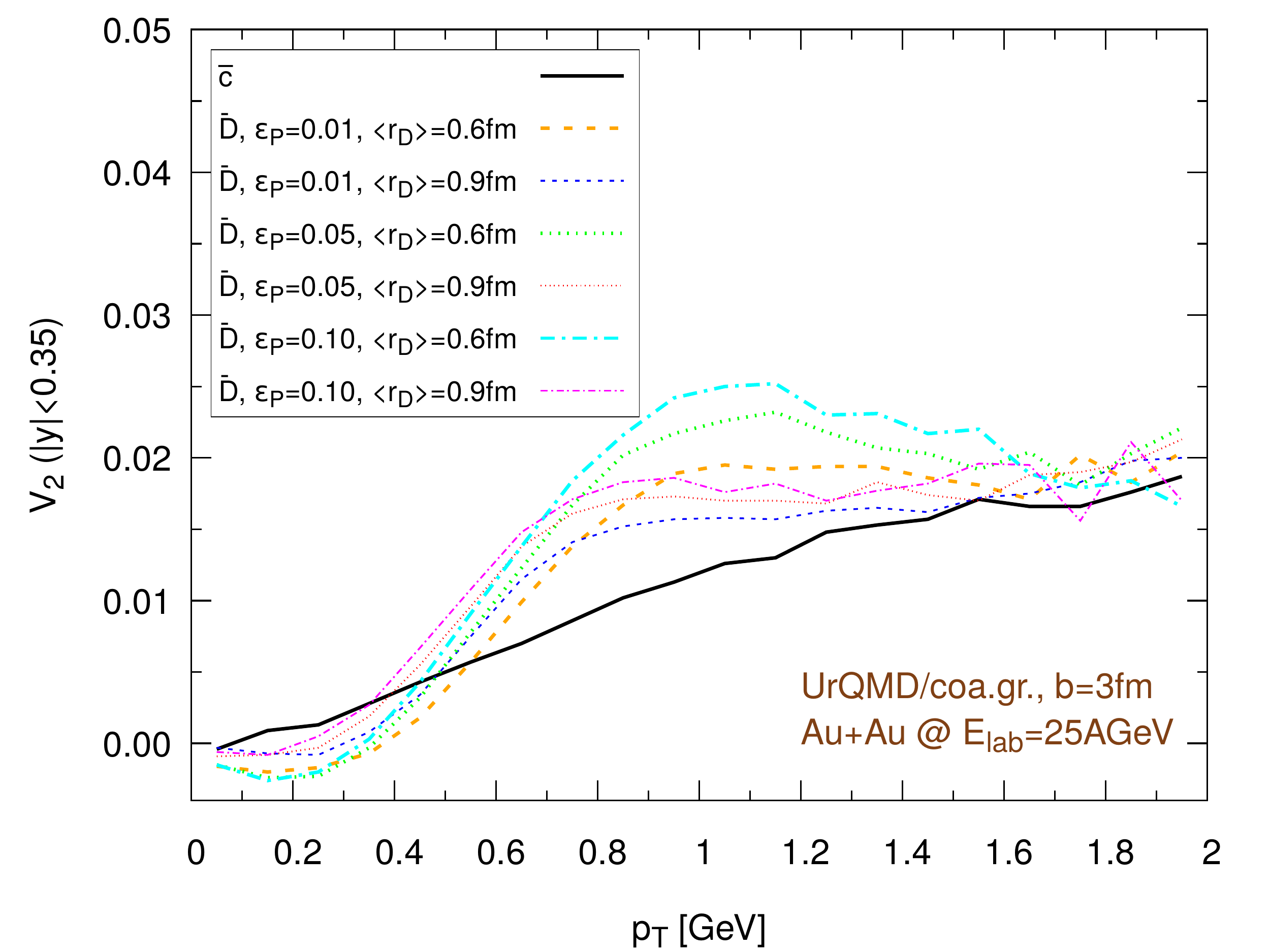}
	\end{minipage} \hspace{3mm}
	\begin{minipage}[b]{0.48\textwidth}
		\includegraphics[width=1\textwidth]{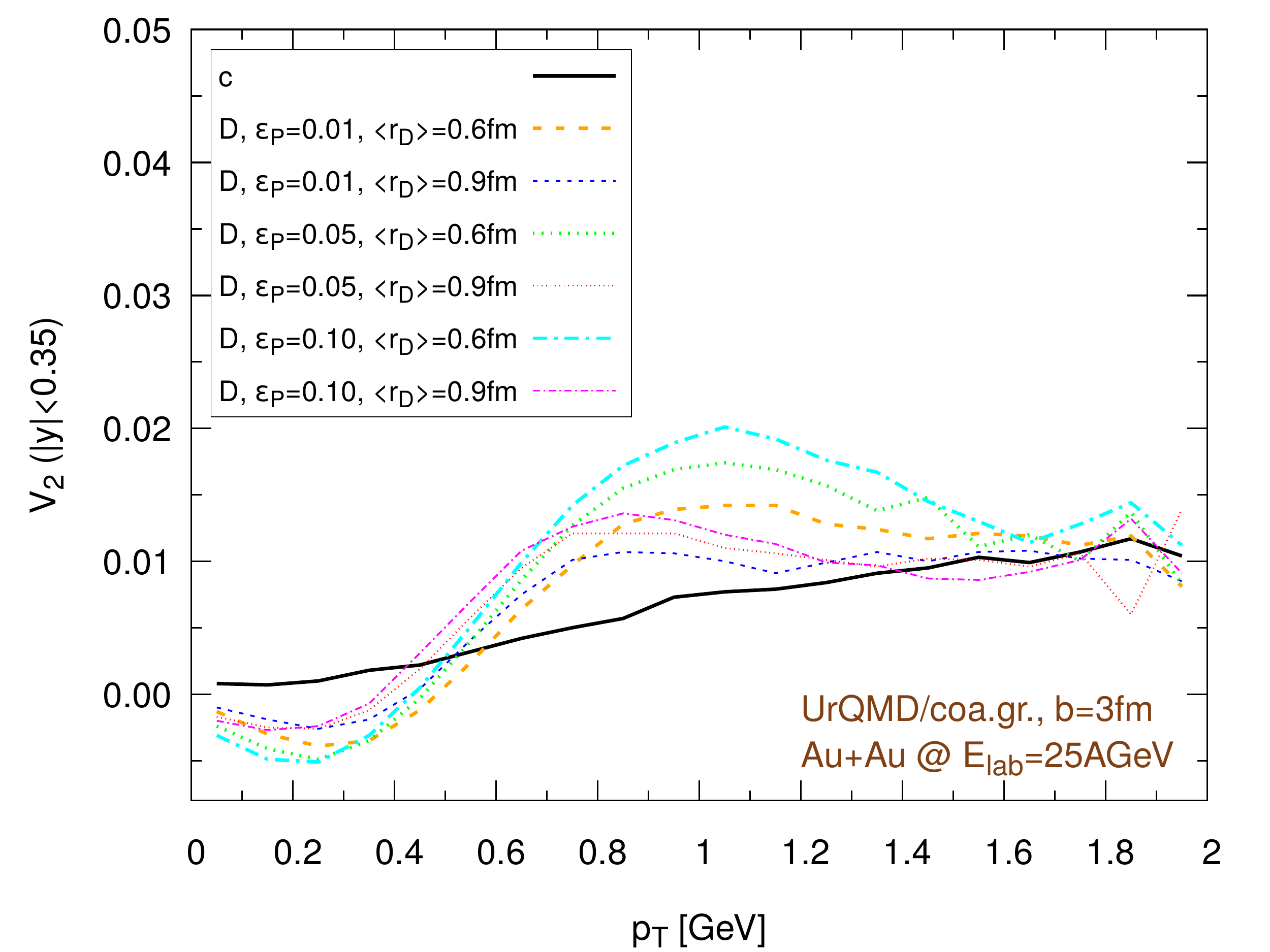}
	\end{minipage}
	\caption{(Color online) Au+Au collisions at
          $E_{\text{lab}}=25\,\AG$, $b=3\,\fm$ in the
          UrQMD/coarse-graining approach. Elliptic flow of charm quarks
          and D-mesons ($|y|<0.35$).  We explore the effect of different
          choices of the hadronization parameters, by performing a
          single hadronization process, without further hadronic
          propagation in the medium. Left: $\bar{c}$ quarks and
          $\bar{\D}$ mesons, right: $c$ quarks and $\D$ mesons.}
	\label{hadro-cg-b3}
\end{figure*}

\begin{figure*}[h]
	\begin{minipage}[b]{0.48\textwidth}
		\includegraphics[width=1\textwidth]{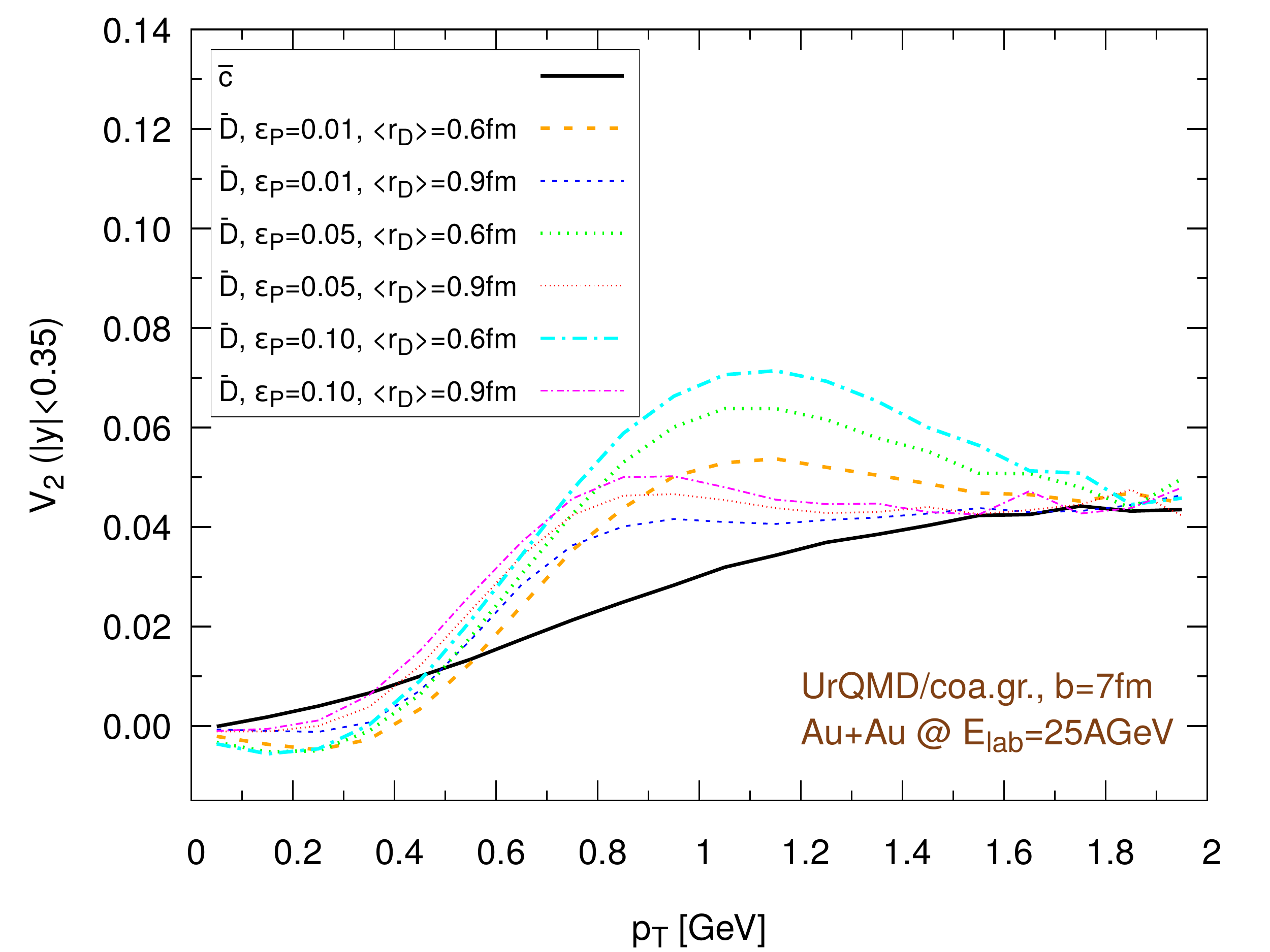}
	\end{minipage} \hspace{3mm}
	\begin{minipage}[b]{0.48\textwidth}
		\includegraphics[width=1\textwidth]{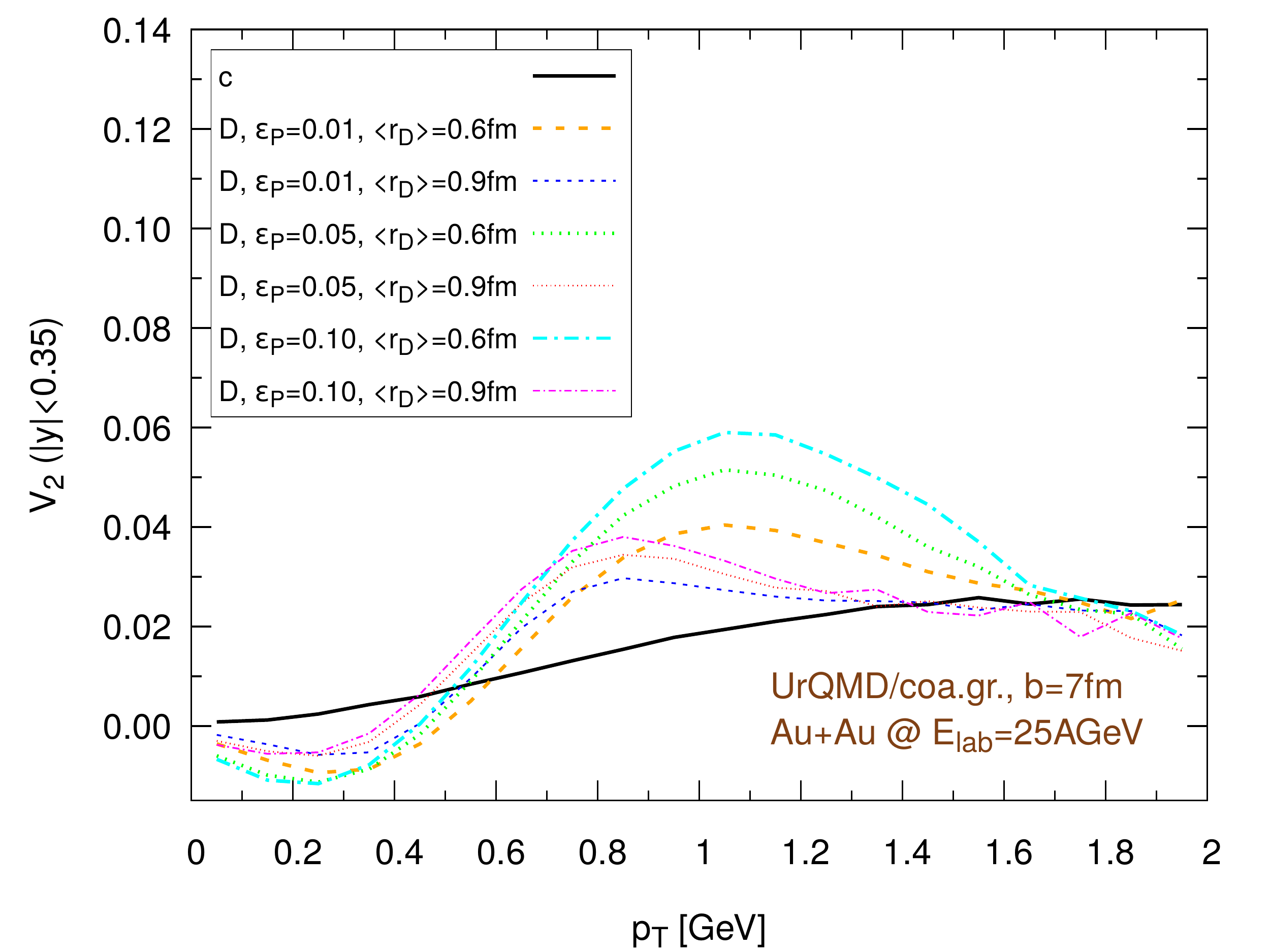}
	\end{minipage}
        \caption{(Color online) Au+Au collisions at
          $E_{\text{lab}}=25\,\AG$, $b=7\,\fm$ in the
          UrQMD/coarse-graining approach. Elliptic flow of charm quarks
          and D-mesons ($|y|<0.35$). We explore the effect of different
          choices of the hadronization parameters, by performing a
          single hadronization process, without further hadronic
          propagation in the medium. Left: $\bar{c}$ quarks and
          $\bar{\D}$ mesons, right: $c$ quarks and $\D$ mesons. }
	\label{hadro-cg-b7}
\end{figure*}

To assess the contribution of the partonic phase and the impact of the
hadronization procedure on the flow, we perform the propagation of charm
quarks until they reach for the first time a cell with temperature
$T=T_c=145\,\MeV$, then, without any further interaction with the
medium, we hadronize the charm quarks. We further explore the effects of
different values of the mean radius of the $\D$ mesons
$\erw{r_{\D_{\mathrm{rms}}}}$ ($0.6\,\fm$ and $0.9\,\fm$) and the Peterson
fragmentation parameter $\epsilon_p$ ($0.01,0.05,0.1$). We recall that the assumptions on the size of the $\D$ mesons play an important role in determining the probability of hadronization by coalescence or fragmentation, so different choices of $\erw{r_{\D_{\mathrm{rms}}}}$ correspond to different contributions of these two hadronization methods to $\D$ meson formation. The results, for Au+Au collisions at $E_{\text{lab}}=25\,\AG$,  are
shown in figures~(\ref{hadro-hy-b3}-\ref{hadro-cg-b7}). More precisely, the results of the UrQMD/hybrid model are shown in Fig.~(\ref{hadro-hy-b3}) for collisions at impact parameter $b=3\,\fm$ and in Fig.~(\ref{hadro-hy-b7}) for collisions at $b=7\,\fm$. The results of the UrQMD/coarse-graining approach are shown in Fig.~(\ref{hadro-cg-b3}) for collisions at $b=3\,\fm$ and in Fig.~(\ref{hadro-cg-b7}) for collisions at $b=7\,\fm$. All figures show the elliptic flow of quarks (solid black lines) at the moment of hadronization and of $\D$ mesons (colored dashed lines) immediately after their formation. The left figures refer to $\bar{c}$ quarks and $\bar{\D}$ mesons, the right figures to $c$ quarks and $D$ mesons. As an expected general trend, the $v_2$ of anti-particles is greater than the $v_2$ of particles. We observe that
most of the flow is built during the partonic phase, a behavior
consistent with the larger values of the transport coefficients at high
temperatures. In addition, the difference in the magnitude of the flow
between the hydro and the coarse-graining approach is clearly visible
even at this stage. This implies that the use of the UrQMD/hybrid model down to temperatures at the limits of QGP existence is not the main responsible of the larger elliptic flow obtained in this model compared to the UrQMD/coarse-graining approach. Therefore, the suspect of an overestimation of $v_2$ due to a misuse of hydrodynamics is strongly reduced. Finally, in all cases, the elliptic flow grows with increasing values of $\epsilon_p$ and it is larger for smaller values of the $\D$ meson radius.\\
It is clear that the details of the hadronization process have a very large impact on the final results, therefore special attention must be paid to a proper treatment of this step in future works. To begin, the probability distribution in Eq. (\ref{prob_to_coalescence}) seems to overestimate of the probability to hadronization by fragmentation with the current choice of the $\D$ meson radius, which might lead to wrong results, in particular when taking into account the formation of resonances with larger radii, especially if the dependence on the mutual spatial distance between the light and the heavy quark was also included\cite{Greco:2003vf,Song:2015sfa}. Apart for an extensive and deep re-checking of the whole procedure and its implementation in the code to better understand the origin of the apparently small percentage of hadronization by coalescence, we might replace Eq. (\ref{prob_to_coalescence}) with a tabulated probability distribution obtained from full transport model simulations. Another possibility might be the adoption of a probability distribution which depends on the module of the relative velocity $|v_r|$ between the heavy quark and the fluid cell, i.e. something like $f(|v_r|)=\exp(-|v_r|/\alpha)$, with $\alpha$ determined by a fit with the elliptic flow measured in experiments at comparable collision energies. In addition, to be consistent with the assumptions made for the computation of the drag and diffusion coefficients in the partonic phase, we should go beyond the naive assumption of instantaneous hadronization and decoupling processes by introducing some probability function depending not only on temperature and chemical potential, but also explicitly on time. Indeed, the survival of $\D$ mesons in the Quark Gluon Plasma might lead to a reduction of the predicted elliptic flow. Then, we should consider the probable formation of intermediate excited states and we should try to constrain the estimates of the $\D$ meson radius, possibly making it also temperature dependent\cite{Satz:2005hx}.  Moreover, we should try to improve the fragmentation process either by constraining the Peterson fragmentation parameter\cite{Chekanov:2008ur} or by adopting other fragmentation models\cite{Bowler:1981sb}, which in some cases have shown a better capability to reproduce the features of experimental data\cite{PhysRevD.73.032002}. Further refinements might include medium modified\cite{Arleo:2008dn} and unfavored\cite{Maciula:2017wov} fragmentation functions. However, unfortunately, at the moment we miss well determined values of fragmentation functions for $\D$ mesons in the low collision energy regime based on robust experimental data. Regarding the coalescence mechanism, the method itself is quite standard and the flow contribution to the final momentum of the open heavy meson is derived from the reliable UrQMD model, therefore the uncertainties are somehow reduced compared to the fragmentation mechanism. Nevertheless, although in this study we did not explore the consequences of different assumptions, the results depend on the estimates of the masses of the constituent quarks, which indirectly enter also in Eq. (\ref{prob_to_coalescence}), therefore, even in this case, different educated choices of the parameters might alter the current predictions.


\section{Discussions and conclusion} 

In this paper we have presented results on $\D$ and $\bar{\D}$ meson
spectra and elliptic flow for Au+Au reactions at
$E_{\text{lab}}=25\,\AG$. These calculations are relevant for the
upcoming FAIR and NICA facilities and for the RHIC BES program. We have used
Pythia\cite{Sjostrand:2014zea,Sjostrand:2006za} to obtain a sample of
correlated charm and anti-charm quarks, then we let the charm quarks
propagate in the medium produced by heavy ion collisions, both in the
partonic and in the hadronic phase, adopting a Langevin approach. In
particular, we have studied Au+Au collisions at two different
centralities, $b=3\,\fm$ and $b=7\,\fm$. The background medium is
modeled either with the UrQMD hybrid model or with the UrQMD coarse
graining approach. The effect of the finite baryon chemical potential is
taken into account in the evaluation of the transport coefficients. The
effect of different hadronization parameters is explored. We have shown
that even at low collision energies the interaction with the medium
produces a sizeable final $\D$ meson elliptic flow, which is larger for
more peripheral collisions. A lower decoupling temperature leads to an
increase of the elliptic flow. This implies that the interaction with
the medium is stronger during the partonic than during the hadronic
phase. This hypothesis is also confirmed by the magnitude of the
elliptic flow of charm quarks immediately before hadronization. The
impact of the later hadronic phase is shown to play a minor
role. One should note that the results are very sensitive to the details
of the hadronization mechanism, i.e. on the probability to hadronize
through coalescence or through Peterson fragmentation and to the choice
of the parameters in each hadronization channel. 

Our study confirms that even at low collision energies the charm quarks
can be an invaluable tool to probe the properties of the
QCD-medium. Nevertheless, there are shortcomings in the present
approach: (I) we rely on Pythia with default \emph{SoftQCD} mode settings to
produce the initial charm-quark momentum distribution in p-p
collisions, but maybe a fine tuning of the settings might produce
noticeable differences. Unfortunately, common models and tools like
FONLL\cite{Cacciari:1998it,Cacciari:2001td,Cacciari:2012ny,Cacciari:2015fta}
or HERWIG\cite{Bellm:2015jjp}, strongly based on pQCD, are not very
reliable in this low-energy range. (II) In the FAIR-energy regime, we
miss one of the main advantages of studying heavy flavors, i.e. precise
pQCD based predictions of the charm-quark initial states, mentioned also
in the introduction. (III) In principle the
coarse-graining approach would allow us to start the Langevin
propagation earlier than in the hydro case, resulting in a clear
improvement of the naive assumption of no interaction at all until full
thermalization. Moreover, we should also introduce a time delay before the spatial separation of the $c - \bar{c}$ couple after its formation is large enough to be considered a ``colored'' object. Since the results obtained so far point toward a major
role of the early dynamics of the system, it is definitely very important to develop a more realistic treatment of this stage. (IV) The hydro model might be improved by taking into account viscous effects, which are not
completely negligible at low collision energies, and possibly
anisotropic hydrodynamics, which would allow to slightly anticipate the
propagation even in the hydro case. (V) To partially take into account the hadronic interactions, in the version of the UrQMD/hybrid model adopted in this work we stop the simulations at temperatures slightly below $T_c$, when, in principle, the fluid description of the medium should be replaced by a transport model, like in the standard UrQMD/hybrid model. We might improve this situation by restoring the full UrQMD/hybrid approach, but neglecting the back-reactions of the $\D$ mesons on the other particles during their mutual interactions. This strategy would provide a more realistic modeling of the hadronic phase, while preserving the possibility of oversampling the $\D$ mesons, which is an almost essential condition to collect a sufficient statistics in an energy regime quite close to the $c-\bar{c}$ production threshold. 
(VI) Another very important limitation of our model
is the hadronization method. As we discussed in Sect. (\ref{hadro_procedure}), here further improvements of the
fragmentation function for low momenta and on the coalescence model are strongly
desired. (VII) We limited our study to $\D^{\pm}$, $\D^0$ and $\bar{\D}^0$ mesons, however, in a more comprehensive study, excited states and strange $\D$ mesons should be included as well.

To conclude, the work that we just presented provides useful
indications about the direction in which further and more refined
studies should focus. Despite their low production rate, the study of
the elliptic flow of charmed mesons carries a wealth of information
about the QGP and the QCD also in the FAIR energy range.

\section*{Acknowledgments} 
We thank the Referee for helping us in improving the robustness of our results and the clarity of their presentation in this article. We gratefully acknowledge Thomas Lang for providing part of the
numerical code which served as a basis for the present work. We also
thank Jan Steinheimer for useful discussions and
suggestions. G. Inghirami was supported by a GSI grant in cooperation
with the J. von Neumann Institute for Computing; he also gratefully
acknowledges support from the H-QM and HGS-HIRe graduate schools. We
thank Laura Tol{\'o}s for providing the meson-baryon scattering amplitudes
to compute the D-meson transport coefficients. J.\ M.\ Torres-Rincon
acknowledges support from US Department of Energy under Contract
No. DE-FG-88ER40388. The computational resources were provided by the
Center for Scientific Computing (CSC) of the Goethe University Frankfurt
and by the Frankfurt Institute for Advanced Studies (FIAS). This work was
supported by the European Cooperation in Science and Technology COST Action CA15213 (THOR).
\bibliography{bibliography}
\end{document}